\documentclass[12 pt]{article}
\usepackage{bbm}
\usepackage{natbib}
\usepackage{amssymb,amsmath,mathtools}
\usepackage{amsthm}
\usepackage[onehalfspacing]{setspace}
\usepackage[colorlinks]{hyperref}
\usepackage{graphicx}
\usepackage{enumitem}
\usepackage{verbatim}
\usepackage{sectsty}
\usepackage[svgnames]{xcolor}
\usepackage{subfiles}
\usepackage[capitalize,nameinlink]{cleveref}
\usepackage[colorinlistoftodos,textsize=tiny]{todonotes}
\usepackage{xcolor}
\usepackage{pgf}
\usepackage{tikz}
\usepackage{xargs}
\usepackage[font={small,it},justification=justified]{caption}
\usepackage[normalem]{ulem}
\usepackage{comment}
\usepackage[margin=.9in]{geometry}
\usepackage{pgfplots}

\def\IntKern{}

\newcommand\Symbolint[3]{%
  {\setbox0=\hbox{$#1{#2#3}{\int}$ }%
  \vcenter{\hbox{$#2#3$ }}\kern\IntKern}
}

\usepackage{subcaption}

\hypersetup{
    pdftitle={Cherry-Picking},
    pdfauthor={Nageeb Ali, Max Mihm, Lucas Siga},
    citecolor=DarkBlue,
    bookmarksnumbered=true,
    urlcolor=Indigo,linkcolor=DarkBlue
}
 \theoremstyle{definition}
 \newtheorem{example}{Example}
\let\exampleWithoutMarker\example
\let\endexampleWithoutMarker\endexample
\renewenvironment{example}{%
  \pushQED{\qed}%
  \exampleWithoutMarker
}{\nobreak\popQED\endexampleWithoutMarker}
 
 \theoremstyle{remark}

\theoremstyle{plain}
\newtheorem{theorem}{Theorem}

\newtheorem{assumption}{Assumption}

\newtheorem{claim}{Claim}

\newtheorem{corollary}{Corollary}

\newtheorem{definition}{Definition}

\newtheorem{lemma}{Lemma}

\newtheorem{proposition}{Proposition}

\renewcommand{\equiv}{:=}

\renewcommand{\epsilon}{\varepsilon}
\newcommandx{\ali}[2][1=]{\todo[linecolor=blue,backgroundcolor=blue!25,bordercolor=blue,#1]{#2}}
\newcommandx{\siga}[2][1=]{\todo[linecolor=red,backgroundcolor=red!25,bordercolor=red,#1]{#2}}  
\newcommandx{\mihm}[2][1=]{\todo[linecolor=green,backgroundcolor=green!25,bordercolor=green,#1]{#2}}

\usetikzlibrary{shapes,arrows,automata,fit}

\tikzstyle{info}=[circle,thick,draw=black,fill=black!25,minimum size=4mm]
\tikzstyle{uninfo}=[circle,thick,draw=black,fill=white,minimum size=4mm]
\tikzstyle{inforecog}=[circle,line width=1mm,draw=black!50,fill=black!25,minimum size=4mm]
\tikzstyle{uninforecog}=[circle,line width=1mm,draw=black!50,fill=white,minimum size=4mm]

\tikzstyle{traded}=[draw, line width=1mm]
\tikzstyle{recog}=[draw=black!50, line width=1mm]

\sectionfont{\color{DarkRed}}
\subsectionfont{\color{DarkRed}}
\subsubsectionfont{\color{DarkRed}}
\crefname{assumption}{assumption}{assumptions}
\newcommand{\appendixref}[1]{\hyperref[#1]{Appendix}}
\newcommand{\suppappendixref}[1]{\hyperref[#1]{Supplementary Appendix}}
\DeclareMathOperator*{\argmax}{arg\,max}

\newcommand{\policy}{\mathcal X}
\newcommand{\congruence}{\policy^+}
\newcommand{\conflict}{\policy^{-}}
\newcommand{\rb}{t}
\newcommand{\info}{\gamma}
\newcommand{\uninf}{\alpha}
\newcommand{\E}{\mathbbm{E}}
\newcommand{\1}{\mathbbm{1}}
\newcommand{\odds}{\mathcal{O}}
\newcommand{\interior}{\text{\normalfont{int}}}
\newcommand{\payoff}{\mathcal P}
\newcommand{\game}{\mathcal G}
\renewcommand{\Re}{\mathbb R}
\newcommand{\supp}{\operatorname{supp}}
\newcommand{\first}{X^{(1)}}
\newcommand{\second}{X^{(2)}}
\newcommand{\breadth}{b}
\newcommand{\basket}{\mathcal B}
\renewcommand{\emptyset}{\varnothing}
\newcommand{\cl}{\operatorname{cl}}

\usepackage{bbm}
\usepackage[english]{babel}
\usepackage{mathtools}
\usepackage{cancel}
\usepackage{multicol}
\usepackage{setspace,xcolor}
\usepackage{color}
\usepackage{comment}

\definecolor{Rust}{RGB}{180,30,43}
\definecolor{Olive}{RGB}{115,126,46}
\definecolor{Aqua}{RGB}{32,129,126}
\definecolor{Forest}{RGB}{75,140,80}
\definecolor{mycolor}{named}{Aqua}

\begin{document}

\begin{titlepage}

\title{The Cherry-Picking Trap\thanks{We thank Arjada Bardhi, Wioletta Dziuda, Navin Kartik, Andreas Kleiner, Benny Moldovanu, Juan Ortner, Mike Powell, Boli Xu, Kun Zhang, Gabriel Ziegler, and various conference and seminar participants.}}
\author{
S. Nageeb Ali\thanks{Department of Economics, Penn State. Email: \href{mailto:nageeb@psu.edu}{nageeb@psu.edu}.}
\and
Maximilian Mihm\thanks{Division of Social Science, NYU-Abu Dhabi. Email: \href{mailto:max.mihm@nyu.edu}{max.mihm@nyu.edu}.}
\and 
Lucas Siga\thanks{Department of Economics, University of Essex. Email: \href{mailto:lucas.siga@essex.ac.uk}{lucas.siga@essex.ac.uk}.}
}
\date{\today}
\maketitle

\bigskip

\begin{abstract}

When a better-informed proposer puts forward a policy proposal, the responder should not take the proposal at face value but ask why it was chosen. We model how this inference can undermine trust even when interests are broadly aligned. A proposer privately observes a ``basket'' of viable policies and proposes one; a responder decides whether to approve the proposal but may lack the expertise to evaluate it. A larger basket makes a Pareto improvement more likely to be available but also gives the proposer greater scope to cherry-pick a policy that favors him at the responder's expense. In the unique equilibrium, Pareto improvements are often proposed and yet rationally rejected. Flexibility that should enable agreement instead breeds distrust.

\end{abstract}
\thispagestyle{empty} 

\end{titlepage}
\newpage

\section{Introduction}

Many political and organizational decisions require one party to approve a proposal put forward by another who is better informed. Legislatures vote on bills drafted behind closed doors, regulators assess projects whose details are difficult to evaluate, CEOs evaluate recommendations from division managers who better understand local conditions, and voters decide on ballot measures drafted by interest groups. While interests may be largely aligned, they need not be perfectly so. The party choosing to approve or reject the proposal, the responder, must then decide whether to trust the proposer. Should she believe that the proposer used his expertise to identify a policy that benefits them both or worry that he cherry-picked a policy that benefits him alone? The responder should not evaluate any proposal in isolation; she should instead reason from the fact that the proposer strategically selected \emph{this} proposal from a broader set of feasible alternatives. We study how this inference itself can undermine trust: the responder may rationally reject the proposal out of fear that it has been cherry-picked. 

This fear manifests in politics. A striking illustration is the US Senate's rejection of the Treaty of Versailles in 1919. Senators' objections went beyond the substance of the treaty to how it was constructed and presented. One objection was that Woodrow Wilson had crafted the treaty with foreign powers without consulting the Senate. A second objection centered on Article X, which committed member countries to come to one another's aid. Critics suspected an ulterior motive: the clause was not a commitment to collective security but a strategic ploy to transfer authority from Congress to the executive branch.  Wilson's insistence that Article X constituted a moral rather than a legal obligation did little to assuage the Senate's concerns, and his refusal to rewrite the treaty deepened suspicion about what the text was designed to accomplish. In his speech against ratification, Henry Cabot Lodge argued that the unwillingness of the treaty's authors to use plain language to describe this obligation made it ``impossible to avoid the conclusion that they are seeking to do by indirection and the use of nebulous phrases what they are not willing to do directly\ldots.'' 

A century later, similar fears about ulterior motives contributed to the demise of Theresa May's Brexit deal, rejected thrice by the UK Parliament. Many MPs did not take the agreement at face value; they reasoned instead from the fact that \textit{this} was the agreement the EU had chosen to offer. Features such as the absence of a unilateral UK exit right and the sequencing of withdrawal before the future relationship were seen as evidence that the proposal had been strategically crafted to preserve EU leverage in subsequent negotiations.    

Such tensions also arise within organizations. Workers and unions often reject deals that freeze wages during austerity, suspecting that management is protecting executive compensation while shifting costs to labor. In merger negotiations, target boards may reject acquisition deals that they suspect transfer risk asymmetrically to the target's shareholders. A CEO may reject a division manager's investment proposal, suspecting that the manager passed over projects that benefit the company in favor of one that expands his empire. A firm may be wary of a supplier's redesign that is billed as cost-saving but may compromise quality.

Underlying the distrust in these examples are two frictions. First, the proposer privately knows which policies are viable, and is able to choose one strategically. Greater flexibility, in terms of more policies being viable, is a double-edged sword because it raises the odds that some Pareto improvement is viable but also gives the proposer greater scope to cherry-pick. A second friction is that the responder may be unable to evaluate the proposed alternative, either because she lacks the relevant expertise or because some of its details are opaque. She would then have to trust that the proposer puts forward Pareto improvements. We identify conditions under which that trust becomes untenable.

In our model, a proposer observes a randomly drawn ``basket'' of policies and the payoffs they deliver to each party. We consider a general policy space that includes Nash bargaining problems, veto bargaining problems \citep{RR:78}, and organizational decisions as studied in the ``project-selection'' literature \citep{armstrong2010model,nocke2013merger,guo2023regret}. The proposer chooses one policy from the basket to propose. A responder then either approves the proposal or rejects it in favor of the status quo. Importantly, the responder may lack the relevant expertise to evaluate the proposal. With probability $\lambda\in (0,1)$, the responder is \textit{informed}:  she observes the payoffs both parties would receive from implementing the proposed policy; with the complementary probability, she is \textit{uninformed} and does not observe any detail. The uninformed responder is in a position similar to the buyer in \cite{akerlof1970market}'s market for lemons: she knows the distribution of policies and can reason about the proposer's selection process, but she is unable to verify whether the actual proposal on the table is a lemon or a peach. 

The game has a unique equilibrium. From each basket, the proposer selects either his favorite policy overall (``cherry-picking'')---hoping to exploit the uninformed responder's trust---or his favorite among Pareto improvements (``moderation''), which is  approved by the informed responder. He faces a dilemma when these policies differ. He is tempted to cherry-pick when the responder is unlikely to be able to evaluate the proposal. The uninformed responder faces a dilemma of her own: should she blindly approve a proposal knowing it was strategically selected? Or should she be wary?\footnote{An informed responder faces no dilemma, as she can judge a proposal on its merits.} We show that she must typically mix between approval and rejection. Were she to always approve, the proposer would be overly tempted to cherry-pick, which would make approval unwise. Were she to always reject, the proposer would always moderate (because cherry-picking would offer no reward), in which case she would do better to approve. Thus, in equilibrium, she must mix, and the proposer's distribution of proposals must render her indifferent.\footnote{In turn, her randomization steers the proposer toward the equilibrium mix of proposals but does not make him indifferent; for generic baskets, he has a strict best response.} This mixing damages welfare: in equilibrium, the proposer often moderates, and yet Pareto improvements are rationally rejected with high probability.

While the cherry-picking trap emerges even when the basket has only a few policies, it is particularly potent in ``large'' baskets. Here, the proposer nearly always moderates, putting forward Pareto improvements with probability converging to one, and yet the responder rejects those proposals with significant probability. Thus, inefficient status quo policies often remain in place even when mutually beneficial alternatives are available and proposed. %

In essence, the cherry-picking trap is a commitment problem. The proposer would often be better off if he could commit to proposing \textit{only} Pareto improvements, or if he could relinquish his ability to select proposals altogether. In that sense, his ability to cherry-pick ends up being self-defeating. Both players can also be better off if the responder is informed with higher probability. The reason is not that better information disciplines the proposer but that more of his proposals pass. This result highlights the value of bipartisan committees, transparency, and oversight. This suggests, for example, that had Woodrow Wilson involved the Senate in designing the Treaty of Versailles, senators might have been less suspicious of his intentions. This finding also rationalizes the significant efforts activists and industry lobby groups undertake to educate regulators, even  though doing so enhances regulators' ability to detect cherry-picked proposals. 

We compare equilibrium behavior to benchmarks in which either side can commit; both can be better off when either commits. If the responder commits, then as in \cite{armstrong2010model}, she will  reject some mutually beneficial policies so as to motivate the proposer to select policies that are even better for her. But her set of approved policies might depart from her favorite ones to curb the proposer's temptation to cherry-pick. If the proposer commits, he offers a lottery over policies that guarantees that the uninformed responder accepts; his optimal commitment corresponds to a concavified payoff function. 

We also evaluate the implications of requiring the proposer to offer a menu of options, not just a single one, to the responder. We show that by delegating some of the choice to the uninformed responder in this way---a form of \textit{in}expertise-based delegation---both parties can benefit if the menu is sufficiently broad relative to the basket. However, a narrow menu can backfire, leading to lower acceptance odds than the single-proposal benchmark. 

The cherry-picking trap also has clear implications for a proposer's incentives to exert costly effort to identify viable policies. We identify how the responder's ability to scrutinize proposals can induce the proposer to work harder. In an optimal stopping variant of our problem, we also document a \textit{productivity paradox}: the parties may be worse off when the proposer is better at identifying viable policies because the responder then fears cherry-picking to a greater degree.

\Cref{Section-Example} illustrates the logic of the cherry-picking trap through a simple example. \Cref{Section-Model} then describes the model, \Cref{Section-MainAnalysis} contains our main results, and \Cref{Section-Extensions} considers extensions. \Cref{Section-Literature} concludes by relating our results to precedents in the literature.

\section{An Example}\label{Section-Example}

A proposer (he) and a responder (she) choose whether to enact a new policy or maintain the status quo. The status quo yields a payoff of $0$ to each party. The proposer privately observes a finite set of viable policies; we refer to this collection as the \textit{basket} of policies. For the example, we assume that both parties commonly know the number of viable policies, $K$; the proposer's informational advantage is that he knows the payoffs associated with each policy. Each policy is characterized by a variable $x\in \Re_{\geq 0}$: a policy with characteristic $x$ delivers payoff $\pi_P(x)=x$ to the proposer and $\pi_R(x)=(2-x)$ to the responder. Each viable policy's characteristic is drawn independently from the exponential distribution with density $e^{-x}$. 

The timeline is as follows. The proposer first observes basket $B$ and proposes a viable policy. Before approving or rejecting the proposal, the responder observes its characteristic with probability $\lambda$ and is then \textit{informed}; otherwise, she is \textit{uninformed}, observing only that some policy has been put forward. 

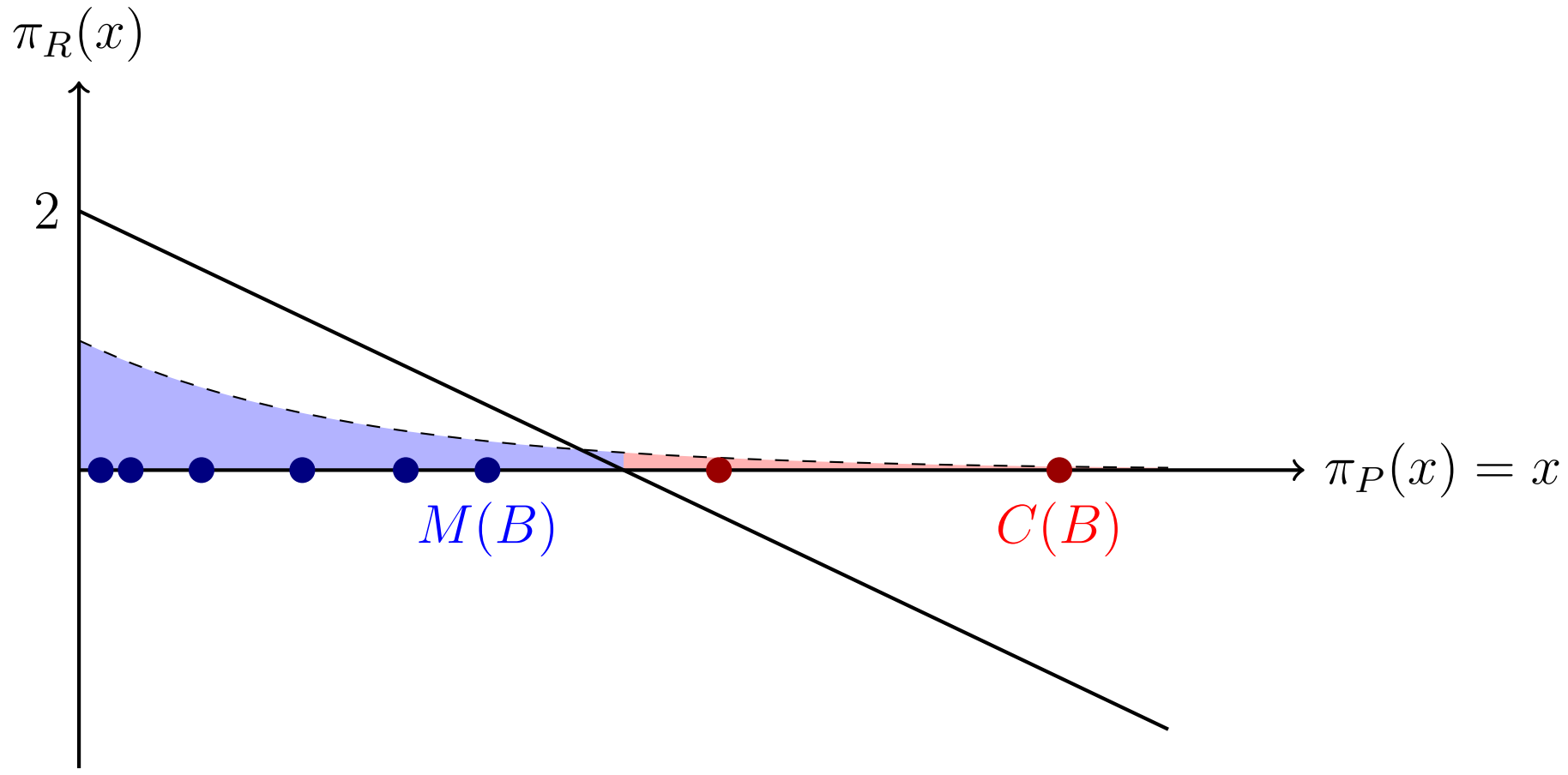
\begin{figure}[b!]
    \centering
   \begin{tikzpicture}[xscale=2.1, yscale=1]
  \fill[blue!30] (0,0)
    -- plot[domain=0:2, smooth, variable=\x] ({\x}, {exp(-\x)})
    -- (2,0) -- cycle;
  \fill[red!30] (2,0)
    -- plot[domain=2:4, smooth, variable=\x] ({\x}, {exp(-\x)})
    -- (4,0) -- cycle;
  \draw[domain=0:4, smooth, variable=\x, dashed]
    plot ({\x}, {exp(-\x)});
  \draw[black, thick] (0,2) -- (4,-2);
  \node[left] at (0,2) {$2$};
  \draw[->, thick] (0,0) -- (4.5,0) node[right] {$\pi_P(x) = x$};
  \draw[->, thick] (0,-2.3) -- (0,3) node[above] {$\pi_R(x)$};
 \node[circle, fill=NavyBlue, inner sep=2pt] at (0.08, 0) {};
  \node[circle, fill=NavyBlue, inner sep=2pt] at (0.19, 0) {};
  \node[circle, fill=NavyBlue, inner sep=2pt] at (0.45, 0) {};
  \node[circle, fill=NavyBlue, inner sep=2pt] at (0.82, 0) {};
  \node[circle, fill=NavyBlue, inner sep=2pt] at (1.2,  0) {};
  \node[circle, fill=NavyBlue, inner sep=2pt] at (1.5, 0) {};
  \node[circle, fill=red!60!black,  inner sep=2pt] at (2.35, 0) {};
  \node[circle, fill=red!60!black,  inner sep=2pt] at (3.6,  0) {};
   \node[below,blue] at (1.5, -0.1) {$M(B)$};
   \node[below,red] at (3.6,  -0.1) {$C(B)$};
\end{tikzpicture}\medskip
    \caption{A basket of eight policies drawn from the exponential distribution. Blue points are mutually beneficial and red points benefit the proposer alone. The proposer's favorite policy overall is $C(B)$ and his favorite Pareto improvement is $M(B)$.}
    \label{Figure-Example}
\end{figure}

Policies whose $x$ lies in $[0,2)$ are mutually beneficial, whereas those whose $x>2$ benefit the proposer alone. In a given basket $B$, the proposer's favorite is denoted  $C(B)\equiv \max_{x\in B} x$, corresponding to the first-order statistic of the $K$ i.i.d. draws. Also relevant is the proposer's favorite among Pareto improvements: if $B\cap [0,2)$ is non-empty, $M(B)\equiv \max_{x\in B\cap [0,2)}x$ denotes his favorite among them; otherwise, we write $M(B)=C(B)$. \Cref{Figure-Example} illustrates these selections for a basket of eight policies.

Ex ante, policies are promising to each party: because $\E[x]=1$, each party  expects to gain from a randomly chosen policy. Moreover, the probability that a randomly chosen policy is mutually beneficial is $1-e^{-2}\approx 0.86$. Yet,  strategic selection stymies trust.

First, suppose that the responder is always uninformed ($\lambda = 0$). Given any behavior by the uninformed responder, the proposer's weakly dominant action is to propose $C(B)$, the first-order statistic among the $K$ draws. Whether the responder approves then depends on how aggressively cherry-picking skews the distribution of proposals. For the exponential distribution,
\begin{align*}%
    \E\left[\max\{x_1,\ldots,x_K\}\right] = \sum_{k=1}^{K} \frac{1}{k}.
\end{align*} 
Selection effects are sufficiently weak for $K\in \{1,2,3\}$ that the responder approves. Once $K\geq 4$, selection effects dominate and the responder opts for the status quo. We see this effect depicted in \Cref{Figure-Example}, where the proposer chooses $C(B)$, which leaves the responder worse off than the status quo. Even limited flexibility thus destroys trust. 

Now suppose the responder is always informed ($\lambda =1$). As she approves only policies that benefit her, the proposer offers his favorite among Pareto improvements (if some Pareto improvement is viable). In \Cref{Figure-Example}, this proposal corresponds to $M(B)$. Flexibility now promotes agreement: a larger basket only raises the odds that a Pareto improvement is viable. 

Our interest is in settings in which the proposer does not know whether the responder is informed ($\lambda\in (0,1)$). Here, the proposer may be tempted to sneak his favorite policy past the (uninformed) responder. But the possibility that the responder is informed exerts a moderating influence. We show that, in the unique equilibrium, the proposer \textit{does} moderate in the vast majority of baskets, and yet the uninformed responder rejects most proposals. To illustrate this phenomenon, suppose $\lambda=1/4$ and $K=8$. 

We first argue that, in equilibrium, the uninformed responder must mix. Suppose, toward a contradiction, that she always approves the proposal. In any basket $B$ where $C(B)$ and $M(B)$ differ, the proposer faces a dilemma: should he propose $M(B)$, which is bound to pass, or propose $C(B)$, which only the uninformed responder approves? He would cherry-pick if $C(B)(1-\lambda)> M(B)$, which can be rewritten as
\begin{align}\label{Equation-Alpha1}
    \rb(B)\equiv \frac{C(B)-M(B)}{M(B)}> \frac{\lambda}{1-\lambda}.
\end{align}
The LHS captures the \textit{temptation ratio}: the proposer's proportional gain from cherry-picking relative to moderation. The RHS is the odds that he faces an informed responder, which discipline his temptation to cherry-pick. For $\lambda = 1/4$, that disciplining force is weak: he cherry-picks often enough that the uninformed responder's expected payoff is negative. The responder would then be better off rejecting the proposal when she cannot evaluate it. 

We also argue that, in equilibrium, the uninformed responder cannot always reject the proposal. Were she to do so, the proposer would have no incentive to cherry-pick, since a policy that harms the responder would never pass. But if the proposer always moderated, the uninformed responder would prefer to accept, which invalidates the putative equilibrium. 

Thus, in equilibrium, the uninformed responder must mix. Let $\uninf\in (0,1)$ denote the probability with which she approves. In a basket $B$, the proposer still faces a dilemma, opting to cherry-pick if $C(B)(1-\lambda)\uninf > M(B)(\lambda+(1-\lambda)\uninf)$, which can be rearranged as
\begin{align}\label{Equation-ExampleEquilibrium}
    \rb(B)> \left(\frac{\lambda}{1-\lambda}\right)\left(\frac{1}{\uninf}\right).
\end{align}
Comparing \eqref{Equation-ExampleEquilibrium} with \eqref{Equation-Alpha1}, we see that the uninformed responder's skepticism disciplines cherry-picking: the proposer now cherry-picks only if the temptation ratio clears a strictly higher bar. The threshold must also leave the uninformed responder indifferent. Given eight options, this indifference pins down a unique threshold. In equilibrium, the  proposer puts forward Pareto improvements in approximately $84\%$ of baskets. Yet, the uninformed responder accepts  with probability $\uninf\approx 0.17$, resulting in an ex ante acceptance probability of only $34\%$ (accounting for the behavior of the informed responder). Although the proposer cherry-picks in just $16\%$ of baskets, the mere specter of cherry-picking precludes agreement two-thirds of the time. 

The parties would benefit if the proposer could commit to proposing only Pareto improvements. He and the responder would then accrue expected payoffs of $1.47$ and $0.53$ respectively, substantially higher than their equilibrium payoffs of $0.58$ and $0.1$ respectively. The proposer would even strictly benefit from relinquishing control altogether, so that a policy is chosen uniformly at random from the basket and put before the responder. Given that policies, ex ante, are beneficial to each party, the uninformed responder would then always accept. Random selection delivers expected payoffs of $0.90$ and $1.03$ to the proposer and responder, respectively. Thus, by giving up control, the proposer would improve his payoff by around $55\%$.
\definecolor{cmod}{HTML}{A83232}
\definecolor{cacc}{HTML}{1F4E79}
\definecolor{cprop}{HTML}{A83232}
\definecolor{cresp}{HTML}{1F4E79}
\definecolor{csurp}{HTML}{2E7D32}

\begin{figure}[t!]
\centering

\begin{subfigure}[t]{0.48\textwidth}
\centering
\begin{tikzpicture}
\begin{axis}[
    width=\linewidth, height=6.5cm,
    xlabel style={yshift=-4pt},
    xlabel={Basket size $K$},
    ylabel={Payoff},
    xmin=0.5, xmax=10.5,
    ymin=0, ymax=2.05,
    xtick={1,2,3,4,5,6,7,8,9,10},
    ytick={0,0.5,1.0,1.5,2.0},
    grid=major,
    grid style={dotted, gray!40},
    axis lines=left,
    tick align=outside,
    label style={font=\small},
    tick label style={font=\small},
]

\addplot[
    color=cprop, thick, mark=*, mark size=1pt,
    mark options={fill=cprop, draw=cprop},
] coordinates {
    (1, 0.897289) (2, 1.308895) (3, 1.564859) (4, 0.816598) (5, 0.614110)
    (6, 0.582841) (7, 0.576438) (8, 0.575568) (9, 0.577191) (10, 0.579971)
};

\addplot[
    color=cresp, thick, mark=square*, mark size=1pt,
    mark options={fill=cresp, draw=cresp},
] coordinates {
    (1, 1.035211) (2, 0.567925) (3, 0.266028) (4, 0.110996) (5, 0.107056)
    (6, 0.103246) (7, 0.099378) (8, 0.095704) (9, 0.092042) (10, 0.088458)
};

\addplot[
    color=csurp, thick, dashed, mark=triangle*, mark size=0pt,
    mark options={fill=csurp, draw=csurp, solid},
] coordinates {
    (1, 1.932500) (2, 1.876820) (3, 1.830887) (4, 0.927594) (5, 0.721165)
    (6, 0.686088) (7, 0.675816) (8, 0.671272) (9, 0.669234) (10, 0.668429)
};

\end{axis}
\node[anchor=north, font=\small] at (current bounding box.south) {\textbf{(a)}};
\end{tikzpicture}
\label{fig:cp-payoffs}
\end{subfigure}%
\hfill
\begin{subfigure}[t]{0.48\textwidth}
\centering
\begin{tikzpicture}
\begin{axis}[
    width=\linewidth, height=6.5cm,
    xlabel style={yshift=-4pt},
    xlabel={Basket size $K$},
    ylabel={Probability},
    xmin=0.5, xmax=10.5,
    ymin=-0.03, ymax=1.05,
    xtick={1,2,3,4,5,6,7,8,9,10},
    ytick={0,0.2,0.4,0.6,0.8,1.0},
    grid=major,
    grid style={dotted, gray!40},
    axis lines=left,
    tick align=outside,
    label style={font=\small},
    tick label style={font=\small},
]

\addplot[
    color=cmod, thick, mark=*, mark size=1pt,
    mark options={fill=cmod, draw=cmod},
] coordinates {
    (1, 0.865001) (2, 0.753640) (3, 0.661774) (4, 0.663836) (5, 0.739582)
    (6, 0.784153) (7, 0.815913) (8, 0.840544) (9, 0.860094) (10, 0.875259)
};

\addplot[
    color=cacc, thick, mark=square*, mark size=1pt,
    mark options={fill=cacc, draw=cacc},
] coordinates {
    (1, 0.966250) (2, 0.938410) (3, 0.915443) (4, 0.463797) (5, 0.360583)
    (6, 0.343044) (7, 0.337908) (8, 0.335636) (9, 0.334617) (10, 0.334215)
};

\end{axis}
\node[anchor=north, font=\small] at (current bounding box.south) {\textbf{(b)}};

\end{tikzpicture}
\label{fig:cp-probs}
\end{subfigure}
\captionsetup{aboveskip=0pt}
\caption{Panel (a) plots ex ante payoffs as a function of the basket size $K$, with {\color{cprop}\textbf{red}} depicting the proposer's payoff, {\color{cresp}\textbf{blue}} depicting the responder's, and {\color{csurp}\textbf{green}} depicting their sum. Panel~(b) plots the ex ante probabilities that a Pareto improvement is proposed ({\color{cmod}\textbf{red}}) and that proposals are  accepted ({\color{cacc}\textbf{blue}}).}
\label{Figure-ExampleWelfare} 
\end{figure}

The parties would also be better off were the basket smaller. \Cref{Figure-ExampleWelfare}(a) depicts their ex ante payoffs for values of $K$ from one to ten. The efficiency loss is small for $K\leq 3$ because strategic selection does not push the uninformed responder to reject proposals. Once $K\geq 4$, the cherry-picking trap pushes the uninformed responder to reject, and more than half the feasible surplus is lost. \Cref{Figure-ExampleWelfare}(b) shows that the proposer moderates in more than $60\%$ of baskets, and yet the ex ante approval rate remains below $50\%$. Matters become worse as $K\rightarrow \infty$: as \Cref{Theorem-LargeBasket} shows,  the approval rate would then converge to $25\%$ despite the proposer putting forward Pareto improvements with probability converging to $1$.

Our baseline analysis treats the size of the basket as a primitive and assumes that the proposer makes a single proposal. In our extensions in \Cref{Section-Extensions}, we model settings in which the proposer can offer a menu of options as well as those in which he  exerts costly effort or searches for policies dynamically. We revisit this example in these extensions to identify the implications of the cherry-picking trap for these settings.

\section{Model}\label{Section-Model}

A proposer and a responder jointly choose whether to enact a new policy or keep the status quo. The proposer privately observes which policies are viable and chooses one to propose. The responder then chooses whether to accept the proposal. Doing so enacts the proposed policy, whereas rejection retains the status quo. The central friction is informational: the proposer knows more about the viable policies than the responder.

\paragraph{Policies and Payoffs:} The basket of policies, both its size and composition, is determined by nature. The number of policies, $k$, is a strictly positive integer drawn according to a distribution $\psi$ with finite mean; we allow $\psi$ to be a Dirac measure. 
Each policy has a {characteristic} $x\in\mathcal X$, where $\mathcal X$ is a connected subset of a Polish space. Characteristics are drawn independently according to an atomless probability distribution $\phi$ with full support on $\mathcal X$.\footnote{In other words, every non-empty relatively open subset $U\subseteq \mathcal{X}$ has strictly positive $\phi$-measure.}  We denote by $\mu$ the probability distribution on the set of possible baskets $\mathcal B$ induced by $\psi$ and $\phi$, and write $\E[\cdot]$ for expectations with respect to this measure.\footnote{\Cref{Appendix-PreliminariesAlternate} provides a formal description of the measure $\mu$ on $\mathcal{B}$.}

Each player accrues a payoff of $0$ from the status quo. A policy with characteristic $x$ delivers payoffs $\pi_P(x)$ and $\pi_R(x)$ to the proposer and responder, respectively. Each payoff function is continuous and $\phi$-integrable. Indifference curves are thin: for every player $i$ and utility $u \in \Re$, the level set $\{x \in \mathcal X : \pi_i(x) = u\}$ has $\phi$-measure $0$. 

We assume that all policies benefit the proposer: $\pi_P(\cdot)\geq 0$; this assumption simplifies the exposition without affecting the analysis. The \textit{congruence region} is the set of policies that also benefit the responder, whereas the \textit{conflict region} is the set of those that harm her; we denote these regions, respectively, by
\begin{align}\label{Equation-CongruenceConflict}
    \mathcal X^+ \equiv \{x \in \mathcal{X} : \pi_R(x) > 0\}\, \text{ and }\, \mathcal X^- \equiv \{x \in \mathcal X : \pi_R(x) < 0\}.
\end{align}
We assume that both regions have strictly positive $\phi$-measure; we also assume that within each region, each player's attainable payoffs form an interval. To isolate the power of strategic selection, we study settings in which players' interests align ex ante.  
\begin{assumption}\label{Assumption-ExAnte}
    On average, policies benefit the responder: $\int_{\mathcal{X}} \pi_R(x)\,\phi(dx)>0$.
\end{assumption}
Under this assumption, the responder would approve a randomly selected proposal; any breakdown in trust stems purely from strategic selection. Because we impose \Cref{Assumption-ExAnte} throughout, we do not refer to it in most of our theorem statements. We summarize a payoff environment by $\payoff\equiv (\mu,\pi_P,\pi_R)$.

\paragraph{Timing and Strategies:} The proposer observes the realized basket and proposes a policy in it. With probability $\lambda\in(0,1)$, the responder is \textit{informed} and observes the characteristic $x$ of the proposed policy; otherwise, she is \textit{uninformed}. We denote the game by $\game(\payoff,\lambda)$.

We now describe strategies in the game. The probability that the proposer selects policy $x$ in basket $B$ is denoted by $\sigma(x | B)$, where $\sigma:\mathcal X \times \mathcal B \to [0,1]$ is measurable and $\sigma(x | B) = 0$ whenever $x \notin B$.  
We summarize the responder's strategy using the tuple $(\info(\cdot),\uninf)$ in which the measurable function  $\info:\policy\rightarrow [0,1]$ denotes the probability $\info(x)$ that she accepts a policy with characteristic $x$ when informed, and the scalar $\uninf\in [0,1]$ denotes the probability that she accepts when uninformed. The strategy profile is then $(\sigma,\info,\uninf)$. We study Bayes--Nash equilibria of this game, which we henceforth call equilibria; \Cref{Theorem-UniqueEqm} establishes that the game has a unique equilibrium (up to   differences on a measure-$0$ set of baskets).\footnote{Accordingly, we call an equilibrium $(\sigma^*, \gamma^*, \alpha^*)$ \emph{essentially unique} if, for any other equilibrium $(\sigma', \gamma', \alpha')$, we have $\alpha' = \alpha^*$ and, for $\mu$-almost every $B \in \mathcal B$, $\gamma'(x)=\gamma^*(x)$ and $\sigma'(x | B) = \sigma^*(x | B)$ for all $x \in B$.}

Below, we describe three settings that fit into this model.

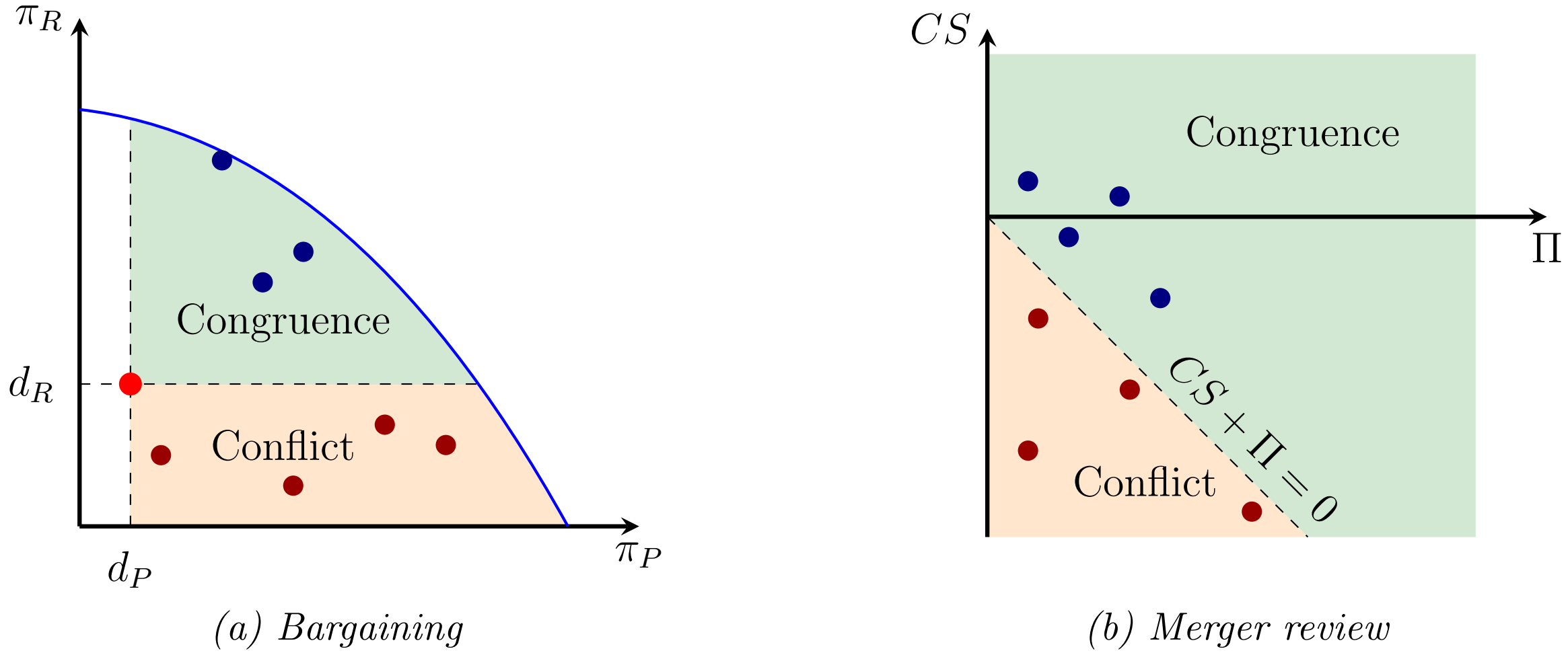
\begin{figure}[h!]
  \centering
  \captionsetup{aboveskip=8pt}
  \captionsetup[subfigure]{font={small,it},skip=4pt}
  \begin{subfigure}[b]{0.48\textwidth}
    \centering
    \begin{minipage}[c][5.8cm][c]{\linewidth}
      \centering
\begin{tikzpicture}[>=stealth,scale=1]

  \def\UtilityCurvePath{(0.0,4.1) .. controls (1.7,3.9) and (3.3,2.7) .. (4.8,0.0)}

  \begin{scope}
    \clip (0.5,0) rectangle (5.2,1.4);
    \fill[orange!20] \UtilityCurvePath -- (4.8,0) -- (0,0) -- cycle;
  \end{scope}

  \begin{scope}
    \clip (0.5,1.4) rectangle (5.2,5.2);
    \fill[ForestGreen!20] \UtilityCurvePath -- (4.8,0) -- (0,0) -- cycle;
  \end{scope}

  \draw[very thick,->] (0,0) -- (0,5.0) node[left] {$\pi_R$};
  \draw[very thick,->] (0,0) -- (5.5,0) node[below] {$\pi_P$};

  \draw[thick,blue] \UtilityCurvePath;

  \draw[dashed] (0.5,0) -- (0.5,4);
  \draw[dashed] (0,1.4) -- (3.9,1.4);

  \filldraw[red] (0.5,1.4) circle (3.0pt);

  \node[below] at (0.5,-0.1) {$d_P$};
  \node[left]  at (-0.1,1.4) {$d_R$};

  \node at (2,2) {Congruence};
  \node at (2,0.8) {Conflict};
    \node[circle, fill=NavyBlue, inner sep=2pt] at (1.8,2.4) {};
     \node[circle, fill=NavyBlue, inner sep=2pt] at (1.4,3.6) {};
     \node[circle, fill=NavyBlue, inner sep=2pt] at (2.2,2.7) {};
    \node[circle, fill=red!60!black,  inner sep=2pt] at (3,1) {};
    \node[circle, fill=red!60!black,  inner sep=2pt] at (3.6,0.8) {};
    \node[circle, fill=red!60!black,  inner sep=2pt] at (2.1,0.4) {};
    \node[circle, fill=red!60!black,  inner sep=2pt] at (0.8,0.7) {};
\end{tikzpicture}
    \end{minipage}
    \caption{Bargaining}
  \end{subfigure}\hfill
  \begin{subfigure}[b]{0.48\textwidth}
    \centering
    \begin{minipage}[c][5.8cm][c]{\linewidth}
      \centering
\begin{tikzpicture}[>=stealth,x=1cm,y=1cm]
    \fill[ForestGreen!20]
      (0,0) -- (0,1.6) -- (4.8,1.6) -- (4.8,-3.15) -- (3.15,-3.15) -- cycle;
    \fill[orange!20]
      (0,0) -- (3.15,-3.15) -- (0,-3.15) -- cycle;

    \draw[very thick,->] (0,-3.15) -- (0,1.85) node[left] {$CS$};
    \draw[very thick,->] (0,0) -- (5.5,0) node[below] {$\Pi$};

    \draw[dashed] (0,0) -- (3.15,-3.15)
      node[pos=0.76,above,sloped] {$CS+\Pi=0$};

    \node at (3.0,0.8) {Congruence};
    \node at (1.55,-2.6) {Conflict};

    \node[circle,fill=NavyBlue,inner sep=2pt] at (0.4,.35) {};
    \node[circle,fill=NavyBlue,inner sep=2pt] at (1.7,-.8) {};
    \node[circle,fill=NavyBlue,inner sep=2pt] at (.8,-.2) {};
    \node[circle,fill=NavyBlue,inner sep=2pt] at (1.3,0.2) {};
    \node[circle,fill=red!60!black,inner sep=2pt] at (1.4,-1.7) {};
    \node[circle,fill=red!60!black,inner sep=2pt] at (2.6,-2.9) {};
    \node[circle,fill=red!60!black,inner sep=2pt] at (.4,-2.3) {};
    \node[circle,fill=red!60!black,inner sep=2pt] at (0.5,-1) {};
    \path (0,-3.7);
  \end{tikzpicture}
    \end{minipage}
    \caption{Merger review}
  \end{subfigure}
  \caption{Panel (a) depicts bargaining with a disagreement point $(d_P,d_R)$. Panel (b) depicts merger review: the firm values profits $\Pi$ while the regulator values total surplus $\Pi+CS$. Blue dots represent feasible policies that benefit both parties while red dots benefit the proposer alone.}
  \label{Figure-BargainingProject}
\end{figure}

\begin{example}\label{Example-Bargaining}
Consider the \textit{classic Nash bargaining problem}: a compact convex set of feasible utility pairs with a disagreement point $(d_P,d_R)$. If we interpret the disagreement point as the status quo and measure payoffs relative to it, our problem corresponds to the proposer selecting from an ex ante uncertain finite set of feasible agreements, and the responder being potentially unable to evaluate the proposed agreement. \Cref{Figure-BargainingProject}(a) depicts this setup.
\end{example}

\begin{example}\label{Example-Project}
We build on the project-selection models of \citet{armstrong2010model} and \citet{nocke2013merger}, which are widely used to model organizational and antitrust decisions. As depicted in \Cref{Figure-BargainingProject}(b), consider a firm that chooses to propose one of a finite set of projects (or mergers) and a regulator who can accept or reject the proposal. The firm cares about its profits while the regulator cares about total surplus. The prior literature characterizes optimal mechanisms in which the regulator commits ex ante to an approval region and can verify ex post whether the proposal lies within it. Our analysis relaxes commitment and studies a regulator who may be unable to evaluate the proposed project. 
\end{example}

\begin{example}\label{Example-Veto}
We turn to the veto-bargaining model of \citet{RR:78}. The proposer, a legislative body, proposes a policy $x \in \Re_{\geq 0}$. The responder is an executive who can veto the proposal---resulting in the status quo $0$---or accept it, resulting in policy $x$. The parties have quadratic loss functions $(x - x_i^*)^2$, with bliss points $x_R^*>0$ for the responder and $x_P^* > 2x_R^*$ for the legislature, so that the legislature's ideal lies in the conflict region. Here, the congruence region is $(0,2x_R^*)$ and the conflict region is $(2x_R^*,2x_P^*]$. Veto bargaining has been modeled with various forms of incomplete information; we study settings in which some policies may not be viable and the responder may be unable to evaluate them.\footnote{For other work on veto bargaining with incomplete information, see \cite{matthews1989veto}, \cite{CameronMcCarty2004}, \cite*{kartik2021delegation}, \cite*{ali2023sequential}, and \cite*{kim2024persuasion}. }
\end{example}

\section{Strategic Selection and Trust}\label{Section-MainAnalysis}

This section contains our main results. \Cref{Section-EqmCharacterization} characterizes the unique equilibrium and describes its welfare implications; \Cref{Section-Theorem1Sketch} sketches the proof for the equilibrium characterization. \Cref{Section-Large} identifies the potency of strategic selection when the proposer likely has many policies to choose from. \Cref{Section-Commitment} compares equilibrium behavior to benchmarks in which either player can commit to a strategy.  

\subsection{Equilibrium Behavior and Welfare Implications}\label{Section-EqmCharacterization}

Herein, we characterize the unique equilibrium of this game: (i) given a basket $B$, the proposer puts forward a policy that harms the responder only if he benefits sufficiently relative to all mutually beneficial policies in $B$; (ii) the informed responder approves any mutually beneficial policy and rejects any harmful one; and (iii) the uninformed responder potentially mixes between approval and rejection. Formalizing these ideas, we call the proposer's favorite element of $B$ the \textit{cherry-picked policy}:
\begin{align*}
C(B)\equiv \argmax\limits_{x\in B} \pi_P(x). 	
\end{align*}
We contrast it with his favorite mutually beneficial policy in $B$ (if one exists), which we call the \emph{moderate policy},
\begin{align*}
M(B)\equiv \begin{cases}
   \argmax\limits_{x\in B\cap\mathcal{X}^+}  \pi_P(x) &\text{if}\; B\cap \mathcal{X}^+\ne \emptyset\\
   C(B)&\text{otherwise}.
\end{cases}
\end{align*}
The two differ exactly when the proposer's favorite viable policy harms the responder but some other viable policy benefits her.\footnote{Our specification sets $M(B)=C(B)$ if no mutually beneficial policy is viable ($B\cap \congruence=\emptyset$). \Cref{Theorem-UniqueEqm} shows that, in equilibrium, the proposer must make this selection $\mu$-a.e.} The proposer then faces a dilemma: should he cherry-pick or moderate? His choice hinges on a simple statistic, which we call the basket's \emph{temptation ratio}:
\begin{align*}
\rb(B)\equiv 
    \frac{\pi_P(C(B))-\pi_P(M(B))}{\pi_P(M(B))}.  
\end{align*}
This term describes the proportional gain from cherry-picking, reflecting his temptation to do so. Whether he succumbs to this temptation will depend on the odds that he faces an informed responder, $\odds_\lambda\equiv\lambda/(1-\lambda)$. We use this notation to describe the equilibrium.

\begin{theorem}\label{Theorem-UniqueEqm}
The following strategy profile $(\sigma^*,\gamma^*,\alpha^*)$ constitutes the essentially unique equilibrium of the game $\game(\payoff,\lambda)$:
   \begin{enumerate}[label=\emph{(\alph*)}]
   	\item The informed responder accepts all mutually beneficial policies and rejects all harmful ones: $\gamma^*(x)=\1[\pi_R(x)\ge 0]$.
   
      \item The uninformed responder accepts with probability 
    $\alpha^*=\alpha_\lambda\equiv \min\{\odds_\lambda / \odds_{\overline\lambda},1\}$
     for some value $\overline\lambda\in [0,1)$.\footnote{We use the convention that $a/0=\infty$ if $a>0$.}

\item The proposer moderates in basket $B$ if its temptation ratio, $t(B)$, is less than $\odds_\lambda /\alpha_\lambda$ and cherry-picks otherwise.   

\end{enumerate}
Every equilibrium coincides with $(\sigma^*,\gamma^*,\alpha^*)$ for $\mu$-almost every $B\in \mathcal B$ and all $x\in B$.
\end{theorem}

We sketch the proof for \Cref{Theorem-UniqueEqm} in \Cref{Section-Theorem1Sketch}, where we also derive the value $\overline\lambda$ referenced in part (b); importantly, $\overline\lambda$ depends only on the payoff environment $\payoff$ and is strictly positive if and only if $\E[\pi_R(C(\cdot))]< 0$. Generally, if the proposer is tempted by policies that harm the responder, this payoff condition holds whenever the basket is (stochastically) sufficiently large. For instance, in the example of \Cref{Section-Example}, $\overline\lambda >0$ whenever the basket has four or more policies. 

\begin{figure}[h!]
\centering
\begin{subfigure}[t]{0.48\textwidth}
\centering
\begin{tikzpicture}[scale=3]
  \draw[->, thick] (0,0) -- (1.2,0);
  \node[anchor=base] at (1.2,-0.16) {$\lambda$};
  \draw[->, thick] (0,0) -- (0,1.2) node[left] {$\alpha_\lambda$};
  \node[anchor=base] at (1,-0.16) {1};
  \node[left] at (0,1) {1};
  \node[anchor=base] at (0.666667,-0.16) {$\overline\lambda$};
  \draw[gray, dashed, very thin] (0.666667,0) -- (0.666667,1);
  \draw[gray, dashed, very thin] (0,1) -- (0.666667,1);
  \draw[gray, dashed, very thin] (1,0) -- (1,1);
  \draw[domain=0:0.666667, samples=100, smooth, variable=\x, cresp, thick]
    plot ({\x}, {\x/(2*(1-\x))});
  \draw[cresp, thick] (0.666667,1) -- (1,1);
  \node[anchor=north, font=\small] at (current bounding box.south) {\textbf{(a)}};
\end{tikzpicture}
\end{subfigure}%
\hfill
\begin{subfigure}[t]{0.48\textwidth}
\centering
\begin{tikzpicture}[x=3cm, y=0.9cm]
  \draw[->, thick] (0,0) -- (1.2,0);
  \node[anchor=base] at (1.2,{-0.48/0.9}) {$\lambda$};
  \draw[->, thick] (0,0) -- (0,4) node[left] {$\rb(B)$};
  \node[anchor=base] at (1,{-0.48/0.9}) {1};
  \node[left] at (0,2) {$\odds_{\overline\lambda}$};
  \node at (0.3,1) {$M(B)$};
  \node at (0.3,3.1) {$C(B)$};
  \node[anchor=base] at ({2/3},{-0.48/0.9}) {$\overline\lambda$};
  \draw[gray, dashed, very thin] ({2/3},0) -- ({2/3},2);
  \draw[gray, dashed, very thin] (1,0) -- (1,4);
  \draw[cprop, thick] (0,2) -- ({2/3},2);
  \begin{scope}
    \clip (0,0) rectangle (1,4);
    \draw[cprop, thick, domain=2/3:0.985, samples=300, smooth, variable=\x]
      plot ({\x}, {2+0.1*(\x/(1-\x)-2)});
  \end{scope}
  \node[anchor=north, font=\small] at (current bounding box.south) {\textbf{(b)}};
\end{tikzpicture}
\end{subfigure}
\captionsetup{aboveskip=0pt}
\caption{Panel (a) plots how the uninformed responder's acceptance probability $\alpha_\lambda$ varies with $\lambda$, the ex ante probability that the responder is informed. Panel (b) depicts the proposer's behavior: given a basket $B$, he cherry-picks if the temptation ratio $\rb(B)$ exceeds $\max\{\odds_{\overline\lambda},\odds_\lambda\}$ and moderates otherwise.}
\label{Figure-EquilibriumBehavior}
\end{figure}
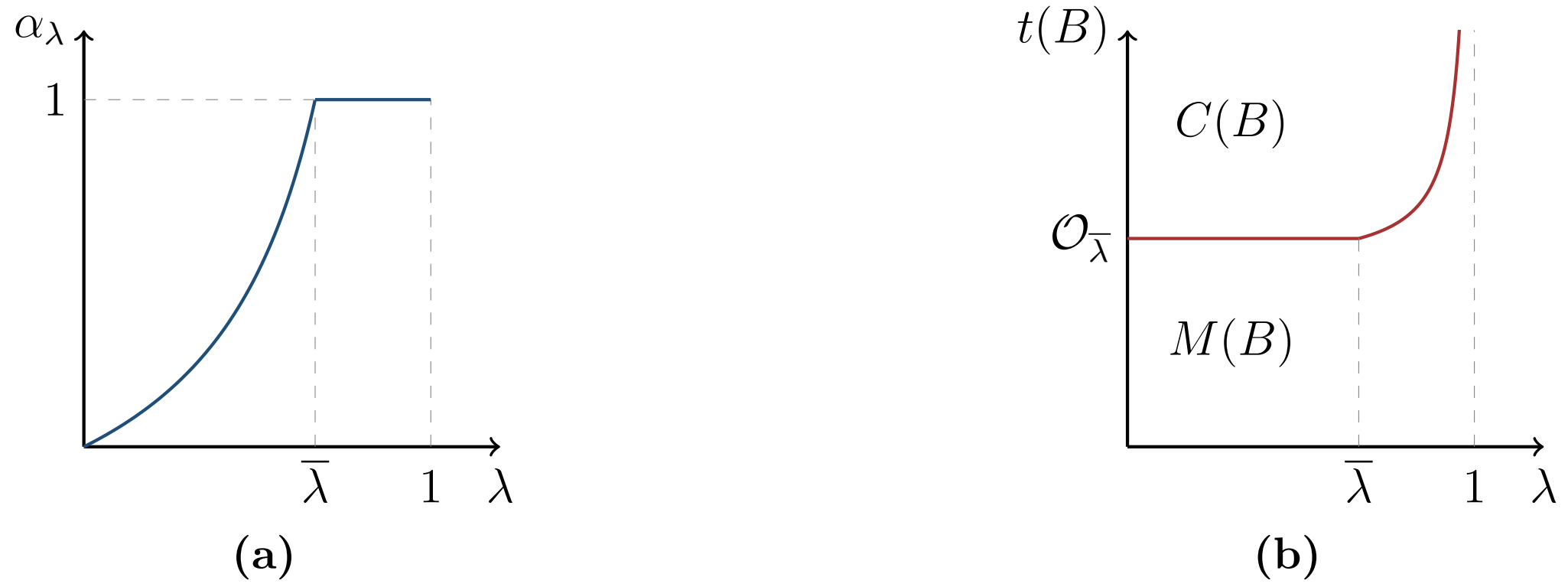

We now describe equilibrium behavior and welfare when $\overline\lambda>0$. The behavior of the informed responder is simple, and so we turn to that of the uninformed responder. As described in (b), she potentially mixes between accepting and rejecting the proposal with her acceptance rate following the curve $\alpha_\lambda$ in \Cref{Figure-EquilibriumBehavior}(a). Her acceptance rate increases in $\lambda$ until it reaches one at $\overline\lambda$, where it stays thereafter. 

This acceptance rate influences the proposer's behavior. When facing a dilemma between moderation and cherry-picking, he moderates if and only if
\begin{align}\label{Equation-ProposerDilemma}
    \underbrace{\bigl[\lambda+(1-\lambda)\alpha_\lambda\bigr]}_{
        \substack{\text{Approval under}\\\text{moderation}}}
    \pi_P(M(B))
    \;\geq\;
    \underbrace{(1-\lambda)\alpha_\lambda}_{
        \substack{\text{Approval under}\\\text{cherry-picking}}}
    \pi_P(C(B)).
\end{align}
Rearranging terms yields the condition in  \Cref{Theorem-UniqueEqm}(c) that the basket's temptation ratio, $\rb(B)$, cannot exceed $\odds_\lambda/\alpha_\lambda$. Observe that the acceptance probability $\alpha_\lambda$ is fine-tuned to ensure that this threshold equals $\odds_{\overline\lambda}$ if $\lambda\leq \overline\lambda$; this threshold value reflects a distribution of proposals that keeps the uninformed responder just indifferent between accepting and rejecting. Hence, as \Cref{Figure-EquilibriumBehavior}(b) depicts, the proposer's behavior does not vary with $\lambda$ below $\overline\lambda$. Past that point, the uninformed responder accepts for sure, and the proposer then sets a threshold that matches the odds of facing an informed responder. 

\Cref{Figure-EquilibriumBehavior}(a) shows that if $\lambda$ is low, the proposer and responder typically end up with the status quo even when Pareto improvements are feasible and proposed. As the example in \Cref{Section-Example} shows, with only a few options and $\lambda=1/4$, the parties fail to agree more than half the time. This disagreement offers a simple account of gridlock in politics. If the legislature cannot discern whether the executive's proposal constitutes a mutually beneficial policy or reflects some ulterior motive (e.g., a power grab), it may rationally reject it, fearing the latter. The same logic applies within organizations. A division manager who privately observes investment opportunities could recommend to headquarters an option that benefits the company as a whole or one that benefits the division alone. If headquarters cannot suitably check the recommendation, proposals that benefit the company will often be rejected. 

\paragraph{Random Versus Strategic Selection:} The cherry-picking trap can be so costly that the proposer might prefer to relinquish his ability to steer the process. Suppose that, instead of the proposer selecting a policy, one were chosen uniformly at random from the basket and put before the responder. As before, the informed responder can evaluate this proposal while the uninformed responder cannot. Because the uninformed responder no longer fears strategic selection, she will accept for sure (given \Cref{Assumption-ExAnte}). This higher acceptance probability can make both parties better off.  

\begin{proposition}\label{Proposition-RandomSelection}
	Suppose $\overline\lambda>0$. The following hold:
   \begin{enumerate}[label=\emph{(\alph*)}]
		\item There exists $\hat\lambda_P\in (0,\overline\lambda)$ such that the proposer prefers random to strategic selection if and only if $\lambda\leq\hat\lambda_P$. 
		\item There exists $\hat\lambda_R>0$ such that the responder prefers random to strategic selection if $\lambda\leq\hat\lambda_R$. 
	\end{enumerate}
\end{proposition}
Thus, both players favor random to strategic selection if $\lambda$ is sufficiently low.\footnote{In the proof, we show that the proposer's marginal gain from strategic over random selection is single-crossing in $\lambda$; by contrast, the responder's marginal gain need not be so. This fact accounts for why (a) has an ``if and only if'' statement and (b) does not.} The proposer may even favor random selection---relinquishing control of the process so as to avoid the cherry-picking trap---while the responder favors strategic selection. 
An alternative interpretation is that the proposer can benefit from committing to remain uninformed---observing a null signal rather than the characteristic of each policy in the basket---when $\lambda\leq\hat\lambda_P$.

\paragraph{Educating the Responder:} The proposer could also mitigate the cherry-picking trap by educating the responder---that is, raising $\lambda$---so that she can better distinguish mutually beneficial policies from those that harm her. While the responder could then better detect cherry-picked policies, equilibrium forces imply that the proposer nevertheless benefits when $\lambda<\overline\lambda$. 
\begin{proposition}\label{Proposition-Education}
	Suppose $\overline\lambda>0$. The following hold:
	  \begin{enumerate}[label=\emph{(\alph*)}]
		\item Each player's equilibrium payoff is linear and strictly increasing in $\lambda$ on $(0,\overline\lambda)$. 
		\item On $[\overline\lambda,1)$, the proposer's equilibrium payoff strictly decreases in $\lambda$ while the responder's strictly increases. 
	\end{enumerate}
\end{proposition}

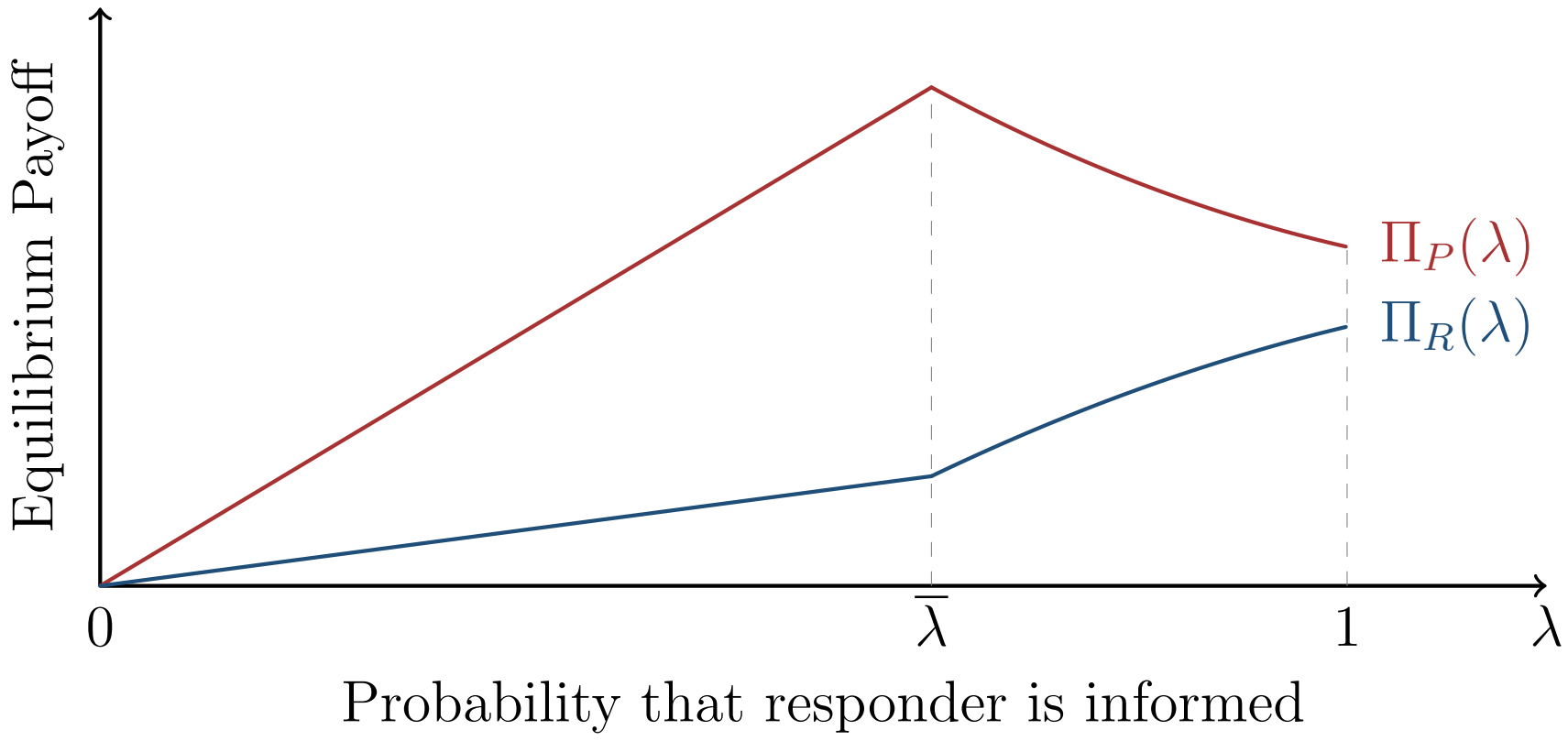
\begin{figure}[h!]
\centering
\begin{tikzpicture}[
  x=9cm,y=3.6cm,
  payoffcurve/.style={thick, line cap=round, line join=round},
  guide/.style={gray, dashed, very thin},
  ticklabel/.style={anchor=base, inner sep=0pt}
]
  \draw[->,thick] (0,0) -- (1.16,0);
  \draw[->,thick] (0,0) -- (0,1.16);
  \node[rotate=90] at (-.05,.58) {Equilibrium Payoff};
  \node[anchor=north] at (.58,-.15) {Probability that responder is informed};
  \node[ticklabel] at (0,-.12) {$0$};
  \node[ticklabel] at ({2/3},-.12) {$\overline\lambda$};
  \node[ticklabel] at (1,-.12) {$1$};
  \node[ticklabel] at (1.16,-.12) {$\lambda$};
  \draw[guide] ({2/3},0) -- ({2/3},1);
  \draw[guide] (1,0) -- (1,.68);
  \draw[cprop,payoffcurve] (0,0) -- ({2/3},1);
  \draw[cprop,payoffcurve,domain=2/3:1,samples=100,variable=\x]
    plot (\x,{1-.45*(3*\x-2)+.13*(3*\x-2)^2});
  \draw[cresp,payoffcurve] (0,0) -- ({2/3},.22);
  \draw[cresp,payoffcurve,domain=2/3:1,samples=100,variable=\x]
    plot (\x,{.22+.40*(3*\x-2)-.10*(3*\x-2)^2});
  \node[right,cprop] at (1.01,.68) {$\Pi_P(\lambda)$};
  \node[right,cresp] at (1.01,.52) {$\Pi_R(\lambda)$};
\end{tikzpicture}
\caption{This figure depicts expected equilibrium payoffs as functions of $\lambda$, with {\color{cprop}\textbf{$\Pi_P(\lambda)$}} showing the proposer's and {\color{cresp}\textbf{$\Pi_R(\lambda)$}} showing the responder's. }
\label{Figure-EducatingResponder}
\end{figure}

\Cref{Figure-EducatingResponder} shows the players' equilibrium payoffs as functions of $\lambda$, highlighting how both rise with $\lambda$ on $(0,\overline\lambda)$; past $\overline\lambda$, the responder's payoff continues to do so while the proposer's falls. 

A first intuition for why the proposer's payoff rises may be that greater scrutiny disciplines him, pushing him to moderate more often. This intuition is false: because the uninformed responder still mixes according to $\alpha_\lambda$, she must remain indifferent between accepting and rejecting the proposal. Consequently, the proposer's equilibrium behavior is independent of $\lambda$. The driving force lies instead in the responder's behavior: as $\lambda$ increases, her acceptance probability when uninformed must adjust to keep the proposer at the same threshold, $\odds_{\overline\lambda}$. Plugging that threshold into \eqref{Equation-ProposerDilemma} on p. \pageref{Equation-ProposerDilemma} yields $(1-\lambda)\alpha_\lambda = \lambda / \odds_{\overline\lambda}$. Hence, a cherry-picked policy, once proposed, is approved with a probability increasing in $\lambda$, despite the higher odds that it is detected. Moderate proposals also pass with a probability that grows in $\lambda$. Thus, while educating the responder invites greater scrutiny, the proposer benefits from the offsetting equilibrium forces.
These gains reverse once $\lambda$ crosses $\overline\lambda$: the uninformed responder then accepts for sure, so greater scrutiny only strengthens the responder's role as a gatekeeper without any countervailing strategic effects. 

The responder's payoff rises for different reasons on either side of $\overline\lambda$. Below it, she gains only when she is informed given that the uninformed responder is indifferent; her payoff rises only because she is informed more often. Above it, a second effect kicks in: a higher $\lambda$ also pushes the proposer to cherry-pick less often.

We interpret this comparative static in light of the practice of lobbying. As described by \cite{grossman/helpman}, a large fraction of lobbying tends to be informational in nature: advocates---activists or industry groups---spend considerable resources to furnish information to a regulator or legislature. The existing literature often interprets this interaction through the lens of a signaling or cheap-talk game \citep[e.g.,][]{potters1992lobbying,austen1993information}. Our finding offers a complementary rationale: improving the regulator's \textit{know-how}, so that she better understands which policies are mutually beneficial, can lead to Pareto gains by raising the probability that policies pass. In this vein, our approach aligns with those of \citet{hall2006lobbying} and \citet{ellis2020strategic}, who view lobbying as a subsidy that enhances the capacity of a regulator to pass policies.  

Within the context of organizations, \Cref{Proposition-Education} identifies the benefits that both a division manager and headquarters accrue when headquarters has a better understanding of technologies, constraints, and tradeoffs relevant to the division before opportunities arise. This better understanding could stem from the division manager educating headquarters or, conversely, from headquarters paying more attention to the division and learning more about the initiatives the division could undertake. We show in \Cref{Section-Effort} that this form of supervisory attention has even stronger benefits when the division manager has to work to identify viable options; attention rewards initiative. This proposition also bears an interesting implication for the measurement of management practices and productivity: improving management practices, in terms of headquarters' understanding, can make both parties better off without improving the quality of proposed or accepted policies.   

\paragraph{Risk Attitudes:} The players' risk attitudes affect equilibrium behavior in an intuitive way. Consider transforming player $i$'s payoff $\pi_i$ by a continuous and strictly increasing function $f:\mathbb{R}\to\mathbb{R}$ satisfying $f(0)=0$, and assume that the transformed payoff $f\circ\pi_i$ remains $\phi$-integrable. We say that player $i$ becomes more risk averse (respectively, risk loving) if $\pi_i$ is replaced by $f\circ \pi_i$, where $f$ is concave (respectively, convex).

\begin{proposition}\label{Proposition-RiskAttitudes}
In equilibrium, the uninformed responder rejects with higher probability if the proposer becomes more risk loving or the responder becomes more risk averse. 
\end{proposition}
Starting with the proposer, he is more tempted to cherry-pick if he becomes more risk loving. In equilibrium, the uninformed responder must then reject with higher probability to curb that temptation. Now suppose the responder becomes more risk averse. In equilibrium, the distribution of proposals generates a lottery whose certainty equivalent is the status quo. The uninformed responder values the lottery less if she becomes more risk averse. Therefore, in equilibrium, the proposer must offer her a more attractive lottery, one in which he moderates more often. He does so only if cherry-picking pays less, which requires the uninformed responder to reject with higher probability. 

\subsection{Proof Sketch for \Cref{Theorem-UniqueEqm} }\label{Section-Theorem1Sketch}

We first describe why, in every equilibrium, the informed responder approves almost every policy that benefits her and rejects almost every policy that harms her. Suppose, toward a contradiction, that she were to reject with positive probability some positive $\phi$-measure set $\tilde{X}$ of mutually beneficial policies. Our sampling process implies that, with positive probability, the entire basket  $B$ is a subset of $\tilde X$. In that event, the proposer must propose a policy in $\tilde X$. Then, for any proposer strategy $\sigma$, the informed responder could profitably deviate to approving policies in $\tilde{X}$ with probability one. A symmetric argument shows that she rejects almost every policy that harms her. Therefore, we take $\gamma^*(x)=\1[\pi_R(x)\ge 0]$ without loss of generality.\footnote{In other words, the informed responder's behavior in an equilibrium must be sequentially rational $\phi$-a.e. }

We now turn to the behavior of the uninformed responder, first explaining how we obtain $\overline\lambda$. Suppose that the selection $C(B)$ harms the responder on average relative to the status quo, i.e., $\E[\pi_R(C(\cdot))]<0$. Moreover, suppose that the proposer uses a threshold strategy $\tau\in [0,\infty]$ whereby, given a basket $B$, he moderates if $\rb(B)\leq \tau$ and cherry-picks otherwise. The uninformed responder's expected payoff from approving is then
\begin{align*}
	V_R(\tau)\equiv \int_{\rb(B)\leq \tau} \pi_R(M(B)) \mu(dB)+\int_{\rb(B)> \tau} \pi_R(C(B)) \mu(dB).
\end{align*}
In the Appendix, \Cref{Lemma-ModerationPositive,Lemma-ThresholdPayoff} use continuity and monotonicity properties to show that $V_R(\cdot)$ has a unique zero, which we denote by $\overline\tau$. This threshold delivers a distribution of proposals that keeps the uninformed responder indifferent between accepting and rejecting; any higher threshold delivers a strictly positive payoff and any lower threshold delivers a strictly negative one. We set $\overline\lambda\equiv \overline\tau/(1+\overline\tau)$ so that $\odds_{\overline\lambda}=\overline\tau$. 

We now argue that if $\lambda<\overline\lambda$, then in equilibrium the proposer uses threshold $\overline\tau$ and the uninformed responder mixes. As in \Cref{Section-Example}, she cannot use a pure strategy. If she always rejected, the proposer would moderate whenever possible ($\tau=\infty$), and because $V_R(\infty)>0$, she would prefer to accept. If she always accepted, the proposer would use the threshold $\odds_\lambda<\overline\tau$, and she would prefer to reject. From \eqref{Equation-ProposerDilemma}, we know that for every $\alpha\in [0,1]$, the proposer uses a threshold strategy; the responder is indifferent only if that threshold equals $\odds_{\overline\lambda}=\overline\tau$. For the proposer to find it optimal to use this threshold, the uninformed responder's acceptance rate must equal $\alpha_\lambda =\odds_\lambda/\odds_{\overline\lambda}$. 

To complete the sketch, suppose $\lambda\geq \overline\lambda$. Then the responder is informed with high enough probability that any acceptance probability $\alpha\in [0,1]$ induces the proposer to use a threshold weakly greater than $\odds_{\overline\lambda}$. This favorable selection of proposals makes it a best response for the uninformed responder to accept with probability one.

\subsection{How Flexibility Breeds Distrust}\label{Section-Large}

We analyze how greater flexibility---in the form of stochastically large baskets---affects the cherry-picking trap. Although a large basket is more likely to contain some mutually beneficial policy, it also gives the proposer greater scope to cherry-pick. Under two additional assumptions, we find that, in equilibrium, the proposer puts forward mutually beneficial policies with probability converging to one. Yet, the uninformed responder still rejects these proposals with probability bounded away from zero; in some cases, that probability can converge to one. Thus, while the proposer's behavior merits trust, the responder rationally distrusts him. 

We start with the additional assumptions. The first asserts that the proposer is tempted by policies that harm the responder. 
\begin{assumption}\label{Assumption-Conflict}
	All of the proposer's near-favorite policies are in the conflict region: there exist $\tilde{\pi}_P\in \interior(\pi_P(\policy))$ and $\delta>0$ such that $\pi_P(x)\geq \tilde\pi_P$ implies $\pi_R(x)\leq -\delta$.\footnote{We denote the interior of $A$ by $\interior(A)$.} 
\end{assumption}  
This assumption reflects conflicts of interest between the two parties: were the proposer able to choose freely from $\policy$, without any scrutiny, he would pick a policy that makes the responder worse off than under the status quo. If $\policy$ is compact, \Cref{Assumption-Conflict} is equivalent to every maximizer of $\pi_P(\cdot)$ lying in the conflict region. 

Our second assumption embodies a temptation to push the responder to her break-even point even when proposing mutually beneficial policies. Recall from \eqref{Equation-CongruenceConflict} that the congruence region, $\policy^+$, comprises the policies that strictly benefit the responder. 
\begin{assumption}\label{Assumption-BreakEven}
The proposer's near-favorite policies within the congruence region leave
the responder with little surplus: for every $\delta>0$, there exists $\tilde\pi_P\in \interior(\pi_P(\policy^+))$ such that for every $x\in \policy^+$, $\pi_P(x)\geq \tilde\pi_P$ implies $\pi_R(x)\leq \delta$.  
\end{assumption}
\Cref{Assumption-BreakEven} says that, in the congruence region, the responder's payoff eventually converges to zero as one moves toward policies that make the proposer better off. 

Many settings satisfy \Cref{Assumption-Conflict,Assumption-BreakEven}. For instance, the assumptions hold in bargaining problems over surplus and policies (e.g., \Cref{Example-Bargaining,Example-Veto}) as well as in project-selection problems (\Cref{Example-Project}) under appropriate support restrictions on the set of feasible projects. 

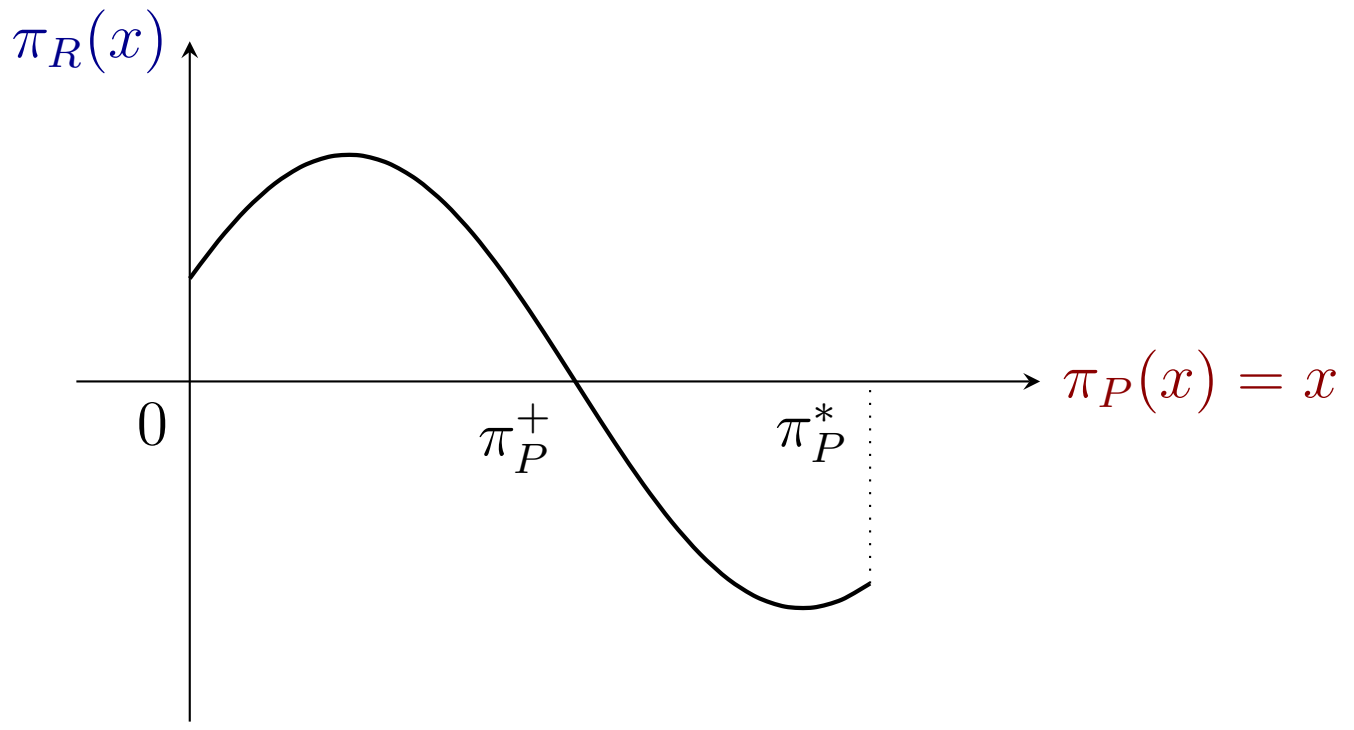
\begin{figure}[t]
\centering
\begin{tikzpicture}[scale=1.5, >=stealth]
\draw[->] (-0.5,0) -- (3.75,0)
  node[right] {$\textcolor{DarkRed}{\pi_P(x)=x}$};
\draw[->] (0,-1.5) -- (0,1.5)
  node[left] {$\textcolor{DarkBlue}{\pi_R(x)}$};
\draw[thick, domain=0:3, smooth] plot (\x, {sin(90*(\x+.3))});
\draw[dotted] (3,{sin(90*(3.3))}) -- (3,0);
\node[below left] at (0,0) {$0$};
\node[below left] at (1.7,0) {$\pi_P^+$};
\node[below left] at (3,0) {$\pi_P^*$};
\end{tikzpicture}
\caption{An illustration of \Cref{Assumption-Conflict,Assumption-BreakEven}.}
\label{Figure-LargeBasketPayoffs}
\end{figure}

Under these assumptions, two payoffs play an important role in our analysis. The first is the proposer's highest feasible payoff, and the second is his highest payoff from a policy in the congruence region:
\begin{align*}
	\pi_P^*\equiv \sup_{x\in \policy}\pi_P(x),\qquad \pi_P^+\equiv \sup_{x\in \congruence} \pi_P(x).
\end{align*} 
Observe that $0<\pi_P^+<\pi_P^*\leq \infty$. \Cref{Figure-LargeBasketPayoffs} illustrates the assumptions and these payoffs. Here, the policy space is $\policy=[0,\pi_P^*]$, $\pi_P(\cdot)$ is the identity function, and the curve shows the responder's payoff as a function of the proposer's. The congruence region is $[0,\pi_P^+)$ and the conflict region is $(\pi_P^+,\pi_P^*]$. Both \Cref{Assumption-Conflict,Assumption-BreakEven} hold: the proposer, were he able to choose freely, would favor a policy in the conflict region and, among policies in the congruence region, those that push the responder to her break-even point.

Using these payoffs, we also identify a critical temptation ratio, 
\begin{align}\label{Equation-TemptationLimit}
	t_\infty \equiv \frac{\pi_P^*-\pi_P^+}{\pi_P^+},
\end{align}
which captures the proportional gain from cherry-picking when the proposer can choose freely. In settings in which the proposer's payoff is unbounded, as in \Cref{Section-Example}, $t_\infty=\infty$. 

Our focus is not on the setting in which all policies are viable but on one in which the set of viable policies is large in a stochastic sense. Formally, we consider a sequence of probability distributions supported on the strictly positive integers, $(\psi_d)_{d=1,2,\ldots}$, each of which has a finite mean. Intuitively, the basket grows large if its size converges to $\infty$ in probability:
\begin{definition}\label{Definition-LargeBaskets}
	The basket \textbf{grows large} if for every integer $K$, $\lim_{d\rightarrow\infty} \sum_{k\leq K}\psi_d(k)= 0$.
\end{definition}
A special case is a deterministic setting, $\psi_d(k)=\mathbbm{1}[k=d]$, as in \Cref{Section-Example}. Within each basket, each policy's characteristic is drawn independently from the same distribution $\phi$.    

\begin{theorem}\label{Theorem-LargeBasket}
Suppose the basket grows large and \Cref{Assumption-ExAnte,Assumption-Conflict,Assumption-BreakEven} hold. For every $\varepsilon>0$, there exists $d_{\varepsilon}$ such that, in every game with $d>d_{\varepsilon}$, the essentially unique equilibrium involves:
\begin{enumerate}[label=\emph{(\alph*)}]
    \item The proposer proposes a mutually beneficial policy with probability greater than $1-\varepsilon$. 
    \item The uninformed responder accepts with probability less than $\odds_{\lambda}/t_\infty+\varepsilon$.
\end{enumerate}
\end{theorem}
\Cref{Theorem-LargeBasket} asserts that in large baskets, the proposer moderates nearly always, and yet the acceptance rates of the uninformed responder can remain low. We observe that as $\lambda\rightarrow 0$ or $t_\infty\rightarrow \infty$, the uninformed responder's acceptance rate converges to zero. Thus, Pareto-improving policies are proposed nearly always and yet often rejected. 

The bound on the uninformed responder's acceptance rate in part (b) is necessarily loose when $\odds_\lambda > t_\infty$; it may also be so when $\odds_\lambda<t_\infty$. The issue is a horse race: baskets whose temptation ratio substantially exceeds $t_\infty$ have vanishing probability but they still matter to the responder because her expected gain from the likely baskets also vanishes as $d\rightarrow \infty$. Accordingly, in some examples, the acceptance rate is substantially lower than the bound. The bound is tight if even the smallest possible basket size tends to infinity, $\min \supp(\psi_d)\rightarrow\infty$. Deterministic growth in basket size is a special case of this stronger condition. 

This upper bound nevertheless delivers two welfare conclusions. Let $\Pi_i^d(\lambda)$ be player $i$'s equilibrium payoff when the basket size has distribution $\psi_d$. 
\begin{corollary}\label{Corollary-LargePayoff}
The following payoff conclusions hold:
\begin{enumerate}[label=\emph{(\alph*)}]
    \item The responder's equilibrium payoff converges to zero: $\lim_{d\rightarrow\infty}\Pi_R^d(\lambda)=0$ for every $\lambda$.
    \item Suppose the proposer's payoffs are bounded. For every $\lambda$ such that $\odds_\lambda<t_\infty$, 
    \begin{align*}
        \limsup_{d\rightarrow\infty}\Pi_P^d(\lambda)\leq \lambda\left(\frac{\pi_P^+\pi_P^*}{\pi_P^*-\pi_P^+}\right)<\pi_P^+.
    \end{align*}
\end{enumerate}
\end{corollary}

\Cref{Corollary-LargePayoff} reveals the self-defeating nature of cherry-picking. In equilibrium, the proposer proposes Pareto improvements with probability converging to one. Yet, his ability to cherry-pick keeps his equilibrium payoff below what he would get if he were able to commit to proposing only Pareto improvements. If the responder is informed with sufficiently low probability, the proposer would even prefer to relinquish control over selection and have a policy chosen uniformly at random, as in \Cref{Proposition-RandomSelection}.\footnote{Because the uninformed responder would accept a randomly selected proposal with probability one---by \Cref{Assumption-ExAnte}---the proposer's payoff would then be $\int_{\congruence} \pi_P(x)\,\phi(dx)+(1-\lambda)\int_{\conflict} \pi_P(x)\,\phi(dx)$. For $\lambda$ sufficiently low, this payoff exceeds the bound in \Cref{Corollary-LargePayoff}(b).} To put it differently, the proposer would be better off with a singleton than with a large basket. 

We now describe the intuition for \Cref{Theorem-LargeBasket}. With high probability, the proposer has the option to offer policies that deliver payoffs relatively close to $\pi_P^*$ and $\pi_P^+$. While the proposer's actual choices depend on his realized basket (and the temptation ratio thereof), the uninformed responder anticipates that when the proposer puts forward a mutually beneficial policy, her payoff is close to zero (by \Cref{Assumption-BreakEven}). At the same time, she recognizes that when he cherry-picks, her payoff is negative and bounded away from zero (by \Cref{Assumption-Conflict}). Thus, she is willing to accept the proposal only if the proposer cherry-picks with vanishing probability. Yet, the uninformed responder cannot accept the proposal with probability substantially higher than $\odds_\lambda/t_\infty$ as that would invite the proposer to cherry-pick too often.

\subsection{The Value of Commitment}\label{Section-Commitment}

This section compares equilibrium behavior and payoffs with those that arise when either player can commit to a strategy in the game. Part (a) of each theorem below holds for every distribution of basket size, imposing only \Cref{Assumption-ExAnte}; for parts (b) and (c), we additionally assume \Cref{Assumption-Conflict,Assumption-BreakEven} and that $\policy$ is compact. 

\paragraph{Responder Commitment:} Suppose the responder commits to a strategy, which specifies the policies she approves (and with what probability) when informed, and the probability with which she approves when uninformed. As in \cite{armstrong2010model} and \cite{nocke2013merger}, she may commit to reject some policies that benefit her so as to steer the proposer toward ones that are even better for her. Doing so, however, strengthens his temptation to cherry-pick, so she may also need to reject proposals when uninformed. Balancing these concerns can lead the responder to target a policy other than her favorite. 

Fix a distribution on the basket size, $\psi_d$. Let $\overline\lambda_d$ be the critical threshold identified in \Cref{Theorem-UniqueEqm} such that the uninformed responder mixes in equilibrium if and only if $\lambda<\overline\lambda_d$. We denote by $\overline\Pi_R^d(\lambda)$ the responder's commitment payoff. In the setting in which all policies are viable, we obtain a sharper characterization of that payoff, which we then denote by $\overline\Pi_R^\infty(\lambda)$.

\begin{theorem}\label{Theorem-ResponderCommitment}
	Suppose the responder commits to a strategy. Then the following hold:
	\begin{enumerate}[label=\emph{(\alph*)}]
		\item The responder values commitment if $\lambda<\overline\lambda_d$. 
        \item If all policies are viable, the responder's value is 
\begin{alignat}{2}
    \overline\Pi_R^\infty(\lambda)={}&
      \max_{x\in\policy,\,\alpha\in[0,1]}
    &\quad&
      \underbrace{\left(\lambda+(1-\lambda)\alpha\right)}_{\text{Approval Probability}}\pi_R(x)
      \notag\\[0.5ex]
    &\text{\normalfont subject to}
    &&\left(\lambda+(1-\lambda)\alpha\right)\pi_P(x)
       \geq(1-\lambda)\alpha\pi_P^*.\tag{O}
\end{alignat}
	\item If the basket grows large, the responder's value converges to the above value: for every $\lambda\in (0,1)$, $\lim_{d\rightarrow\infty}\overline\Pi_R^d(\lambda)=\overline\Pi_R^\infty(\lambda)$. 
	\end{enumerate}
\end{theorem}  
We begin with part (a). It identifies how the responder profits from commitment whenever she mixes, when uninformed, in equilibrium.  A commitment to turning down proposals when uninformed would make her strictly better off because that would induce the proposer to moderate more often.\footnote{When $\lambda>\overline\lambda_d$, whether the responder benefits from commitment depends on details of the payoff environment.} In some examples, the responder's commitment also makes the proposer better off because it curbs his temptation to cherry-pick. 

Part (b) characterizes the optimal commitment strategy when all policies are viable. That strategy couples a recommended policy $x$---the only policy that the responder approves when informed---with an acceptance probability $\alpha$ for when she is uninformed. The proposer must find it incentive-compatible to obey the recommendation, which captures the obedience constraint. If the recommended policy is mutually beneficial, the responder gains from increasing that acceptance probability but doing so tightens the obedience constraint. Therefore, in any solution, either the obedience constraint binds or $\alpha =1$. This observation, together with the fact that a target outside $\congruence$ gives the responder a non-positive payoff, allows us to rewrite the maximization problem as
\begin{align*}
	\max_{x\in \congruence}\min\left\{\frac{\lambda\pi_P^*}{\pi_P^*-\pi_P(x)},1\right\}\pi_R(x).
\end{align*}
If the first term above is less than one, the proposer's obedience constraint binds. Then, if $\policy\subset \Re^n$ and $\pi_i(\cdot)$ is differentiable for each player $i$, we can write the first-order condition for an interior optimum $x^*$ as 
\begin{align*}
	\left(\pi_P^*-\pi_P(x^*)\right)\nabla\pi_R(x^*)+\pi_R(x^*)\nabla\pi_P(x^*)=0. 
\end{align*}

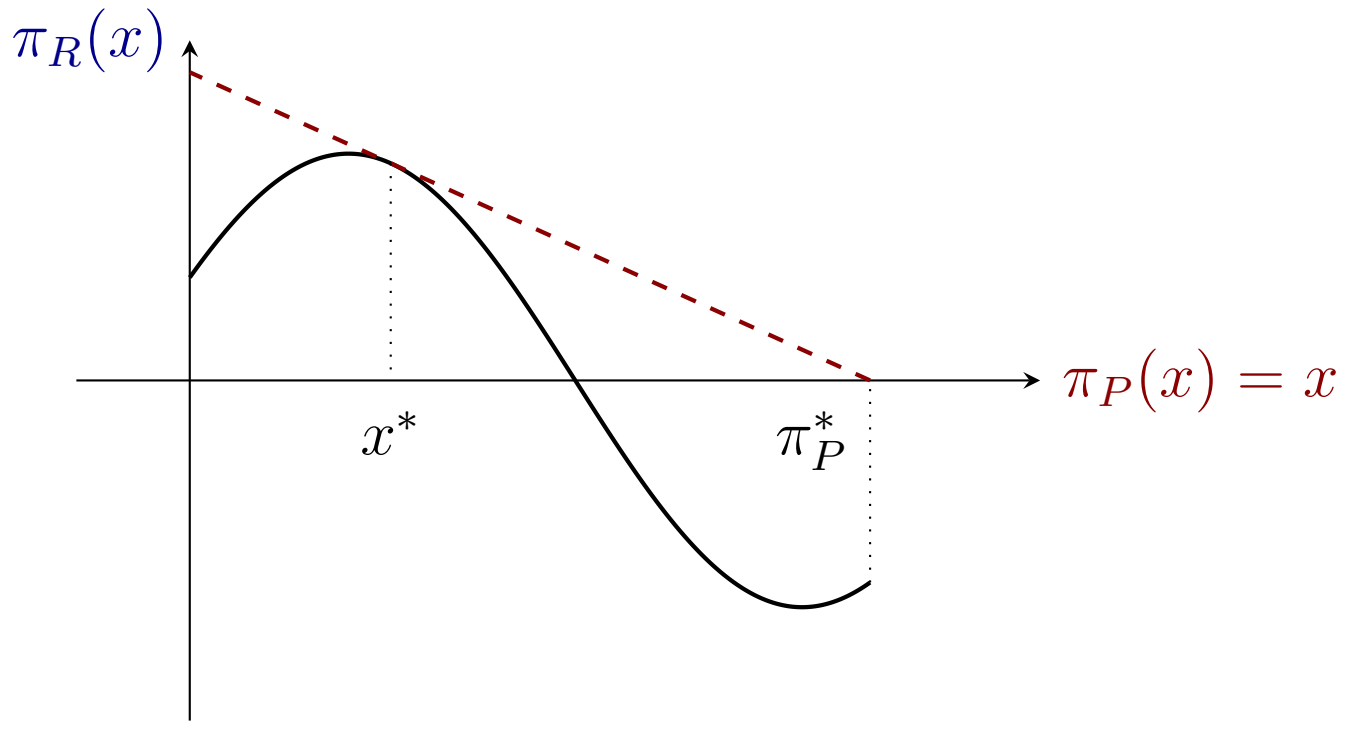
\begin{figure}[h!]
\centering
\begin{tikzpicture}[scale=1.5, >=stealth]
\draw[->] (-0.5,0) -- (3.75,0)
  node[right] {$\textcolor{DarkRed}{\pi_P(x)=x}$};
\draw[->] (0,-1.5) -- (0,1.5)
  node[left] {$\textcolor{DarkBlue}{\pi_R(x)}$};
\draw[thick, domain=0:3, samples=201, smooth] plot (\x, {sin(90*(\x+.3))});
\draw[dotted] (3,{sin(90*(3.3))}) -- (3,0);
\def\tangentx{0.8862350246}
\draw[thick, dashed, DarkRed] (3,0) -- (0,{3*sin(90*(\tangentx+.3))/(3-\tangentx)});
\draw[dotted, black] (\tangentx,{sin(90*(\tangentx+.3))}) -- (\tangentx,0);
\node[below] at (\tangentx,0) {$\vphantom{\pi_P^+}x^{*}$};
\node[below left] at (3,0) {$\vphantom{\pi_P^+}\pi_P^*$};
\end{tikzpicture}
\caption{Optimal responder commitment when the obedience constraint binds. The dashed ray from $(\pi_P^*,0)$ is tangent to the payoff curve at the target $x^*$. The responder chooses $\alpha$ to induce that target.}
\label{Figure-ResponderCommitment}
\end{figure}

\Cref{Figure-ResponderCommitment} illustrates this condition: when (O) binds, $x^*$ locally maximizes $\pi_R(x)/(\pi_P^*-\pi_P(x))$, so the ray from $(\pi_P^*,0)$ is tangent to the payoff curve at $x^*$. The intuition is that for any target $x^*$, the responder deters cherry-picking by committing to  reject with sufficiently high probability when uninformed. This deterrence makes it costly for the responder to set a target that is unattractive to the proposer. The optimal target therefore tilts toward the proposer's favorite policy, as the tangency condition shows.\footnote{An optimally chosen policy can remain optimal over a range of $\lambda$, reflecting the idea that the responder maintains the obedience constraint by adjusting the acceptance probability $\alpha$. } 

Part (c) establishes that the value of this limit problem approximates her optimal commitment payoff in large baskets. While a particular target may not be viable, a large basket contains nearby policies with high probability. Accepting policies in a small neighborhood of a suitable target can therefore approximate the limit payoff. At the same time, large baskets likely contain policies close to the proposer's favorite, which implies that the limiting obedience constraint remains relevant. In these large baskets, commitment can be highly valuable to the responder relative to her equilibrium payoff, which is close to zero.

\paragraph{Proposer Commitment:} Suppose instead that the proposer commits to a strategy. We argued after \Cref{Corollary-LargePayoff} that he would do better if he could commit to proposing only mutually beneficial policies. His optimal commitment strategy builds on this idea: by limiting how often he puts forward harmful policies, he can induce the uninformed responder to approve proposals. The optimal commitment strategy involves a form of \textit{persuasion}, where the proposer chooses a lottery over beneficial and harmful policies such that the uninformed responder always accepts and the informed responder rejects harmful policies.  

For this benchmark, we impose an additional assumption satisfied by our examples: every policy that gives the responder a non-negative payoff can be approached by a sequence of policies that give her a strictly positive payoff. Formally, in terms of the congruence and conflict regions, the assumption is $\policy\setminus\conflict \subseteq \text{cl}(\congruence)$. Given that players have continuous payoff functions, this assumption implies that the best payoff the proposer can attain while leaving the responder just indifferent can be approximated by policies that give her a strictly positive payoff. 

Given a distribution on basket size, $\psi_d$, let $\widehat\Pi_P^d(\lambda)$ denote the proposer's commitment value; as before $\overline\lambda_d$ denotes the critical threshold identified in \Cref{Theorem-UniqueEqm}. We obtain a sharper characterization of that commitment value when all policies are viable, which we denote by $\widehat\Pi_P^\infty(\lambda)$. 

\begin{theorem}\label{Theorem-ProposerCommitment}
Suppose the proposer commits to a strategy. Then the following hold:
\begin{enumerate}[label=\emph{(\alph*)}]
    \item The proposer values commitment if and only if $\lambda<\overline\lambda_d$.
    \item If all policies are viable, the proposer's value is
    \begin{alignat}{2}
        \widehat\Pi_P^\infty(\lambda)={}&
          \max_{\nu\in\Delta(\policy)}
        &\quad&\int_\policy\left(1-\lambda\1[\pi_R(x)<0]\right)\pi_P(x)\,\nu(dx)
          \notag\\[0.5ex]
        &\text{\normalfont subject to}
        &&\int_\policy\pi_R(x)\,\nu(dx)\geq0.\tag{P}
    \end{alignat}
    An optimal lottery exists. It can be chosen to be either a single policy that delivers payoff $\pi_P^+$ or a lottery over two policies, one in $\conflict$ and one in $\congruence$, that makes {\normalfont (P)} bind and delivers a payoff strictly higher than $\pi_P^+$. 
    \item If the basket grows large, the proposer's value converges to the above value: for every $\lambda\in(0,1)$,
    $\lim_{d\rightarrow\infty}\widehat\Pi_P^d(\lambda)=\widehat\Pi_P^\infty(\lambda)$.
\end{enumerate}
\end{theorem}

We begin with part (a). It identifies a sharp condition under which the proposer strictly benefits from commitment: whenever equilibrium behavior involves randomization by the uninformed responder, the proposer could strictly improve his payoff by committing to an alternative strategy. That alternative strategy involves slightly more moderation, which results in the uninformed responder accepting for sure. Once $\lambda$ crosses $\overline\lambda_d$, the uninformed responder already accepts for sure, and so the proposer no longer values commitment.\footnote{Even if he cannot commit to a strategy, he can benefit from committing to learn less. Suppose he commits to observe a Blackwell garbling of the basket's characteristics but not to his strategy. There exists a single garbling such that, for every $\lambda<\overline\lambda_d$, the resulting game has an equilibrium in which he does strictly better.}

Part (b) characterizes the optimal commitment when all policies are viable. The proposer accrues at most $\lambda\pi_P^+$ from a lottery that the uninformed responder rejects for sure (given that the informed responder rejects all policies that harm her). By committing to a maximizer of $\pi_P$ on $\policy\setminus\conflict$, the proposer could do strictly better and obtain $\pi_P^+$. Therefore, it suffices to consider lotteries that ensure that the uninformed responder accepts, which leads to the persuasion constraint (P) and the objective function above. 

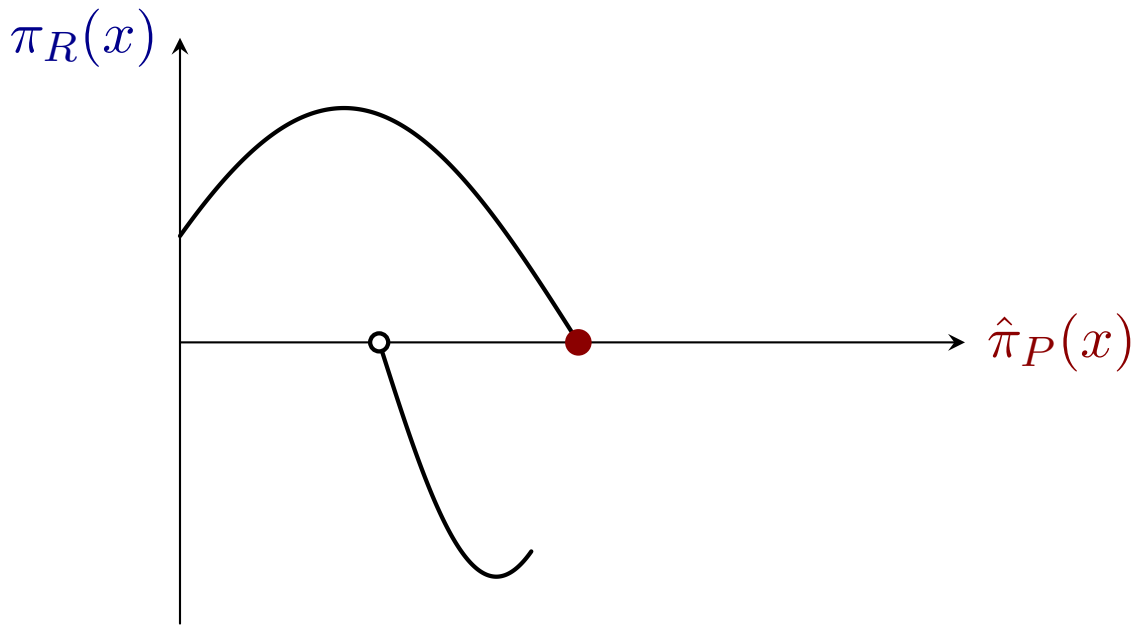
\begin{figure}[h!]
\centering
\newcommand{\ProposerCommitmentPanel}[3]{%
  \begingroup
  \pgfmathsetmacro{\q}{1-#1}
  \def\croot{1.7}
  \pgfmathsetmacro{\loss}{-sin(90*(3+.3))}
  \pgfmathsetmacro{\vend}{3*\q}
  \ifnum#3=0
    \def\xtan{#2}
    \pgfmathsetmacro{\ytan}{sin(90*(\xtan+.3))}
    \pgfmathsetmacro{\bet}{\loss/(\loss+\ytan)}
    \pgfmathsetmacro{\vstar}{\bet*\xtan+(1-\bet)*\vend}
  \fi
  \begin{tikzpicture}[x=1.55cm,y=1.55cm,>=stealth,font=\small,
    line cap=round,line join=round]
    \path[use as bounding box] (-0.55,-1.3) rectangle (4.1,1.45);
    \draw[->] (0,-1.2) -- (0,1.3)
      node[left] {$\textcolor{DarkBlue}{\pi_R(x)}$};
    \draw[->] (0,0) -- (3.35,0)
      node[right] {$\textcolor{DarkRed}{\hat{\pi}_P(x)}$};
    \draw[thick,domain=0:\croot,samples=121,smooth]
      plot (\x,{sin(90*(\x+.3))});
    \draw[thick,domain=\croot:3,samples=101,smooth]
      plot ({\q*\x},{sin(90*(\x+.3))});
    \draw[fill=white,thick] ({\q*\croot},0) circle (1.7pt);
    \fill (\croot,0) circle (1.7pt);
    \ifnum#3=1
      \fill[DarkRed] (\croot,0) circle (2.5pt);
    \else
      \draw[DarkRed,thick,dashed] (\xtan,\ytan) -- (\vend,-\loss);
      \fill (\xtan,\ytan) circle (2pt);
      \fill (\vend,-\loss) circle (2pt);
      \fill[DarkRed] (\vstar,0) circle (2.5pt);
    \fi
  \end{tikzpicture}%
  \endgroup
}
\begin{subfigure}[t]{0.48\textwidth}
    \centering
    \ProposerCommitmentPanel{0.10}{1.174741670}{0}
    \caption{Low $\lambda$}
\end{subfigure}\hfill
\begin{subfigure}[t]{0.48\textwidth}
    \centering
    \ProposerCommitmentPanel{0.50}{1.7}{1}
    \caption{High $\lambda$}
\end{subfigure}
\caption{Optimal proposer commitment. The red dashed line extends the graph to depict the upper concave envelope and the red dot marks the commitment payoff. }
\label{Figure-ProposerCommitment}
\end{figure}

The solution to this problem takes a simple form: either a single policy that yields $\pi_P^+$ or a two-point lottery that leaves the uninformed responder indifferent. \Cref{Figure-ProposerCommitment} depicts this solution, building on \Cref{Figure-LargeBasketPayoffs}, but with a horizontal axis that measures
\begin{align*}
	\hat{\pi}_P(x)\equiv \left(1-\lambda\1[\pi_R(x)<0]\right)\pi_P(x).
\end{align*}
The payoff function $\hat\pi_P(\cdot)$ coincides with $\pi_P(\cdot)$ for all mutually beneficial policies; the distinction arises for policies that harm the responder. Here, $\hat\pi_P(x)$ internalizes the rejection from the informed responder. A lottery over policies then corresponds to payoffs in the convex hull of this graph. Given the constraint (P), the optimal proposer commitment picks the rightmost point of the upper concave envelope on or above the horizontal axis. As \Cref{Figure-ProposerCommitment} shows, that point lies past $\pi_P^+$ when $\lambda$ is low and coincides with $\pi_P^+$ when $\lambda$ is high. 

When all policies are viable, the proposer's commitment strategy can choose \textit{any} lottery $\nu\in \Delta(\policy)$; in our baseline setup in which the proposer chooses from randomly drawn baskets, his commitment strategy selects instead from a restricted set of lotteries. Part (c) shows that this feasibility constraint becomes inconsequential as the basket grows large.

We close with two observations about this benchmark. First, for a fixed basket-size distribution $\psi_d$, the proposer's commitment payoff falls in the probability that the responder is informed, unlike his equilibrium payoff. Because his commitment strategy already ensures that the uninformed responder accepts with probability one, raising $\lambda$ reduces his ability to cherry-pick without offering any countervailing rewards. Second, in large baskets and relative to equilibrium payoffs, the responder benefits when the proposer's commitment strategy mixes over policies in $\congruence$ and $\conflict$, the case in \Cref{Figure-ProposerCommitment}(a). When informed, the responder can accept only those in the congruence region, which gives her a payoff bounded away from zero. By contrast, her equilibrium payoff vanishes as the basket grows large.

\section{Extensions}\label{Section-Extensions}
\subsection{When Menus Restore Trust}\label{Section-Menus}

Suppose that, rather than offering a single proposal, the proposer were required to offer the responder a menu of options. Such requirements are common. For instance, in judicial appointments across the world, a nominating body is often required to put forward several candidates from whom another authority chooses an appointee.\footnote{Examples include court appointments in various US states, Argentina, Brazil, Colombia, and South Africa, as well as the election of judges to the European Court of Human Rights.} Analogously, under the ``rule of three'' that governed federal hiring decisions in the United States, the Office of Personnel Management would put forward three candidates to the hiring manager; this rule has since been broadened to a ``rule of many.'' Mortgage regulations often induce brokers to present clients with a menu of options meeting specified criteria, to address the concern that brokers may otherwise steer clients toward loans that increase their compensation.

Our setup rationalizes these practices: when the proposer presents a menu of options, he is less able to cherry-pick, thereby making it easier for the responder to trust him. Strikingly, a menu partially delegates the choice to a potentially \textit{worse}-informed party; our finding is that this form of delegation can help. This rationale thus contrasts with the standard expertise-based perspective on delegation \citep{holmstrom1984delegation,dessein2002authority}, which emphasizes the benefits of delegating choice to a better-informed party.\footnote{Analogously, the veto-based delegation rationale \citep*{kartik2021delegation} also involves delegation to a privately informed party; there, the menu serves to screen the responder's preferences. Studying worst-case regret, \cite{guo2023regret} also identify the benefits of allowing the proposer to put forward a menu of options rather than a single one; doing so gives the responder a better fallback option than rejection.} Here, the cherry-picking trap pushes toward \textit{in}expertise-based delegation because such delegation mitigates adverse selection. We illustrate this phenomenon using a simple example. 
\begin{example}\label{Example-MenuUninformed}
In the example of \Cref{Section-Example}, if $\lambda=0$, the responder would reject whenever the basket had four or more policies. The reason is that she anticipates that the proposer selects the first-order statistic from $K$ draws, which means that the proposed policy is adversely selected from her perspective. Now suppose the proposer were required to offer not one but two options in an unordered menu. Because she cannot observe the characteristic of either option, if she accepts, she picks each with equal probability. In response, the proposer offers his two favorite options. Thus, the responder faces a 50-50 lottery between the first-order statistic, $\first_K$, and the second-order statistic, $\second_K$:
\begin{align*}
	\E\left[\frac{X^{(1)}_K+X^{(2)}_K}{2}\right]=\sum_{k=1}^K\frac{1}{k}-\frac{1}{2}.
\end{align*}
This term is lower than two for $K=2,\ldots,6$, and so the responder accepts for these basket sizes, whereas she would do so only for $K\leq 3$ in the single-proposal benchmark. 
\end{example}
\Cref{Example-MenuUninformed} shows that requiring two options can assuage the responder's fear that proposals have been \textit{extremely} cherry-picked and thereby lead to a higher acceptance rate. However, our analysis also identifies a countervailing force: a menu that is narrow relative to the basket of viable policies---such as three candidates drawn from a large pool of applicants---can backfire and lead to \textit{lower} acceptance rates than a single proposal. We illustrate this possibility below.

\begin{example}\label{Example-MenuExponential}
Suppose the responder is informed with probability $\lambda=1/4$, as in \Cref{Section-Example}. The informed responder observes the characteristic of each option in the menu and can choose to accept either one or reject both. The proposer now faces a dilemma: should he pack the menu with two cherry-picked policies or include a mutually beneficial option so as to court the informed responder? The uninformed responder, in turn, must decide whether to pick an option from the menu blindly or to go with the status quo.

\Cref{Figure-MenuWelfare} illustrates equilibrium behavior and welfare for basket sizes $K=2,\ldots,10$: the uninformed responder accepts for sure if $K\leq 6$, with interior probability if $K=7$, and rejects for sure once $K\geq 8$. Thus, once $K\geq 8$, the ex ante acceptance rate is roughly $1/4$ whereas it is roughly $1/3$ in the single-proposal benchmark (see \Cref{Figure-ExampleWelfare}). Therefore, relative to that benchmark, requiring a menu of two options raises acceptance in small baskets but lowers it in large ones. Comparing \Cref{Figure-MenuWelfare}(a) to \Cref{Figure-ExampleWelfare}(a) shows that the responder is better off with the menu at every basket size shown, whereas the proposer gains only when $4\leq K\leq 7$.  
\end{example}
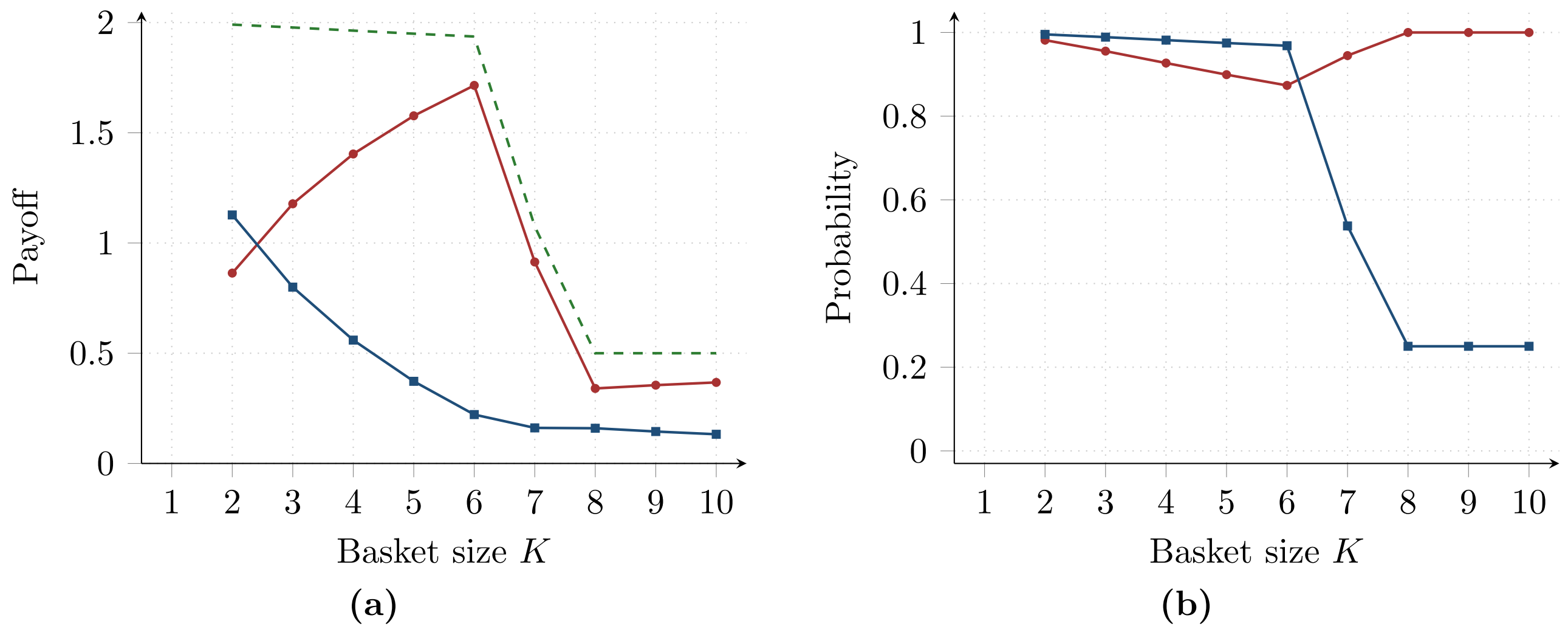
\begin{figure}[t!]
\centering

\begin{subfigure}[t]{0.48\textwidth}
\centering
\begin{tikzpicture}
\begin{axis}[
    width=\linewidth, height=6.5cm,
    xlabel style={yshift=-4pt},
    xlabel={Basket size $K$},
    ylabel={Payoff},
    xmin=0.5, xmax=10.5,
    ymin=0, ymax=2.05,
    xtick={1,2,3,4,5,6,7,8,9,10},
    ytick={0,0.5,1.0,1.5,2.0},
    grid=major,
    grid style={dotted, gray!40},
    axis lines=left,
    tick align=outside,
    label style={font=\small},
    tick label style={font=\small},
]

\addplot[
    color=cprop, thick, mark=*, mark size=1pt,
    mark options={fill=cprop, draw=cprop},
] coordinates {
    (2, 0.863552725695) (3, 1.177846123687) (4, 1.403917847657) (5, 1.576993373132) (6, 1.715057835867)
    (7, 0.913938937439) (8, 0.340288523537) (9, 0.355174806655) (10, 0.367598746712)
};

\addplot[
    color=cresp, thick, mark=square*, mark size=1pt,
    mark options={fill=cresp, draw=cresp},
] coordinates {
    (2, 1.127289454861) (3, 0.799865927389) (4, 0.559553238664) (5, 0.372596961623) (6, 0.221674125703)
    (7, 0.161305590651) (8, 0.159711420196) (9, 0.144825185730) (10, 0.132401252258)
};

\addplot[
    color=csurp, thick, dashed, mark=triangle*, mark size=0pt,
    mark options={fill=csurp, draw=csurp, solid},
] coordinates {
    (2, 1.990842180556) (3, 1.977712051076) (4, 1.963471086320) (5, 1.949590334755) (6, 1.936731961570)
    (7, 1.075244528090) (8, 0.499999943732) (9, 0.499999992385) (10, 0.499999998969)
};

\end{axis}
\node[anchor=north, font=\small] at (current bounding box.south) {\textbf{(a)}};
\end{tikzpicture}
\label{fig:menu-payoffs}
\end{subfigure}%
\hfill
\begin{subfigure}[t]{0.48\textwidth}
\centering
\begin{tikzpicture}
\begin{axis}[
    width=\linewidth, height=6.5cm,
    xlabel style={yshift=-4pt},
    xlabel={Basket size $K$},
    ylabel={Probability},
    xmin=0.5, xmax=10.5,
    ymin=-0.03, ymax=1.05,
    xtick={1,2,3,4,5,6,7,8,9,10},
    ytick={0,0.2,0.4,0.6,0.8,1.0},
    grid=major,
    grid style={dotted, gray!40},
    axis lines=left,
    tick align=outside,
    label style={font=\small},
    tick label style={font=\small},
]

\addplot[
    color=cmod, thick, mark=*, mark size=1pt,
    mark options={fill=cmod, draw=cmod},
] coordinates {
    (2, 0.981684361111) (3, 0.955424102151) (4, 0.926942172641) (5, 0.899180669510) (6, 0.873463923140)
    (7, 0.944746182856) (8, 0.999999887465) (9, 0.999999984770) (10, 0.999999997939)
};

\addplot[
    color=cacc, thick, mark=square*, mark size=1pt,
    mark options={fill=cacc, draw=cacc},
] coordinates {
    (2, 0.995421090278) (3, 0.988856025538) (4, 0.981735543160) (5, 0.974795167377) (6, 0.968365980785)
    (7, 0.537622264045) (8, 0.249999971866) (9, 0.249999996193) (10, 0.249999999485)
};

\end{axis}
\node[anchor=north, font=\small] at (current bounding box.south) {\textbf{(b)}};

\end{tikzpicture}
\label{fig:menu-probs}
\end{subfigure}
\captionsetup{aboveskip=0pt}
\caption{Panel (a) plots ex ante payoffs when the proposer must offer two options and $\lambda=1/4$, with {\color{cprop}\textbf{red}} depicting the proposer's payoff, {\color{cresp}\textbf{blue}} depicting the responder's, and {\color{csurp}\textbf{green}} depicting their sum. Panel~(b) plots the ex ante probability that the menu contains at least one Pareto improvement ({\color{cmod}\textbf{red}}) and the ex ante probability that an option is accepted ({\color{cacc}\textbf{blue}}).}
\label{Figure-MenuWelfare} 
\end{figure}

The example shows the importance of a menu's breadth relative to the basket. To formalize this implication, we consider a deterministic basket size, $K$, and require the proposer to put forward a menu of $n_K\in \{2,\ldots,K\}$ options. Our interest is in $n_K/K$, the menu's \textit{relative breadth}. The timing is that the proposer observes the basket and then chooses the menu of $n_K$ options. The responder then accepts an option in the menu or rejects them all, resulting in the status quo. If she is informed, she observes the characteristic of each option in the menu; otherwise, she does not observe the characteristic of any of them.\footnote{This dichotomy captures a setting in which the informed responder can distinguish mutually beneficial policies from those that harm her while the uninformed responder cannot. An alternative setup is one in which the responder can evaluate each item in the menu with independent probability $\lambda$. That setup yields different conclusions because the probability that the responder cannot evaluate any item vanishes in the menu size.} Thus, if the uninformed responder accepts, she picks an item in the menu with uniform probability. 
We study Perfect Bayesian equilibria (PBE), thus imposing sequential rationality off the equilibrium path.\footnote{Our definition of PBE follows \cite{fudenberg1991perfect}. We impose a stronger solution concept here because a Bayes--Nash equilibrium would allow the informed responder to reject all policies in an ``off-path'' menu that contains a mutually beneficial policy. } 
\begin{theorem}\label{Theorem-Menu}
    Suppose \Cref{Assumption-ExAnte,Assumption-Conflict,Assumption-BreakEven} hold. There exist $\underline\breadth$ and $\overline\breadth$, where $0<\underline\breadth\leq \overline\breadth <1$, such that the following hold for every $\lambda\in (0,1)$:
    \begin{enumerate}[label=\emph{(\alph*)}]
        \item Broad menus promote trust: If $ 	\liminf_{K\rightarrow\infty}n_K/K>\overline \breadth$, then, for all $K$ sufficiently large, the uninformed responder accepts with probability one in every PBE.
        \item Narrow menus hurt trust: If $ 	\limsup_{K\rightarrow\infty}n_K/K<\underline \breadth$, then, for all $K$ sufficiently large, the uninformed responder rejects with probability one in every PBE. 
    \end{enumerate}    
    Moreover, $\underline\breadth=\overline\breadth$ if the players have opposing rankings over policies.\footnote{That is, if for every $x$ and $x'$, $\pi_P(x)>\pi_P(x')$ implies $\pi_R(x)<\pi_R(x')$.}
\end{theorem}
\Cref{Theorem-Menu} describes the equilibrium implications of both relatively broad and narrow menus. The logic for part (a) is that a broad menu limits the proposer's ability to cherry-pick. Thus, the uninformed responder has less to fear and is more willing to accept. This finding demonstrates the aforementioned benefits of \textit{in}expertise-based delegation: when preferences are misaligned, an expert can counter a non-expert's suspicion by delegating some of the choice to her. 

Part (b) identifies the countervailing force that applies with narrow menus. Here, the uninformed responder rejects with probability one, whereas with a single proposal she accepts with positive probability for every $\lambda>0$. Thus, a narrow menu backfires. 

Why does a narrow menu backfire? The issue is that in the single-proposal benchmark, the proposer has to choose between securing the informed responder's approval and exploiting the uninformed responder. With a narrow menu of two or more options, he can do both: if the uninformed responder accepts with positive probability, the proposer's best response would be to pack the menu, whenever he can, with at most one mutually beneficial policy and at least $(n_K-1)$ cherry-picked ones. The responder's gain from the former vanishes as the basket grows while her losses remain bounded away from zero. The resulting lottery over these policies therefore cannot rationalize acceptance. In effect, the menu pairs several poison pills with at most a single sweetener. The informed responder can choose the sweetener, but the uninformed responder would steer clear for fear of drawing a poison pill.

We obtain the critical breadths, $\underline\breadth$ and $\overline\breadth$, using the following logic. When the uninformed responder accepts, she draws an option from the menu at random. The lower cutoff $\underline\breadth$ is the smallest breadth $\breadth$ at which a random draw from the top $\breadth$-quantile of the proposer's payoffs gives her a nonnegative expected payoff; a narrower menu, which the proposer fills with his favorites, cannot win her approval. The upper cutoff $\overline\breadth$ is the breadth at which a random draw from the bottom $\breadth$-quantile of her own payoffs---the worst menu that the proposer could assemble---gives her an expected payoff of zero; a broader menu wins her approval no matter what it contains. The two cutoffs generally differ but coincide if the players have opposing rankings over policies, as in \Cref{Section-Example}. In that example, $\underline\breadth=\overline\breadth=1/e$. There, the proposer would have to offer a relatively broad menu (at least $37\%$) to obtain the uninformed responder's approval.

Our discussion above focuses on how a menu's breadth affects acceptance rates. Below, we obtain implications for payoffs. 
\begin{proposition}\label{Proposition-MenuWelfare}
Suppose \Cref{Assumption-ExAnte,Assumption-Conflict,Assumption-BreakEven} hold and fix $\lambda\in(0,1)$. For all $K$ sufficiently large, the following ex ante payoff comparisons hold in every PBE of the menu game:
\begin{enumerate}[label=\emph{(\alph*)}]
    \item The proposer strictly prefers the single-proposal benchmark if the menu is narrow, i.e., $\limsup_{K\rightarrow\infty}n_K/K<\underline\breadth$.
    \item The responder strictly prefers the menu if at least one of the following holds:
    \begin{enumerate}[label=\emph{(\roman*)},nolistsep]
        \item the menu is broad, i.e., $\liminf_{K\rightarrow\infty}n_K/K>\overline\breadth$;
        \item $\odds_\lambda<t_\infty$, so that she mixes when uninformed in the single-proposal benchmark.\footnote{Recall that $t_\infty$, defined on p. \pageref{Equation-TemptationLimit}, is the limiting temptation ratio in a large basket.}
    \end{enumerate}
\end{enumerate}
\end{proposition}
Part (a) highlights how a narrow menu backfires on the proposer. By \Cref{Theorem-Menu}(b), the uninformed responder rejects for sure, so the proposer obtains approval only from the informed responder. By contrast, with a single proposal, he would obtain at least some approval from the uninformed responder. Part (b) identifies conditions under which the responder benefits from the menu. Under condition (i), the responder's payoff is bounded away from zero in every PBE of the menu game because a broad menu limits how adversely selected the option she draws can be. By contrast, her payoff converges to zero in the equilibrium of the single-proposal game. Condition (ii) is more subtle: even a narrow menu, which she rejects when uninformed, can make her better off. The idea is that in the single-proposal benchmark, if she mixes when uninformed, then her surplus comes only from approving moderate proposals when informed. With a narrow menu, she still gains nothing when uninformed, but when informed she can choose the sweetener from the menu, including in baskets where, with a single proposal, the proposer would have cherry-picked.

\subsection{Search and Initiative}\label{Section-Effort}

Our baseline model takes the basket's size and composition as given. Yet, in many applications, the proposer identifies viable options through costly effort or search. A division manager scouts for acquisitions and capital projects before recommending any to headquarters; a politician drafts and redrafts a bill before trying to get the support of others. The proposer's willingness to engage in this costly process then hinges on how often the responder accepts a proposal, which in turn depends on her beliefs about whether the proposal was cherry-picked. 

This section describes the implications that the cherry-picking trap has for the search process. We consider two  setups. The first is a game in which the proposer's effort raises the probability of identifying viable policies. The second is an optimal stopping problem in which the proposer chooses when to stop searching and what to propose. 

\paragraph{Attention Rewards Initiative:} Suppose that, before learning anything about the basket's size or composition, the proposer chooses how much effort to  put into identifying viable policies. We model this choice as a probability $q\in[0,1]$ of discovering a basket, which costs him $c(q)$, where $c:[0,1]\rightarrow\Re_{\geq 0}$ is differentiable and satisfies $c(0)=c'(0)=0$. He bears this cost whether or not his effort succeeds. With probability $q$, he identifies a basket $B$ drawn from the baseline distribution $\mu$; otherwise, no policy is available and the status quo prevails. If he identifies a basket, the game proceeds as before. For concreteness, we assume that the responder does not observe the proposer's effort; our results below would continue to hold if she did. We study Perfect Bayesian equilibria of this game.\footnote{PBE imposes two important restrictions here: first, the responder is sequentially rational, and second, the ``you-can't-signal-what-you-don't-know'' property implies that an unexpected proposal does not change the uninformed responder's belief about the composition of the basket.} 

We evaluate how the proposer's willingness to exert effort varies with the probability that the responder is informed. Recall that, in the unique equilibrium of the baseline model, the uninformed responder accepts for sure if and only if $\lambda\geq\overline\lambda$, and that $\overline\lambda>0$ whenever cherry-picking is a concern (i.e., $\E[\pi_R(C(\cdot))]<0$).
\begin{proposition}\label{Proposition-Effort}
	Suppose $\overline\lambda>0$. For any selection of PBE, the following hold:
	  \begin{enumerate}[label=\emph{(\alph*)}]
		\item The proposer's equilibrium effort increases in $\lambda$ on $(0,\overline\lambda]$ and decreases in $\lambda$ on $[\overline\lambda,1)$.
		\item The proposer's equilibrium payoff is strictly single-peaked in $\lambda$ and attains a unique maximum at $\lambda=\overline\lambda$.  
		\item The responder's equilibrium payoff strictly increases in $\lambda$ on $(0,\overline\lambda]$.  
	\end{enumerate}
\end{proposition}
The proposition asserts that raising the probability that the responder is informed within the interval $(0,\overline\lambda)$ induces the proposer to work harder to identify viable options and makes both players better off. Beyond that point, the proposer's effort and payoff decline; the responder's payoff may continue to increase or decline depending on primitives. Equilibrium effort is maximized at $\overline\lambda$ regardless of the shape of the cost function. 

These equilibrium effects are rooted in \Cref{Proposition-Education}, where we described the benefits of educating the responder. Recall that in the unique equilibrium of the baseline game, raising the probability that the responder is informed within $(0,\overline\lambda)$ raises the acceptance rates of both moderate and cherry-picked policies. Therefore, with an endogenous effort choice, the proposer expects to see the fruits of his labor with higher probability, which makes him more willing to work at a higher $\lambda$. This force disappears once $\lambda$ crosses $\overline\lambda$ because the uninformed responder then accepts for sure; further increases in $\lambda$ only reduce the returns to cherry-picking. As for the responder, she benefits from both the incentive effect and her own greater ability to scrutinize proposals so long as $\lambda\leq \overline\lambda$. Once $\lambda$ crosses $\overline\lambda$, these effects push in opposite directions, so the net effect is ambiguous.

\Cref{Proposition-Effort} identifies how greater scrutiny can attract and reward initiative from the proposer. Suppose headquarters invests in its capacity to supervise a division, so as to better understand its needs and potential. A division manager is then more willing to search for ideas to put before headquarters because he expects more of his proposals to be trusted. This finding connects to \citet{halacprat2016}, where workers' incentives to exert effort depend on their expectation that good performance will be recognized. It contrasts with \cite{aghiontirole1997}, where greater scrutiny from the principal reduces the agent's effective control and thereby suppresses his incentive to acquire information; a similar force manifests in our setting only once $\lambda$ crosses $\overline\lambda$. This contrast highlights the distinction between a technical review team that helps headquarters understand a division's proposal and a central product team that develops alternatives that can displace that proposal; our analysis identifies the benefits of the former whereas \cite{aghiontirole1997} highlights the costs of the latter. 

\Cref{Proposition-Effort} also implies that when choosing how to allocate scarce attention across divisions, headquarters may find it optimal to concentrate attention on some---spurring initiative from those divisions---while neglecting others. For example, suppose there are $n\geq 2$ symmetric divisions whose projects are approved separately. Each manager incurs the effort cost $c(q)=\kappa q^2/2$. Headquarters allocates a total ``attention capacity'' $\Lambda$ across the $n$ divisions, choosing $(\lambda_1,\ldots,\lambda_n)$ such that $\sum_{i} \lambda_i =\Lambda$. Suppose attention is scarce ($\Lambda <n\overline\lambda$) and that effort is sufficiently costly that no manager chooses full effort. Then, the optimal allocation is unequal because the equilibrium payoff that headquarters accrues from devoting capacity to division $i$ is strictly convex on $[0,\overline\lambda]$.\footnote{In fact, if $\Lambda< \overline\lambda$, the optimal allocation is to concentrate attention on one division and neglect all others.}

\paragraph{Optimal Stopping and the Productivity Paradox:} Suppose that the proposer searches for policies over time, choosing when to stop and what to propose. We consider this optimal stopping problem within the context of the example of \Cref{Section-Example}. Search begins at time zero, and both players observe elapsed time. Viable policies arrive according to a Poisson process at rate $\eta>0$, and each has a characteristic that is an independent draw from the unit exponential distribution. We interpret $\eta$ as the searcher's productivity. The proposer privately observes each arrival and can choose to stop or continue; if he stops, he can propose any policy that has arrived by then, or abandon search. If he proposes a policy, the game proceeds as before: the responder, who is informed with probability $\lambda\in(0,1)$, chooses whether to accept it. Acceptance implements the policy permanently; if she rejects the proposal or the proposer abandons search, then the game ends with the status quo left in place.

Search delays the gains from policy. Both players discount at rate $r>0$, and the status quo continues while the proposer searches. If policy $x$ is adopted at time $T$, from then on, the proposer and responder, respectively, obtain the flow payoffs $\pi_P(x)=x$ and $\pi_R(x)=2-x$. We normalize players' discounted utility so that the adoption yields a payoff of $e^{-r T}\pi_i(x)$. 

We characterize the essentially unique stationary Perfect Bayesian equilibrium, i.e., a PBE in which no player's equilibrium strategy conditions on time.\footnote{The game has no equilibrium in which the proposer abandons search altogether, since the informed responder accepts any mutually beneficial policy. However, it may have non-stationary PBE.} The proposer's stopping decision depends on how likely each policy is to be accepted. An informed responder accepts if and only if $x\leq2$, where we break ties in favor of acceptance. Suppose that the uninformed responder accepts with probability $\uninf$. Then the proposer's expected normalized gain from proposing $x$ is
\begin{align*}
\hat\pi_P(x,\uninf)=
\begin{cases}
[\lambda+(1-\lambda)\uninf]x,&x\leq2,\\
(1-\lambda)\uninf x,&x>2.
\end{cases}
\end{align*}
This payoff function determines the proposer's stopping rule. Writing $X$ for a unit exponential draw, his \textit{reservation value} $v$ equates the cost of waiting with the option value of another arrival:
\begin{equation}\label{Equation-DiscountReservation}
r v=\eta\E\bigl[\max\{\hat\pi_P(X,\uninf)-v,0\}\bigr].
\end{equation}
The proposer stops and proposes a policy once its expected gain exceeds his reservation value. The distribution of a proposal's characteristic is therefore independent of when it is made. 

We evaluate how the proposer's productivity affects equilibrium behavior and welfare. We focus on the case in which the proposer is more likely to face an uninformed responder ($\lambda<1/2$), as that delivers the sharpest conclusions.
\begin{proposition}\label{Proposition-OptimalStopping}
Suppose $\lambda<1/2$. There exist $\underline\eta>0$ and $\overline\eta> \underline\eta$ such that the game admits an essentially unique stationary PBE with the following behavior and payoffs:
\begin{enumerate}[label=\emph{(\alph*)}]
	\item If $\eta \leq\underline\eta$, the uninformed responder accepts for sure. The proposer offers the first policy with characteristic $x\geq x_+(\eta)$, where $x_+(\eta)\leq 1$.
	\item If $\eta\in (\underline\eta,\overline\eta)$, the uninformed responder mixes. The proposer offers the first policy with characteristic $x\geq 1$. In this range, the uninformed responder's acceptance probability $\alpha_\eta$ and the proposer's equilibrium payoff strictly fall in $\eta$. 
	\item If $\eta \geq \overline\eta$, the uninformed responder mixes. The proposer offers the first policy whose characteristic lies in $[x_+(\eta),2]\cup [x_{-}(\eta),\infty)$, where $x_+(\eta)< 2\leq x_{-}(\eta)$.
\end{enumerate}	
\end{proposition}

The proposition identifies three regions of interest. First, if the proposer is sufficiently unproductive, he is not so picky. In particular, he uses a cutoff below one, which guarantees that the uninformed responder accepts.\footnote{For the unit exponential distribution, $\E[X\mid X\geq x]=x+1$.} 

Once his productivity crosses $\underline\eta$, the cherry-picking trap ensues: now, were the uninformed responder to accept for sure, the proposer would become pickier---passing over modest mutually beneficial policies---so that his proposals would harm her on average. Therefore, in equilibrium, the uninformed responder accepts with lower probability. In response, the proposer uses a cutoff rule ($x\geq 1$) that guarantees that the responder is willing to mix. In this region, productivity backfires for the proposer: his equilibrium payoff falls in $\eta$ because the rise in rejection rates more than offsets the benefit of shorter delay. The responder, by contrast, gains from greater productivity because she receives proposals from the same distribution, only sooner. 

In the third region, the uninformed responder again mixes, but the proposer's selection rule takes a more complicated form: he proposes mutually beneficial policies close to the responder's break-even point and cherry-picked policies that are sufficiently tempting, and passes over the rest. This behavior mirrors \Cref{Theorem-UniqueEqm}, in which the proposer cherry-picks when doing so is sufficiently lucrative and otherwise moderates. Here, both the proposer and responder can suffer from local improvements in productivity.

\Cref{Proposition-OptimalStopping} shows that the cherry-picking trap generates a \textit{productivity paradox}: greater productivity can backfire because it makes waiting for more favorable policies more attractive to the proposer, inducing the responder to trust him less. Productivity can be a curse for both parties. For example, suppose $\lambda=1/4$, as in \Cref{Section-Example}, and the discount rate, $r=1$. Both players are better off with $\eta=3$ than with the proposer being twice as productive. If $\eta=3$, both parties incur greater delay but the uninformed responder accepts for sure, resulting in fewer disagreements. At twice the productivity, the uninformed responder rejects with positive probability, and her rational distrust causes the proposer's and responder's equilibrium payoffs to fall by 44\% and 35\%, respectively.\footnote{If $\eta=3$, the proposer's equilibrium payoff is $0.907$ and the responder's is $0.097$; if $\eta=6$, these values are $0.504$ and $0.063$, respectively.} Thus, a division manager who is better at generating ideas may be trusted less by headquarters, and both could be worse off.\footnote{An implication of the productivity paradox is that holding $\eta$ fixed, both parties could benefit from a random expiration deadline, which reduces the proposer's temptation to wait for a policy that benefits him alone.} 

\section{Discussion and Related Literature}\label{Section-Literature}

The cherry-picking trap identifies a challenge fundamental to organizations and politics: when a proposer chooses to put forward one specific policy among many, the sheer fact that he has selected it might reflect negatively on it. This inference is reminiscent of \cite{akerlof1970market}'s lemons problem where a seller's willingness to part with an object makes the buyer beware. We show that the ability to cherry-pick damages trust, even when mutually beneficial policies are frequently available and proposed. Parties may favor random over strategic selection and both may prefer the responder to be better informed. Offering a menu of options delegates some of the choice to a potentially uninformed responder and can make both parties better off by limiting the scope for strategic selection. However, a menu that is too narrow relative to the basket can backfire and lead to lower acceptance rates than a single proposal. Finally, equilibrium distrust reduces the proposer's incentive to search for viable policies, and a more productive proposer can be worse off.  

We obtain these conclusions by combining two interrelated but distinct frictions. First, the responder does not observe which policies are viable and must rely on the proposer's expertise. Second, the responder may be unable to evaluate the policy that the proposer puts forward. Combining these frictions allows us to study how the scope for strategic selection affects trust. Were there a single viable policy, the uninformed responder would approve, and giving the proposer more options can undermine that trust, making  both parties worse off. Each of these frictions has been studied in isolation in prior work. 

The first friction is the focus of the project-selection literature initiated by \cite{armstrong2010model} and \cite{nocke2013merger}.\footnote{This analysis has been extended in numerous directions; for example, see \cite*{li2017power}, \cite{guo2023regret}, \cite{mallick2025art}, and \cite{li2026full}.} In their work, an agent privately knows which projects are viable and chooses one to propose to a principal who can always evaluate proposed projects. In an optimal mechanism,  the principal may reject some mutually beneficial projects so as to steer the agent toward projects that benefit her even more. This logic manifests in our responder-commitment benchmark with the twist that the responder must also curb the proposer's temptation to cherry-pick, which tilts her optimal target toward his favorite policy.

The second friction that we study is that the responder cannot always evaluate the proposed policy. This friction features in an early literature on inspections, auditing, and verification \citep[e.g.,][]{border1987samurai,khalil1997auditing}. When the principal cannot commit to inspections, equilibrium typically features random inspection, which requires the agent to cheat some of the time.\footnote{Recent work in this vein includes \cite{alonso2024organizing}, who study a designer who can choose an experiment, a sender who can misrepresent it at a cost, and a receiver who can audit it, learning the truth with some probability. They find that the principal can benefit from auditing less, because lax auditing induces the designer to choose a more informative experiment. Also see \cite{figueroa2024benefit}, who consider how an applicant can strategically obfuscate her proposal in a patent approval process.} This friction also manifests in recent work on bargaining and trade when offers are not freely observed; see \cite{ravid2020ultimatum}, \cite*{denti2022experimental}, \cite{wolitzky2023unobserved}, and \cite{fabbri2025attention}. These settings and ours share the strategic logic that unconditional acceptance invites opportunism, so rejection may be necessary to keep the proposer in line. Our setting differs in two respects. The proposer selects from a basket that the responder cannot see. And our setting delivers an essentially unique equilibrium, which has a positive acceptance rate. Prior work, by contrast, features multiple equilibria, some of which involve no trade.

\cite*{che2013pandering} show how strategic selection can affect trust through pandering. In their model, a sender observes the value of each project and recommends one to the receiver; the receiver, who is otherwise uninformed, is free to adopt any one. The parties' only disagreement is on how to rank projects relative to the outside option. If the receiver's outside option is attractive enough, the sender may have to pander toward a project that looks better conditional on being recommended. By contrast, a full-trust equilibrium---in which the sender recommends his favorite and the receiver follows that recommendation---exists if the receiver expects to benefit from each project ex ante, i.e., the analogue of \Cref{Assumption-ExAnte} holds. There, more projects would only make both parties better off, whereas a larger basket breeds distrust in our setting. Also on cheap-talk communication, \cite{li2026learning} consider a setting in which a sender commits to how much to learn prior to communication; they show that the sender optimally chooses to learn coarsely so that his communication is more credible.

Our results in \Cref{Section-Effort} contrast with \citet{aghiontirole1997}. Therein, an uninformed principal rubber-stamps the agent's recommendation, which gives the agent incentives to acquire information. When the principal is better informed, her tendency to overrule the agent crowds out his initiative. In our model, greater scrutiny spurs initiative: the proposer works harder to identify viable options when the responder is better informed because his proposal then passes more often. In this regard, our work identifies the value of supervisory attention, a point made in a different context by \cite{halacprat2016}. We also observe how supervisory attention can have a convex reward, which implies that when headquarters is dividing its attention across multiple otherwise identical divisions, its optimal scheme concentrates on some divisions and neglects others. In this regard, our analysis complements \cite*{dessein2016rational}, who  identify the benefits of concentrating attention on a few tasks so as to facilitate better coordination; our setup does not feature a coordination motive. 

Parallel to the problem we study is the concern that a researcher who accesses a rich data set can selectively choose what to disclose, which has been studied in various forms by \citet*{ditillio2017persuasion,ditillio2021strategic}, \cite{bardhi2026anecdotes}, \cite*{dai2026bayesian}, and  \cite{gao2026inference}. Therein too, flexibility enhances the ability to cherry-pick although the strategic selection of data leads to a fundamentally different framework and inference problem. 

Our work speaks to how inefficient status quo policies persist even when Pareto improvements are available. \cite{fernandez1991resistance} study how ex ante uncertainty about who gains and who loses creates resistance to reform; \Cref{Assumption-ExAnte} shuts down this force by ensuring that the responder benefits, on average, from a randomly selected policy. \cite{strulovici2010learning} and \cite{dziuda2016dynamic} identify dynamics that engender a status quo bias. Our mechanism is static and stems from strategic selection alone: the responder rejects proposals that are likely Pareto improvements because she cannot rule out that they were cherry-picked. Were the proposer able to commit to put forward only Pareto improvements, the status quo would be overturned whenever one is available. 

Our findings also offer a strategic account of zero-sum thinking in politics, which features prominently in populist rhetoric today. While such thinking is sometimes viewed as a psychological trait \citep{meegan2010zero,davidai2019politics},  recent work identifies rational underpinnings: \cite*{ali2025political} show that voters may be wary of policies that others support, and \cite{dziuda2025traps} identify how politicians may forgo common-interest policies to keep voters pessimistic about cooperation.\footnote{Less related, \cite*{ali2021adverse} show theoretically and experimentally that zero-sum thinking persists when people do not learn about counterfactuals. \cite{DelgadoVegaLee2025} identify how growth can eventually lead to zero-sum conflict.} Our model delivers the complementary rationale that strategic selection can lead the responder to treat a proposal with prejudice and reject it, even when both parties are likely to benefit. Despite abundant opportunities for mutual gain, the cherry-picking trap pushes behavior toward that of a zero-sum game.

\begin{singlespace}
    \addcontentsline{toc}{section}{References}
    \bibliographystyle{ecta}
    \bibliography{com}

\begin{thebibliography}{50}
\newcommand{\enquote}[1]{``#1''}
\expandafter\ifx\csname natexlab\endcsname\relax\def\natexlab#1{#1}\fi

\bibitem[\protect\citeauthoryear{Aghion and Tirole}{Aghion and
  Tirole}{1997}]{aghiontirole1997}
\textsc{Aghion, P. and J.~Tirole} (1997): \enquote{Formal and Real Authority in
  Organizations,} \emph{Journal of Political Economy}, 105, 1--29.

\bibitem[\protect\citeauthoryear{Akerlof}{Akerlof}{1970}]{akerlof1970market}
\textsc{Akerlof, G.~A.} (1970): \enquote{The Market for ``Lemons": Quality
  Uncertainty and the Market Mechanism,} \emph{Quarterly Journal of Economics},
  84, 488--500.

\bibitem[\protect\citeauthoryear{Ali, Kartik, and Kleiner}{Ali
  et~al.}{2023}]{ali2023sequential}
\textsc{Ali, S.~N., N.~Kartik, and A.~Kleiner} (2023): \enquote{Sequential veto
  bargaining with incomplete information,} \emph{Econometrica}, 91, 1527--1562.

\bibitem[\protect\citeauthoryear{Ali, Mihm, and Siga}{Ali
  et~al.}{2025}]{ali2025political}
\textsc{Ali, S.~N., M.~Mihm, and L.~Siga} (2025): \enquote{The Political
  Economy of Zero-Sum Thinking,} \emph{Econometrica}, 93, 41--70.

\bibitem[\protect\citeauthoryear{Ali, Mihm, Siga, and Tergiman}{Ali
  et~al.}{2021}]{ali2021adverse}
\textsc{Ali, S.~N., M.~Mihm, L.~Siga, and C.~Tergiman} (2021): \enquote{Adverse
  and advantageous selection in the laboratory,} \emph{American Economic
  Review}, 111, 2152--78.

\bibitem[\protect\citeauthoryear{Alonso and C{\^a}mara}{Alonso and
  C{\^a}mara}{2024}]{alonso2024organizing}
\textsc{Alonso, R. and O.~C{\^a}mara} (2024): \enquote{Organizing data
  analytics,} \emph{Management Science}, 70, 3123--3143.

\bibitem[\protect\citeauthoryear{Armstrong and Vickers}{Armstrong and
  Vickers}{2010}]{armstrong2010model}
\textsc{Armstrong, M. and J.~Vickers} (2010): \enquote{A model of delegated
  project choice,} \emph{Econometrica}, 78, 213--244.

\bibitem[\protect\citeauthoryear{Austen-Smith}{Austen-Smith}{1993}]{austen1993information}
\textsc{Austen-Smith, D.} (1993): \enquote{Information and Influence: Lobbying
  for Agendas and Votes,} \emph{American Journal of Political Science}, 37,
  799--833.

\bibitem[\protect\citeauthoryear{Bardhi and Bobkova}{Bardhi and
  Bobkova}{2026}]{bardhi2026anecdotes}
\textsc{Bardhi, A. and N.~Bobkova} (2026): \enquote{Anecdotes or Statistics?}
  Working Paper.

\bibitem[\protect\citeauthoryear{Border and Sobel}{Border and
  Sobel}{1987}]{border1987samurai}
\textsc{Border, K.~C. and J.~Sobel} (1987): \enquote{Samurai accountant: A
  theory of auditing and plunder,} \emph{Review of Economic Studies}, 54,
  525--540.

\bibitem[\protect\citeauthoryear{{Bueno de Mesquita} and Dziuda}{{Bueno de
  Mesquita} and Dziuda}{2025}]{dziuda2025traps}
\textsc{{Bueno de Mesquita}, E. and W.~Dziuda} (2025): \enquote{Partisan
  Traps,} Working Paper.

\bibitem[\protect\citeauthoryear{Cameron and McCarty}{Cameron and
  McCarty}{2004}]{CameronMcCarty2004}
\textsc{Cameron, C. and N.~McCarty} (2004): \enquote{Models of Vetoes and Veto
  Bargaining,} \emph{Annual Review of Political Science}, 7, 409--435.

\bibitem[\protect\citeauthoryear{Che, Dessein, and Kartik}{Che
  et~al.}{2013}]{che2013pandering}
\textsc{Che, Y.-K., W.~Dessein, and N.~Kartik} (2013): \enquote{Pandering to
  persuade,} \emph{American Economic Review}, 103, 47--79.

\bibitem[\protect\citeauthoryear{Dai, Fudenberg, and Pei}{Dai
  et~al.}{2026}]{dai2026bayesian}
\textsc{Dai, Y., D.~Fudenberg, and H.~Pei} (2026): \enquote{Bayesian Persuasion
  with Selective Disclosure,} Working Paper.

\bibitem[\protect\citeauthoryear{Davidai and Ongis}{Davidai and
  Ongis}{2019}]{davidai2019politics}
\textsc{Davidai, S. and M.~Ongis} (2019): \enquote{The politics of zero-sum
  thinking: The relationship between political ideology and the belief that
  life is a zero-sum game,} \emph{Science Advances}, 5, eaay3761.

\bibitem[\protect\citeauthoryear{Delgado-Vega and Lee}{Delgado-Vega and
  Lee}{2025}]{DelgadoVegaLee2025}
\textsc{Delgado-Vega, {\'A}. and B.~E. Lee} (2025): \enquote{When Growth Leads
  to Zero-Sum Conflict,} Working paper.

\bibitem[\protect\citeauthoryear{Denti, Marinacci, and Rustichini}{Denti
  et~al.}{2022}]{denti2022experimental}
\textsc{Denti, T., M.~Marinacci, and A.~Rustichini} (2022):
  \enquote{Experimental cost of information,} \emph{American Economic Review},
  112, 3106--3123.

\bibitem[\protect\citeauthoryear{Dessein}{Dessein}{2002}]{dessein2002authority}
\textsc{Dessein, W.} (2002): \enquote{Authority and Communication in
  Organizations,} \emph{Review of Economic Studies}, 69, 811--838.

\bibitem[\protect\citeauthoryear{Dessein, Galeotti, and Santos}{Dessein
  et~al.}{2016}]{dessein2016rational}
\textsc{Dessein, W., A.~Galeotti, and T.~Santos} (2016): \enquote{Rational
  Inattention and Organizational Focus,} \emph{American Economic Review}, 106,
  1522--1536.

\bibitem[\protect\citeauthoryear{Di~Tillio, Ottaviani, and
  S{\o}rensen}{Di~Tillio et~al.}{2017}]{ditillio2017persuasion}
\textsc{Di~Tillio, A., M.~Ottaviani, and P.~N. S{\o}rensen} (2017):
  \enquote{Persuasion Bias in Science: Can Economics Help?} \emph{Economic
  Journal}, 127, F266--F304.

\bibitem[\protect\citeauthoryear{Di~Tillio, Ottaviani, and
  S{\o}rensen}{Di~Tillio et~al.}{2021}]{ditillio2021strategic}
---\hspace{-.1pt}---\hspace{-.1pt}--- (2021): \enquote{Strategic Sample
  Selection,} \emph{Econometrica}, 89, 911--953.

\bibitem[\protect\citeauthoryear{Dziuda and Loeper}{Dziuda and
  Loeper}{2016}]{dziuda2016dynamic}
\textsc{Dziuda, W. and A.~Loeper} (2016): \enquote{Dynamic collective choice
  with endogenous status quo,} \emph{Journal of Political Economy}, 124,
  1148--1186.

\bibitem[\protect\citeauthoryear{Ellis and Groll}{Ellis and
  Groll}{2020}]{ellis2020strategic}
\textsc{Ellis, C.~J. and T.~Groll} (2020): \enquote{Strategic Legislative
  Subsidies: Informational Lobbying and the Cost of Policy,} \emph{American
  Political Science Review}, 114, 179--205.

\bibitem[\protect\citeauthoryear{Fabbri}{Fabbri}{2026}]{fabbri2025attention}
\textsc{Fabbri, F.} (2026): \enquote{Attention Holdup,} Working Paper.

\bibitem[\protect\citeauthoryear{Fernandez and Rodrik}{Fernandez and
  Rodrik}{1991}]{fernandez1991resistance}
\textsc{Fernandez, R. and D.~Rodrik} (1991): \enquote{Resistance to Reform:
  Status Quo Bias in the Presence of Individual-Specific Uncertainty,}
  \emph{American Economic Review}, 81, 1146--1155.

\bibitem[\protect\citeauthoryear{Figueroa and Lemus}{Figueroa and
  Lemus}{2024}]{figueroa2024benefit}
\textsc{Figueroa, N. and J.~Lemus} (2024): \enquote{The Benefit of the Doubt:
  Patent Examination Under Strategic Obfuscation,} \emph{Management Science},
  70, 5853--5870.

\bibitem[\protect\citeauthoryear{Fudenberg and Tirole}{Fudenberg and
  Tirole}{1991}]{fudenberg1991perfect}
\textsc{Fudenberg, D. and J.~Tirole} (1991): \enquote{Perfect {Bayesian}
  Equilibrium and Sequential Equilibrium,} \emph{Journal of Economic Theory},
  53, 236--260.

\bibitem[\protect\citeauthoryear{Gao}{Gao}{2026}]{gao2026inference}
\textsc{Gao, Y.} (2026): \enquote{Inference from Selectively Disclosed Data,}
  Working Paper.

\bibitem[\protect\citeauthoryear{Grossman and Helpman}{Grossman and
  Helpman}{2001}]{grossman/helpman}
\textsc{Grossman, G. and E.~Helpman} (2001): \emph{Special Interest Politics},
  Cambridge, MA: MIT Press.

\bibitem[\protect\citeauthoryear{Guo and Shmaya}{Guo and
  Shmaya}{2023}]{guo2023regret}
\textsc{Guo, Y. and E.~Shmaya} (2023): \enquote{Regret-Minimizing Project
  Choice,} \emph{Econometrica}, 91, 1567--1593.

\bibitem[\protect\citeauthoryear{Halac and Prat}{Halac and
  Prat}{2016}]{halacprat2016}
\textsc{Halac, M. and A.~Prat} (2016): \enquote{Managerial Attention and Worker
  Performance,} \emph{American Economic Review}, 106, 3104--3132.

\bibitem[\protect\citeauthoryear{Hall and Deardorff}{Hall and
  Deardorff}{2006}]{hall2006lobbying}
\textsc{Hall, R.~L. and A.~V. Deardorff} (2006): \enquote{Lobbying as
  Legislative Subsidy,} \emph{American Political Science Review}, 100, 69--84.

\bibitem[\protect\citeauthoryear{Hoeffding}{Hoeffding}{1963}]{hoeffding1963probability}
\textsc{Hoeffding, W.} (1963): \enquote{Probability Inequalities for Sums of
  Bounded Random Variables,} \emph{Journal of the American Statistical
  Association}, 58, 13--30.

\bibitem[\protect\citeauthoryear{Holmstr{\"o}m}{Holmstr{\"o}m}{1984}]{holmstrom1984delegation}
\textsc{Holmstr{\"o}m, B.} (1984): \enquote{On the Theory of Delegation,} in
  \emph{Bayesian Models in Economic Theory}, ed. by M.~Boyer and R.~Kihlstrom,
  Amsterdam: North-Holland, 115--141.

\bibitem[\protect\citeauthoryear{Kartik, Kleiner, and Van~Weelden}{Kartik
  et~al.}{2021}]{kartik2021delegation}
\textsc{Kartik, N., A.~Kleiner, and R.~Van~Weelden} (2021): \enquote{Delegation
  in veto bargaining,} \emph{American Economic Review}, 111, 4046--4087.

\bibitem[\protect\citeauthoryear{Khalil}{Khalil}{1997}]{khalil1997auditing}
\textsc{Khalil, F.} (1997): \enquote{Auditing without commitment,} \emph{RAND
  Journal of Economics}, 28, 629--640.

\bibitem[\protect\citeauthoryear{Kim, Kim, and Van~Weelden}{Kim
  et~al.}{2025}]{kim2024persuasion}
\textsc{Kim, J.~S., K.~Kim, and R.~Van~Weelden} (2025): \enquote{Persuasion in
  veto bargaining,} \emph{American Journal of Political Science}, 69,
  1115--1127.

\bibitem[\protect\citeauthoryear{Li, Matouschek, and Powell}{Li
  et~al.}{2017}]{li2017power}
\textsc{Li, J., N.~Matouschek, and M.~Powell} (2017): \enquote{Power Dynamics
  in Organizations,} \emph{American Economic Journal: Microeconomics}, 9,
  217--241.

\bibitem[\protect\citeauthoryear{Li}{Li}{2026}]{li2026full}
\textsc{Li, W.} (2026): \enquote{Full discretion is inevitable,} \emph{RAND
  Journal of Economics}.

\bibitem[\protect\citeauthoryear{Li and Xu}{Li and Xu}{2026}]{li2026learning}
\textsc{Li, Y. and B.~Xu} (2026): \enquote{Learning and Communication Towards
  Unanimous Consent,} Working Paper.

\bibitem[\protect\citeauthoryear{Mallick and Teoman}{Mallick and
  Teoman}{2025}]{mallick2025art}
\textsc{Mallick, V. and E.~Teoman} (2025): \enquote{The Art of Waiting,}
  Working Paper.

\bibitem[\protect\citeauthoryear{Matthews}{Matthews}{1989}]{matthews1989veto}
\textsc{Matthews, S.~A.} (1989): \enquote{Veto threats: Rhetoric in a
  bargaining game,} \emph{Quarterly Journal of Economics}, 104, 347--369.

\bibitem[\protect\citeauthoryear{Meegan}{Meegan}{2010}]{meegan2010zero}
\textsc{Meegan, D.~V.} (2010): \enquote{Zero-sum bias: Perceived competition
  despite unlimited resources,} \emph{Frontiers in Psychology}, 1, 191.

\bibitem[\protect\citeauthoryear{Nocke and Whinston}{Nocke and
  Whinston}{2013}]{nocke2013merger}
\textsc{Nocke, V. and M.~D. Whinston} (2013): \enquote{Merger policy with
  merger choice,} \emph{American Economic Review}, 103, 1006--1033.

\bibitem[\protect\citeauthoryear{Potters and Van~Winden}{Potters and
  Van~Winden}{1992}]{potters1992lobbying}
\textsc{Potters, J. and F.~Van~Winden} (1992): \enquote{Lobbying and asymmetric
  information,} \emph{Public Choice}, 74, 269--292.

\bibitem[\protect\citeauthoryear{Ravid}{Ravid}{2020}]{ravid2020ultimatum}
\textsc{Ravid, D.} (2020): \enquote{Ultimatum bargaining with rational
  inattention,} \emph{American Economic Review}, 110, 2948--2963.

\bibitem[\protect\citeauthoryear{Romer and Rosenthal}{Romer and
  Rosenthal}{1978}]{RR:78}
\textsc{Romer, T. and H.~Rosenthal} (1978): \enquote{Political Resource
  Allocation, Controlled Agendas, and the Status Quo,} \emph{Public Choice},
  33, 27--43.

\bibitem[\protect\citeauthoryear{Strulovici}{Strulovici}{2010}]{strulovici2010learning}
\textsc{Strulovici, B.} (2010): \enquote{Learning While Voting: Determinants of
  Collective Experimentation,} \emph{Econometrica}, 78, 933--971.

\bibitem[\protect\citeauthoryear{Winkler}{Winkler}{1988}]{winkler1988extreme}
\textsc{Winkler, G.} (1988): \enquote{Extreme Points of Moment Sets,}
  \emph{Mathematics of Operations Research}, 13, 581--587.

\bibitem[\protect\citeauthoryear{Wolitzky}{Wolitzky}{2023}]{wolitzky2023unobserved}
\textsc{Wolitzky, A.} (2023): \enquote{Unobserved-offers bargaining,}
  \emph{American Economic Review}, 113, 136--173.

\end{thebibliography}
\end{singlespace} 

\appendix

\section{Main Appendix}

\Cref{Appendix-Preliminaries} exposits some preliminary notation and ideas. \Cref{Appendix-Theorem1Proof} proves \Cref{Theorem-UniqueEqm}, \Cref{Appendix-Propositions123} proves \Crefrange{Proposition-RandomSelection}{Proposition-RiskAttitudes}, and \Cref{Appendix-LargeBasket} proves \Cref{Theorem-LargeBasket}. All other results are proven in the Supplementary Appendix.

\subsection{Preliminaries}\label{Appendix-Preliminaries}

\label{Appendix-PreliminariesAlternate}

Conditional on its size $k$, a basket consists of $k$ independent draws from $\phi$. Thus, for every measurable collection $A$ of baskets,\footnote{The sigma-algebra on $\basket$ is generated by the sets $\{B:B\subseteq D\}$, where $D$ ranges over the Borel subsets of $\policy$. For each $k$, the preimage of $\{B:B\subseteq D\}$ under $(x_1,\ldots,x_k)\mapsto\{x_1,\ldots,x_k\}$ is the measurable set $D^k$, so the basket map is measurable.}
\begin{align*}
\mu(A)=\sum_{k\ge1}\psi(k)\phi^k\bigl\{(x_1,\ldots,x_k):\{x_1,\ldots,x_k\}\in A\bigr\}.
\end{align*}
Because $\phi$ is atomless, the draws are distinct almost surely, so $|B|=K$, where $K$ has distribution $\psi$. We call a basket \emph{regular} if each player's payoffs are nonzero and distinct across its policies. Thin indifference curves imply that both players' sampled payoffs are nonzero almost surely. Moreover, conditional on any draw, an independent draw gives either player the same payoff with probability zero. Since each basket is finite, it is regular almost surely. On such baskets, $C(B)$ and $M(B)$ are uniquely determined, $t(B)$ is finite, and $t(B)=0$ if and only if $C(B)=M(B)$. If these policies differ, $C(B)\in\conflict$ and $M(B)\in\congruence$, so moderation strictly benefits the responder. Throughout, statements about individual baskets concern regular baskets. On exceptional baskets, we break ties measurably and set $t(B)=0$ when its denominator is zero. Whenever we need to select one of several maximizers, we take the first under a fixed measurable ordering of policies and the induced ordering of menus.

Our assumptions also guarantee that the temptation ratio has no atom at any strictly positive value. Moreover, all relevant payoffs are integrable. For $i\in \{P,R\}$, let $H_i(B)\equiv\sum_{x\in B}|\pi_i(x)|$ and observe that
\begin{align}\label{Equation-PreliminariesAlternateBound}
 \E[H_i(B)]\le\E[K]\,\E_\phi[|\pi_i(X)|]<\infty.
\end{align}
The absolute payoff from any selected policy and the absolute average payoff of any menu are bounded by $H_i(B)$. Given a proposal strategy $\sigma$, we write $\pi_R(\sigma(B))\equiv\sum_{x\in B}\sigma(x\mid B)\pi_R(x)$ for the responder's expected payoff from accepting in basket $B$, and $\E[\pi_R(\sigma)]$ for its expectation over baskets.

Finally, for $0\le\tau<\infty$, let $\sigma_\tau$ denote the strategy that moderates when $t(B)\le\tau$ and cherry-picks otherwise:
\begin{align}\label{Equation-PreliminariesAlternateThreshold}
\sigma_\tau(x\mid B)\equiv\1_{\{x=M(B)\}}\1_{\{t(B)\le\tau\}}+\1_{\{x=C(B)\}}\1_{\{t(B)>\tau\}}.
\end{align}
Observe that $\E[\pi_R(\sigma_\tau)]=V_R(\tau)$, with $V_R$ defined in \Cref{Section-Theorem1Sketch}.

\subsection{Proof of \Cref{Theorem-UniqueEqm}}\label{Appendix-Theorem1Proof}

We begin with some preliminary results. 
Throughout our analysis, we impose \Cref{Assumption-ExAnte}, namely that the responder's ex ante payoff from a randomly selected policy is strictly positive ($\E_\phi[\pi_R(X)]> 0$). Our first step establishes that if the proposer selects a mutually beneficial policy whenever one is available, the responder obtains a strictly positive payoff. Therefore, she accrues a strictly positive payoff from the strategy that selects  $M(B)$ from $\mu$-a.e. basket $B$. 

\begin{lemma}\label{Lemma-ModerationPositive}
Consider a proposer strategy $\sigma$ that puts forward a mutually beneficial policy whenever one is viable: $\supp \sigma(\cdot\mid B)\subseteq B\cap \congruence$  for all baskets $B$ such that $B\cap\congruence\neq\varnothing$.
Then the responder accrues a strictly positive payoff: $\E[\pi_R(\sigma)]>0$. Therefore, in particular, $\E[\pi_R(M)]>0$. 
\end{lemma}
\begin{proof}
Suppose the basket has $k$ elements, labeled as $X_j$ for $j=1,\ldots,k$, and is regular. If only one mutually beneficial policy is viable, then $\sigma$ must select it; if two or more are viable, then the responder's payoff is strictly positive. Finally, if none are viable, then her payoff can be no lower than the sum of her payoffs from all $k$ policies. Hence, we can write
\begin{align*}
\pi_R(\sigma(B))\ge\sum_{j=1}^k\left(\pi_R(X_j)\prod_{\ell\ne j}\1_{\{\pi_R(X_\ell)<0\}}\right),
\end{align*}
where the $j^{th}$ term counts her payoff from policy $X_j$ only if every other policy harms her. Since each policy in the basket is drawn independently according to measure $\phi$, the expectation of each term on the right is $\E_\phi[\pi_R(X)]\left(\phi(\conflict)\right)^{k-1}$. Taking expectations over baskets yields
\begin{align*}
 \E[\pi_R(\sigma)]
 \ge\E_\phi[\pi_R(X)]
       \sum_{k\ge1}\psi(k)k(\phi(\conflict))^{k-1}>0,
\end{align*}
where the strict inequality follows from \Cref{Assumption-ExAnte} and $\phi(\conflict)>0$. 
\end{proof}

The previous step shows that if the proposer moderates always, the uninformed responder would accept. The next result identifies when the uninformed responder benefits from accepting the proposal if the proposer uses a threshold $\tau\geq 0$ for deciding when to cherry-pick. Recall from \Cref{Section-Theorem1Sketch} that her expected payoff given a threshold $\tau$ is
\begin{align*}
 V_R(\tau)&\equiv\int_{t(B)\le\tau}\pi_R(M(B))\,\mu(dB)
 +\int_{t(B)>\tau}\pi_R(C(B))\,\mu(dB).
\end{align*}

\begin{lemma}\label{Lemma-ThresholdPayoff}
The following hold:
\begin{enumerate}[label=\emph{(\alph*)}]
	\item If $\E[\pi_R(C)]\ge0$, then $V_R(\tau)>0$ for every $\tau>0$.
	\item  If $\E[\pi_R(C)]<0$, there exists a unique $\overline\tau>0$ such that $V_R(\overline\tau)=0$. Moreover, for $\tau\in (0,\overline\tau)$, $V_R(\tau)<0$, and for $\tau>\overline\tau$, $V_R(\tau)>0$.
\end{enumerate}
\end{lemma}
Part (a) says that if cherry-picking doesn't push the responder to choose the status quo, then she accepts the proposal given any threshold rule. Part (b) pertains to settings in which cherry-picking pushes her to choose the status quo: then there exists a unique threshold $\overline\tau$ that leaves her indifferent, and any more (respectively, less) cherry-picking induces her to reject (respectively, accept). 

\begin{proof}

Given a basket $B$, let $\Delta(B)\equiv\pi_R(M(B))-\pi_R(C(B))$ denote the responder's gain from moderation. Observe that $\Delta(B)$ is non-negative for every basket $B$ and strictly positive for any basket $B$ in which $M(B)\neq C(B)$ (or equivalently, $t(B)>0$). Observe that if $\tau'>\tau$, 
\begin{align}\label{Equation-ValueMonotonicity}
	V_R(\tau')-V_R(\tau)=\int_{t(B)\in (\tau,\tau']} \Delta(B)\mu(dB),
\end{align}
which is non-negative by definition. Therefore, $V_R(\tau)$ is non-decreasing in $\tau$. Below, we prove the following claim:
\begin{claim}\label{Claim-StrictMonotone}
If $V_R(\tau)\leq 0$, then $V_R(\tau')>V_R(\tau)$ for every $\tau'>\tau$. 	
\end{claim}
To prove \Cref{Claim-StrictMonotone}, fix $\tau\ge0$ and suppose $V_R(\tau)\le0$. Observe that
\begin{align*}
V_R(\tau)=\E[\pi_R(M)]- \int_{t(B)>\tau}\Delta(B)\,\mu(dB).
\end{align*}
\Cref{Lemma-ModerationPositive} establishes that the first term is strictly positive. Therefore, $V_R(\tau)\leq 0$ implies that $\mu(\{B\in \basket:t(B)>\tau\})>0$.  We show that for every $\tau' >\tau$, $\mu(\{B\in\basket:t(B)\in (\tau,\tau')\})>0$: in other words, there is a positive $\mu$-measure set of baskets with temptation ratios in $(\tau,\tau')$.

First, choose a regular basket $\hat B$ that has $k\ge2$ elements with $\psi(k)>0$ and whose temptation ratio exceeds $\tau$. Let $\hat\pi_P^-$ and $\hat\pi_P^+$ be the proposer's payoffs from cherry-picking and moderation in this basket. By construction of $\hat B$, $\hat\pi_P^->(1+\tau)\hat\pi_P^+$.

Second, we show that there exists $\pi_*>0$ that is in the closure of each of $\pi_P(\congruence)$ and $\pi_P(\conflict)$. Observe that full support and thin indifference curves imply that $\cl(\congruence)\cup \cl(\conflict)=\policy$, where $\cl(\cdot)$ identifies the closure of a set. Given that both the congruence and conflict regions are nonempty and $\policy$ is connected, $\cl(\congruence)$ and $\cl(\conflict)$ have a nonempty intersection; choose $x$ in that intersection. Because $\pi_P(\cdot)$ is continuous, it follows that $\pi_P(x)\in \cl(\pi_P(\congruence))\cap \cl(\pi_P(\conflict))$. If $\pi_P(x)>0$, then we set $\pi_*=\pi_P(x)$. If $\pi_P(x)=0$, then observe that $\pi_P(\congruence)$ includes the interval $(0,\hat\pi_P^+]$ and $\pi_P(\conflict)$ includes the interval $(0,\hat\pi_P^-]$, where $\hat\pi_P^-$ and $\hat\pi_P^+$ are the proposer's cherry-picking and moderation payoffs in basket $\hat B$.\footnote{Recall our assumption that $\pi_P(\congruence)$ and $\pi_P(\conflict)$ each form an interval.} Because $\hat\pi_P^->\hat\pi_P^+$, it then follows that we can set $\pi_*=\hat\pi_P^+$. 

Given that $\pi_*>0$ is in the closure for each, for every $\theta\in (0,1)$, there exist policies $x^-\in\conflict$ and $x^+\in\congruence$ such that $ \pi_P(x^-)=(1-\theta)\pi_*+\theta \hat\pi_P^-$ and $ \pi_P(x^+)=(1-\theta)\pi_*+\theta \hat\pi_P^+$. Observe that given a choice between $x^-$ and $x^+$, the proposer's relative benefit from choosing $x^-$ is
\begin{align*}
 \frac{\pi_P(x^-)-\pi_P(x^+)}{\pi_P(x^+)}
 &=\frac{\theta(\hat\pi_P^--\hat\pi_P^+)}{(1-\theta)\pi_*+\theta \hat\pi_P^+}.
\end{align*}
This expression is continuous in $\theta$, approaches zero as $\theta\rightarrow 0$, and approaches $t(\hat B)>\tau$ as $\theta\uparrow1$. Thus, for every $\tau'>\tau$, there exists $\theta$ such that the ratio lies in $(\tau,\tau')$. Fix the resulting policies $x^-$ and $x^+$. 

We now argue that there exists a positive $\mu$-measure set of baskets whose temptation ratios lie in $(\tau,\tau')$. Because $\pi_P(\cdot)$ is continuous, one can choose an open neighborhood $O^-$ in $\conflict$ around $x^-$ and one in $\congruence$ around $x^+$, $O^+$, such that the temptation ratio when selecting between any policy in $O^+$ and any in $O^-$ is in $(\tau,\tau')$. Because $\phi$ satisfies independence and full support and $\psi(k)>0$, there exists a positive measure of baskets that has one policy in $O^+$ and $(k-1)$ policies in $O^-$; any such basket has a temptation ratio in $(\tau,\tau')$. Observe also that for any such basket, $\Delta(B)>0$. Therefore, by \eqref{Equation-ValueMonotonicity}, $V_R(\tau')>V_R(\tau)$, establishing \Cref{Claim-StrictMonotone}. 

Now observe that we can also write for $\tau>0$ that
$$V_R(\tau)=\E[\pi_R(C)]+\int_{0<t(B)\le\tau}\Delta(B)\,\mu(dB).$$
Because $\Delta$ is integrable and $\mu(\{B\in\basket:t(B)=\tau\})=0$, it follows that $V_R(\cdot)$ is continuous at every $\tau>0$. It is also continuous at $0$ as $\tau\rightarrow  0$ because the set $\{B\in \basket:t(B)\in (0,\tau]\}$ shrinks to the empty set. Given that every regular basket has a finite temptation ratio, it also follows that $\lim_{\tau\to\infty}V_R(\tau)=\E[\pi_R(M)]>0$. 

Using the above, we can now deduce the statements in the lemma. For part (a), suppose that $V_R(0)=\E[\pi_R(C)]\geq 0$. If the inequality hold strictly, then $V_R(\tau)\geq V_R(0)>0$ holds immediately; if $V_R(0)=0$, then the conclusion follows from \Cref{Claim-StrictMonotone}. For part (b), suppose that $V_R(0)=\E[\pi_R(C)]<0$. Continuity and the strictly positive limit as $\tau\to \infty$ imply that there exists a zero $\overline\tau>0$. \Cref{Claim-StrictMonotone} implies that $V_R(\tau)<0$ for $\tau<\overline\tau$ and $V_R(\tau)>0$ for $\tau>\overline\tau$.
\end{proof}

Using these lemmas, we now prove \Cref{Theorem-UniqueEqm}.

\begin{proof}[Proof of \Cref{Theorem-UniqueEqm}]
	We first show that every equilibrium has the stated form. We then verify that these strategies constitute an equilibrium. Consider any equilibrium $(\sigma,\gamma,\alpha)$.

First, we show that the informed responder accepts almost every policy that benefits her and rejects almost every policy that harms her. Suppose, toward a contradiction, that there is a positive $\phi$-measure set $\tilde X\subseteq\congruence$ on which $\gamma(x)<1$. For any $k$ with $\psi(k)>0$, independence implies that all $k$ policies lie in $\tilde X$ with strictly positive probability. In that event, the proposer must select a policy in $\tilde X$, whatever his strategy. The informed responder could therefore profitably deviate to accepting every policy in $\tilde X$ with probability one: this deviation never lowers her payoff and strictly raises it on that event. A symmetric argument shows that she rejects almost every policy in $\conflict$. Because the zero-payoff level has $\phi$-measure zero, $\gamma(x)=\gamma^*(x)$ for $\phi$-almost every $x$. Since each basket contains finitely many draws from $\phi$, this equality holds for every $x\in B$ in $\mu$-almost every basket $B$.

Next, we show that the uninformed responder accepts with strictly positive probability. Suppose instead toward a contradiction that $\alpha=0$. A mutually beneficial policy then gives the proposer a strictly positive payoff, whereas a harmful policy delivers zero. Therefore, in equilibrium, the proposer must select a mutually beneficial policy whenever one is available, for $\mu$-almost every basket. \Cref{Lemma-ModerationPositive} then implies that the uninformed responder accrues a strictly positive payoff from accepting, which implies that she has a profitable deviation.

Given that $\alpha>0$, we now determine the proposer's choice. If $C(B)=M(B)$, this policy uniquely maximizes his payoff. Otherwise, if $C(B)\neq M(B)$, the proposer's optimal choice must lie in $\{C(B),M(B)\}$.\footnote{Every mutually beneficial policy is approved with probability $\lambda+(1-\lambda)\alpha$, and every harmful policy with probability $(1-\lambda)\alpha$. Within each group, the proposer therefore selects the policy that gives him the highest payoff, namely $M(B)$ and $C(B)$, respectively.} The gain from cherry-picking is 
\begin{align*}
(1-\lambda)\alpha\pi_P(C(B))-[\lambda+(1-\lambda)\alpha]\pi_P(M(B))=
(1-\lambda)\alpha\pi_P(M(B))
 \left[t(B)-\frac{\odds_\lambda}{\alpha}\right].
\end{align*}
Thus, he moderates if $t(B)<\odds_\lambda/\alpha$ and cherry-picks if $t(B)>\odds_\lambda/\alpha$. At equality, we break ties in favor of moderation. Because $\odds_\lambda/\alpha>0$, $\mu(\{B\in\basket:t(B)=\odds_\lambda/\alpha\})=0$. Hence, $\sigma(\cdot\mid B)=\sigma_{\odds_\lambda/\alpha}(\cdot\mid B)$ for $\mu$-a.e. $B$, where $\sigma_{\odds_\lambda/\alpha}$ is defined in \eqref{Equation-PreliminariesAlternateThreshold} on p. \pageref{Equation-PreliminariesAlternateThreshold}.

The remaining step determines $\alpha$. Given the proposer's behavior, the uninformed responder's payoff from accepting is $V_R(\odds_\lambda/\alpha)$. If $\E[\pi_R(C)]\ge0$, \Cref{Lemma-ThresholdPayoff}(a) implies that this payoff is strictly positive. Therefore, $\alpha=1$, and we set $\overline\lambda=0$. Suppose instead that $\E[\pi_R(C)]<0$. Let $\overline\tau>0$ be the unique zero from \Cref{Lemma-ThresholdPayoff}(b), and set $\overline\lambda=\overline\tau/(1+\overline\tau)$, so that $\odds_{\overline\lambda}=\overline\tau$.

If $\lambda<\overline\lambda$, the uninformed responder cannot always accept: doing so induces the threshold $\odds_\lambda<\overline\tau$, at which her payoff from accepting is strictly negative. Because we have already ruled out always rejecting, she must mix. Her indifference requires $V_R(\odds_\lambda/\alpha)=0$. By \Cref{Lemma-ThresholdPayoff}(b), this is possible only if $\odds_\lambda/\alpha=\overline\tau$, or equivalently, $\alpha=\odds_\lambda/\odds_{\overline\lambda}$. If $\lambda\ge\overline\lambda$, any $0<\alpha<1$ induces a threshold strictly above $\odds_\lambda\ge\overline\tau$. The same lemma then implies that accepting gives her a strictly positive payoff, contradicting mixing. Therefore, $\alpha=1$, including when $\lambda=\overline\lambda$.

Combining these cases, every equilibrium has: (i) $\gamma(x)=\gamma^*(x)$ for $\phi$-a.e. $x$, (ii) $\sigma=\sigma_{\odds_\lambda/\alpha_\lambda}$ $\mu$-a.s., and (iii) $\alpha=\alpha_\lambda
\equiv\min\left\{\frac{\odds_\lambda}{\odds_{\overline\lambda}},1\right\}$. 
The cutoff $\overline\lambda$ depends only on the payoff environment and is strictly positive exactly when $\E[\pi_R(C)]<0$.

To establish that an equilibrium exists, consider the stated profile $(\sigma^*,\gamma^*,\alpha_\lambda)$. The proposer's payoff comparison above shows that $\sigma^*$ is a best response, and $\gamma^*$ is optimal whenever the responder is informed. If $\lambda<\overline\lambda$, the induced threshold is $\overline\tau$, so the uninformed responder is indifferent and her mixing is optimal. If $\lambda\ge\overline\lambda$, the induced threshold is $\odds_\lambda$ and her payoff from accepting is nonnegative by \Cref{Lemma-ThresholdPayoff}, so accepting with probability one is optimal. These strategies therefore constitute an equilibrium. 
\end{proof}

\subsection{Proofs of \Crefrange{Proposition-RandomSelection}{Proposition-RiskAttitudes}}\label{Appendix-Propositions123}

For $i\in\{P,R\}$, let $\Pi_i(\lambda)$ denote player $i$'s ex ante equilibrium payoff in the baseline game when the responder is informed with probability $\lambda$.

\begin{proof}[Proof of \Cref{Proposition-RandomSelection}]
Under random selection, a uniformly selected policy has distribution $\phi$, regardless of the basket's size. Therefore, by \Cref{Assumption-ExAnte}, the uninformed responder accepts with probability one. The resulting payoff to player $i\in \{P,R\}$ is
\begin{align*}
\Pi_i^{\mathrm{rand}}(\lambda)
&=\int_{\policy}\pi_i(x)\,\phi(dx)
-\lambda\int_{\conflict}\pi_i(x)\,\phi(dx),
\end{align*}
where the second term accounts for harmful policies rejected by the informed responder. \Cref{Theorem-UniqueEqm} implies that, for $\lambda\le\overline\lambda$, the proposer uses the same selection rule, $\sigma_{\odds_{\overline\lambda}}$, while $(1-\lambda)\alpha_\lambda=\lambda/\odds_{\overline\lambda}$. Hence each player's equilibrium payoff is proportional to $\lambda$:
\begin{align}\label{Equation-Proportionality}
\Pi_i(\lambda)=\left(\lambda/\overline\lambda\right)\Pi_i(\overline\lambda).
\end{align}

For part (a), the proposer's gain from strategic over random selection for $\lambda\in (0,\overline\lambda]$ is
\begin{align*}
\Pi_P(\lambda)-\Pi_P^{\mathrm{rand}}(\lambda)
&=\lambda\left[\frac{\Pi_P(\overline\lambda)}{\overline\lambda}
+\int_{\conflict}\pi_P(x)\,\phi(dx)\right]
-\int_{\policy}\pi_P(x)\,\phi(dx).
\end{align*}
Observe that this expression is affine and strictly increasing in $\lambda$, since $\Pi_P(\overline\lambda)\ge0$ and $\int_{\conflict}\pi_P(x)\,\phi(dx)>0$. Moreover, its limit as $\lambda\to 0$ is strictly negative.

To locate the cutoff, we now show that the proposer strictly prefers strategic selection whenever $\lambda\ge\overline\lambda$. In this parameter region, the uninformed responder accepts with probability one under either selection rule. Because proposing randomly is a feasible strategy for the proposer, $\Pi_P(\lambda)\ge\Pi_P^{\mathrm{rand}}(\lambda)$. Moreover, if equality held, random selection coupled with the responder's best response would constitute another equilibrium, which contradicts the essential uniqueness claim in \Cref{Theorem-UniqueEqm}.\footnote{Observe that $\overline\lambda>0$ implies that, with strictly positive probability, the basket has at least two policies. On $\mu$-a.e. non-singleton basket, random selection differs from the equilibrium choice described in \Cref{Theorem-UniqueEqm}.} Therefore, $\Pi_P(\lambda)>\Pi_P^{\mathrm{rand}}(\lambda)$ for every $\lambda\ge\overline\lambda$.

Because the proposer's gain from strategic selection is strictly positive at $\overline\lambda$, the affine expression above has a unique zero $\hat\lambda_P\in(0,\overline\lambda)$, and the gain remains strictly positive above $\overline\lambda$. Thus, the gain is single-crossing in $\lambda$.

For part (b), the proposer accrues at least $\E_\phi[\pi_R(X)]>0$ from random selection, given that she can always approve the proposal when informed. By contrast, \eqref{Equation-Proportionality} implies that her payoff from strategic selection converges to zero as $\lambda\rightarrow 0$. Hence, there exists $\hat\lambda_R\in(0,\overline\lambda)$ such that she strictly prefers random selection whenever $0<\lambda\le\hat\lambda_R$.
\end{proof}

\begin{proof}[Proof of \Cref{Proposition-Education}] 
For part (a), \eqref{Equation-Proportionality} establishes linearity, so it remains to show that $\Pi_P(\overline\lambda)$ and $\Pi_R(\overline\lambda)$ are strictly positive. The proposer's payoff is strictly positive because the proposed policy benefits him almost surely and is accepted with probability at least $(1-\overline\lambda)$. At $\overline\lambda$, the responder is indifferent when uninformed, so her expected payoff comes entirely from informed acceptance. When informed, she never receives a negative payoff and receives a strictly positive payoff whenever every policy in the basket is mutually beneficial. Such baskets have strictly positive probability. 

For part (b), fix $\overline\lambda\le\lambda<\lambda'<1$. By \Cref{Theorem-UniqueEqm}, the uninformed responder accepts with probability one at both information levels. The proposer therefore chooses the policy in each basket that maximizes his expected payoff:
\begin{align*}
\Pi_P(\lambda)
&=\E\left[\max_{x\in B}\left\{[1-\lambda\1_{\conflict}(x)]\pi_P(x)\right\}\right].
\end{align*}
Each policy's expected payoff falls as $\lambda$ increases, so the maximum does as well. On a basket that contains only harmful policies---which occurs with strictly positive probability---this maximum is $(1-\lambda)\pi_P(C(B))$, which falls strictly in $\lambda$. Taking expectations delivers that  $\Pi_P(\lambda')<\Pi_P(\lambda)$.

For the responder, greater information reduces the acceptance of harmful policies and can only change the proposer's choice from a harmful policy to a mutually beneficial one. Therefore,
\begin{align*}
\Pi_R(\lambda')-\Pi_R(\lambda)
&\ge(\lambda'-\lambda)\E\left[-\pi_R(C(B))\1_{\{B\subseteq\conflict\}}\right]>0,
\end{align*}
where the strict inequality follows because baskets containing only harmful policies occur with strictly positive probability.
\end{proof}
\begin{proof}[Proof of \Cref{Proposition-RiskAttitudes}]
Fix $\lambda\in(0,1)$ and let $\widetilde\alpha_\lambda$ denote uninformed acceptance in any equilibrium after either preference change. Also, let $\widetilde V_R(\cdot)$ denote the responder's expected payoff given the proposer's threshold under either preference change. 

We show that $\widetilde\alpha_\lambda\le\alpha_\lambda$.\footnote{Observe that the payoff transformations do not affect the behavior of the informed responder, who will continue using $\gamma^*$ almost surely.} If $\alpha_\lambda=1$, there is nothing to show. Therefore, we suppose that $\alpha_\lambda<1$ and, toward a contradiction, that $\widetilde\alpha_\lambda>\alpha_\lambda$. Observe that without any preference change, the proposer's best response to an acceptance probability $\widetilde\alpha_\lambda$ is to use the threshold $\tau\equiv\odds_\lambda/\widetilde\alpha_\lambda$, which is strictly less than $\odds_{\overline\lambda}$. By \Cref{Lemma-ThresholdPayoff}(b), we know that $V_R(\tau)<0$.

First, suppose that the proposer becomes more risk loving. Because $f$ is strictly increasing, his preferred policies $C(B)$ and $M(B)$ are unchanged. Convexity and $f(0)=0$ imply that his temptation ratio increases:
\begin{align*}
\widetilde t(B)\equiv\frac{f(\pi_P(C(B)))}{f(\pi_P(M(B)))}-1
&\ge\frac{\pi_P(C(B))}{\pi_P(M(B))}-1=t(B).
\end{align*}
Therefore, the temptation ratio for each basket increases, which raises his propensity to cherry-pick. It then follows that
\begin{align*}
\widetilde V_R(\tau)
=\E\left[\pi_R(C(B))+\Delta(B)\1_{\{\widetilde t(B)\le\tau\}}\right]\le\E\left[\pi_R(C(B))+\Delta(B)\1_{\{t(B)\le\tau\}}\right]=V_R(\tau),
\end{align*}
which is strictly negative as established above. The responder therefore strictly prefers to reject, contradicting $\widetilde\alpha_\lambda>0$.

Suppose instead that the responder becomes more risk averse. Because the proposer still uses the threshold $\tau$, it then follows from Jensen's inequality and $f$ being concave and strictly increasing with $f(0)=0$ that
\begin{align*}
\widetilde V_R(\tau)=\E\sum_{x\in B}\sigma_\tau(x\mid B)f(\pi_R(x))
&\le f\left(\E\sum_{x\in B}\sigma_\tau(x\mid B)\pi_R(x)\right)
=f(V_R(\tau))<0.
\end{align*}
Again, she strictly prefers to reject, contradicting $\widetilde\alpha_\lambda>0$.\footnote{Concavity can cause \Cref{Assumption-ExAnte} to fail, but the argument above does not require this assumption to hold for the transformed payoffs. If $\widetilde\alpha_\lambda=0$, the desired comparison holds immediately.}
\end{proof}

\subsection{Proofs of \Cref{Theorem-LargeBasket} and \Cref{Corollary-LargePayoff}}\label{Appendix-LargeBasket}
We use $\mu_d$ and $\E_d$ for the basket law and expectation when the basket-size distribution is $\psi_d$. Throughout this analysis, we consider a sequence of baskets growing large, in the sense of \Cref{Definition-LargeBaskets}. We first show that the expected positive part of the responder's payoff from moderation vanishes. Because her payoffs may be unbounded, we must control their contribution on the rare baskets that miss a given set of policies. 

\begin{lemma}\label{Lemma-LargeBasketAbsence}
For every measurable set $D$ with $\phi(D)>0$, and every measurable selection $S(B)\in B$, which may depend on $d$, $\mu_d(B\cap D=\varnothing)\rightarrow 0$ and $\E_d\left[|\pi_R(S(B))|\1_{\{B\cap D=\varnothing\}}\right]\rightarrow 0$ as $d\rightarrow\infty$. 
\end{lemma}
\begin{proof}
If $\phi(D)=1$, every basket intersects $D$ almost surely, so both quantities are zero. Otherwise, let $q=1-\phi(D)$. Conditional on a basket of size $k$, the probability that $B$ misses $D$ is $q^k$, which converges to $0$ as $k\to\infty$. Given that it misses $D$, each draw has the distribution of $X$ conditional on $X\notin D$. Since the selected policy's absolute payoff is bounded by the sum of the absolute payoffs from all $k$ draws,
\begin{align*}
\E_{\phi^k}\!\left[|\pi_R(S(B))|\,\1_{\{B\cap D=\varnothing\}}\right]
&\leq q^k k\,\E_\phi\!\left[|\pi_R(X)|\mid X\notin D\right].
\end{align*}
The conditional expectation on the right is finite by integrability. The term $kq^k$ is bounded and tends to zero as $k$ grows. Since basket size tends to infinity in probability, averaging over basket sizes gives both conclusions.
\end{proof}

We apply this bound below to moderation. We write $m_d\equiv\E_d[(\pi_R(M(B)))^+]$, where $z^+\equiv\max\{z,0\}$.
\begin{lemma}\label{Lemma-ModerationSurplus}
If \Cref{Assumption-BreakEven} holds, the responder's expected gains from moderation disappear in large baskets: $\lim_{d\rightarrow\infty}m_d=0$.
\end{lemma}
\begin{proof}
Fix $\eta>0$ and choose $\tilde\pi_P<\pi_P^+$ so that every policy in $\congruence$ that gives the proposer at least $\tilde\pi_P$ delivers at most $\eta$ to the responder. \Cref{Assumption-BreakEven} ensures that such a cutoff exists. Let $D_\eta\equiv\{x\in\congruence:\pi_P(x)>\tilde\pi_P\}$. This set is nonempty and open, so full support implies that $\phi(D_\eta)>0$. If $B$ contains a policy in $D_\eta$, then the proposer's favorite mutually beneficial policy $M(B)$ must also give him more than $\tilde\pi_P$. It therefore gives the responder at most $\eta$. Therefore, $m_d
\le\eta+\E_d\left[|\pi_R(M(B))|\1_{\{B\cap D_\eta=\varnothing\}}\right]$.
Holding $\eta$ fixed, the second term converges to zero by \Cref{Lemma-LargeBasketAbsence}. Therefore, $\limsup_{d\to\infty}m_d\le\eta$. The conclusion then follows from taking $\eta\to 0$.
\end{proof}

We use these preliminary results to establish \Cref{Theorem-LargeBasket}.

\begin{proof}[Proof of \Cref{Theorem-LargeBasket}]
Fix $\lambda\in(0,1)$. By \Cref{Theorem-UniqueEqm}, the uninformed responder accepts with probability $\alpha_\lambda^d>0$, and the proposer selects either $C(B)$ or $M(B)$ almost surely. For each $d$, let $\sigma^*$ denote the proposer's equilibrium strategy under basket law $\mu_d$.

For part (a), observe that whenever the proposed policy benefits the responder, it is $M(B)$. Hence $(\pi_R(\sigma^*(B)))^+\le(\pi_R(M(B)))^+$. Moreover, because the uninformed responder accepts with positive probability, $\E_d[\pi_R(\sigma^*(B))]\ge0$.
Her expected losses therefore cannot exceed her expected gains:
\begin{align}\label{Equation-LargeBasketLosses}
\E_d[(\pi_R(\sigma^*(B)))^-]\le\E_d[(\pi_R(\sigma^*(B)))^+]\le m_d\rightarrow0,
\end{align}
where $z^-\equiv\max\{-z,0\}$ and the limit follows from \Cref{Lemma-ModerationSurplus}.

Using \Cref{Assumption-Conflict}, we argue that this payoff bound implies a bound on the probability of a harmful proposal. Choose $\tilde\pi_P<\pi_P^*$ and $\delta>0$ such that $\pi_P(x)\ge\tilde\pi_P$ implies $\pi_R(x)\le-\delta$, and let $D_-\equiv\{x\in\policy:\pi_P(x)>\tilde\pi_P\}$. Full support implies that $\phi(D_-)>0$. If a basket contains a policy in $D_-$, then $C(B)$ gives the responder a payoff at most $-\delta$. On such a basket, any equilibrium proposal with a negative responder payoff must be $C(B)$ and therefore imposes a loss of at least $\delta$. It follows that
\begin{align*}
\mu_d(\pi_R(\sigma^*(B))\le0)
&\le\frac{\E_d[(\pi_R(\sigma^*(B)))^-]}{\delta}+\mu_d(B\cap D_-=\varnothing)
\le\frac{m_d}{\delta}+\mu_d(B\cap D_-=\varnothing)
\rightarrow0,
\end{align*}
where the last term vanishes by \Cref{Lemma-LargeBasketAbsence}, which proves part (a).

For part (b), fix a finite $u\in(\tilde\pi_P,\pi_P^*)$ and let $D_u\equiv\{x\in\policy:\pi_P(x)>u\}$. Full support again implies that $\phi(D_u)>0$. Part (a) and \Cref{Lemma-LargeBasketAbsence} imply that $\mu_d(\pi_R(\sigma^*(B))>0,\ B\cap D_u\ne\varnothing)\rightarrow1$.
Therefore, for every sufficiently large $d$, there is a regular basket on which the proposer's optimal choice is $M(B)\in\congruence$ even though $C(B)$ gives him more than $u$, which implies that
\begin{align*}
\left(\lambda+(1-\lambda)\alpha_\lambda^d\right)\pi_P(M(B))
&\ge(1-\lambda)\alpha_\lambda^d\pi_P(C(B)).
\end{align*}
Because $\pi_P(M(B))\le\pi_P^+\le\tilde\pi_P<u<\pi_P(C(B))$, it follows that
\begin{align*}
\alpha_\lambda^d
&\le\odds_\lambda\frac{\pi_P(M(B))}{\pi_P(C(B))-\pi_P(M(B))}
\le\odds_\lambda\frac{\pi_P^+}{u-\pi_P^+}.
\end{align*}
The acceptance probability does not depend on the realized basket. For each fixed $u$, the bound therefore holds for all sufficiently large $d$, so it also bounds $\limsup_{d\to\infty}\alpha_\lambda^d$. Letting $u$ increase to $\pi_P^*$, or tend to infinity if $\pi_P^*=\infty$, gives 
\begin{align*}
\limsup_{d\to\infty}\alpha_\lambda^d\le\frac{\odds_\lambda}{t_\infty},
\end{align*}
with the right side equal to zero when $t_\infty=\infty$. 
\end{proof}

\begin{proof}[Proof of \Cref{Corollary-LargePayoff}]
Fix $\lambda\in(0,1)$. For part (a), the responder's payoff when informed is $(\pi_R(\sigma^*(B)))^+$, while her payoff from accepting when uninformed is $\pi_R(\sigma^*(B))$. Because $\E_d[\pi_R(\sigma^*(B))]\le\E_d[(\pi_R(\sigma^*(B)))^+]\le m_d$ by \eqref{Equation-LargeBasketLosses}, her equilibrium payoff satisfies
\begin{align*}
0\le\Pi_R^d(\lambda)
&=\lambda\E_d[(\pi_R(\sigma^*(B)))^+]+(1-\lambda)\alpha_\lambda^d\E_d[\pi_R(\sigma^*(B))]
\le[\lambda+(1-\lambda)\alpha_\lambda^d]m_d
\le m_d\rightarrow0.
\end{align*}

For part (b), suppose that proposer payoffs are bounded and $\odds_\lambda<t_\infty$. A mutually beneficial proposal gives the proposer at most $\pi_P^+$ and is accepted with probability $\lambda+(1-\lambda)\alpha_\lambda^d$. A harmful proposal gives him at most $\pi_P^*$ and is accepted with probability $(1-\lambda)\alpha_\lambda^d$. Therefore,
\begin{align*}
\Pi_P^d(\lambda)
&\le\max\left\{[\lambda+(1-\lambda)\alpha_\lambda^d]\pi_P^+,\ (1-\lambda)\alpha_\lambda^d\pi_P^*\right\}.
\end{align*}
Both terms are continuous and increasing in $\alpha_\lambda^d$. Applying \Cref{Theorem-LargeBasket}(b) delivers
\begin{align*}
\limsup_{d\to\infty}\Pi_P^d(\lambda)
&\le\max\left\{\lambda(1+t_\infty^{-1})\pi_P^+,\ \lambda t_\infty^{-1}\pi_P^*\right\}
=\lambda\frac{\pi_P^+\pi_P^*}{\pi_P^*-\pi_P^+}
<\pi_P^+,
\end{align*}
where the equality follows from the definition of $t_\infty$ and the strict inequality follows because $\odds_\lambda<t_\infty$ implies $\lambda\pi_P^*<\pi_P^*-\pi_P^+$.
\end{proof}

\newpage

\section{Supplementary Appendix}\label{Appendix-Supplementary}

\Cref{Appendix-ResponderCommitment} considers responder commitment and \Cref{Appendix-ProposerCommitment} considers proposer commitment. \Cref{Appendix-Menus} proves the results on menus whereas \Cref{Appendix-Effort} and \Cref{Appendix-Search}, respectively, proves those on effort and optimal search. 

\subsection{Responder Commitment}\label{Appendix-ResponderCommitment}

\begin{proof}[Proof of \Cref{Theorem-ResponderCommitment}]
Fix $\lambda\in(0,1)$. Part (a) uses the baseline assumptions for a fixed basket-size distribution $\psi_d$. Suppose that $\lambda<\overline\lambda_d$. By \Cref{Theorem-UniqueEqm}, the uninformed responder mixes in equilibrium. Let $\sigma^*$ denote the proposer's equilibrium strategy. The responder's equilibrium payoff comes from approving mutually beneficial policies when she is informed: $$\Pi_R^d(\lambda)=\lambda\E_d[(\pi_R(\sigma^*(B)))^+].$$ 
Now suppose that the responder commits to rejecting when uninformed while continuing to use $\gamma^*$ when informed. Every best response by the proposer involves him choosing $M(B)$ almost surely whenever a mutually beneficial policy is available, which results in the expected payoff, $\lambda\E_d[(\pi_R(M(B)))^+]$. Observe that
\begin{align*}
\lambda\E_d[(\pi_R(M(B)))^+]-\Pi_R^d(\lambda)
&=\lambda\int_{\{B:t(B)>\overline\tau_d\}}
\pi_R(M(B))\,\mu_d(dB)>0,
\end{align*} 
where the strict inequality follows from the facts that equilibrium cherry-picking occurs with strictly positive probability---$\mu_d(\{B\in \basket:t(B)>\overline\tau_d\})>0$---and that $\pi_R(M(B))>0$ whenever $t(B)>0$. 

For parts (b) and (c), we additionally assume compactness and \Cref{Assumption-Conflict,Assumption-BreakEven}. For part (b), suppose that all policies are viable. Continuity and compactness deliver that
\begin{align*}
\pi_P^*=\max_{x\in\policy}\pi_P(x)<\infty.
\end{align*}
We first show that every feasible pair $(x,\alpha)$ can be implemented, and then show that no commitment gives the responder more than the program's maximum. 

Take any pair $(x,\alpha)$ satisfying (O), and let the responder approve only $x$ when informed and accept with probability $\alpha$ when uninformed. The proposer obtains $(\lambda+(1-\lambda)\alpha)\pi_P(x)$ from $x$, whereas every deviation gives him at most $(1-\lambda)\alpha\pi_P^*$. Thus, $x$ is a best reply by (O), and the responder obtains the program's objective. The program's value is strictly positive because, for any $x\in\congruence$, the feasible pair $(x,0)$ gives her $\lambda\pi_R(x)>0$. The feasible set is nonempty and closed in compact $\policy\times[0,1]$, so the continuous objective attains a maximum that can be implemented. 

We now show that no commitment strategy can deliver a higher payoff. Consider a commitment strategy $(\gamma,\alpha)$ and any proposer best reply $x$. Because he can choose any $x'\in\policy$,
\begin{align*}
(\lambda+(1-\lambda)\alpha)\pi_P(x)
\ge(\lambda\gamma(x)+(1-\lambda)\alpha)\pi_P(x)\ge(\lambda\gamma(x')+(1-\lambda)\alpha)\pi_P(x')
\ge(1-\lambda)\alpha\pi_P(x').
\end{align*}
Since the proposer can choose $x'$ so that $\pi_P(x')=\pi_P^*$, $(x,\alpha)$ must satisfy (O). If $\pi_R(x)>0$, her actual payoff is at most the program's objective at $(x,\alpha)$; otherwise, her payoff is nonpositive and hence below the program's strictly positive value. Thus, no commitment strategy delivers more than the program's value. This proves part (b).

For part (c), we first show that large baskets allow the responder to approximate the value in (b). Fix a small $\epsilon>0$. Choose a pair $(x,\alpha)$ whose objective exceeds $\overline\Pi_R^\infty(\lambda)-\varepsilon$ and whose obedience constraint holds strictly. By continuity, there is an open neighborhood $U\subseteq\congruence$ of $x$ such that, for every $x'\in U$, $(x',\alpha)$ satisfies (O) strictly and the responder obtains a payoff that exceeds $\overline\Pi_R^\infty(\lambda)-2\varepsilon$. Consider a commitment to $(\1_U,\alpha)$. For any basket $B$ such that $B\cap U\ne\varnothing$, every proposer best reply lies in $U$: choosing a policy there gives him more than $(1-\lambda)\alpha\pi_P^*$, an upper bound on his payoff from a policy outside $U$. On the remaining baskets, we bound the responder's payoff from below by $-\max_{z\in\policy}|\pi_R(z)|>-\infty$, given that $\policy$ is compact. Therefore,
\begin{align*}
\overline\Pi_R^d(\lambda)\ge \mu_d(B\cap U\ne\varnothing)\left(\overline\Pi_R^\infty(\lambda)-2\varepsilon\right)-\mu_d(B\cap U=\varnothing)\max_{z\in\policy}|\pi_R(z)|.
\end{align*}
Because $\phi(U)>0$, \Cref{Lemma-LargeBasketAbsence} implies that $\mu_d(B\cap U\ne\varnothing)\rightarrow 1$. So taking $d\to\infty$ and then taking $\epsilon\rightarrow 0$ delivers $\liminf_{d\to\infty}\overline\Pi_R^d(\lambda)\ge\overline\Pi_R^\infty(\lambda)$.

We now bound the value from above. Fix $\varepsilon>0$. By compactness and continuity, there exists $\eta>0$ such that every pair $(x,\alpha)\in\policy\times[0,1]$ satisfying
\begin{align*}
(\lambda+(1-\lambda)\alpha)\pi_P(x)\ge(1-\lambda)\alpha(\pi_P^*-\eta)
\end{align*}
gives $(\lambda+(1-\lambda)\alpha)\pi_R(x)\le\overline\Pi_R^\infty(\lambda)+\varepsilon$. Consider any commitment strategy $(\gamma,\alpha)$ and any proposer best reply $x\in B$. Since he can choose $C(B)$,
\begin{align*}
(\lambda+(1-\lambda)\alpha)\pi_P(x)
&\ge(\lambda\gamma(x)+(1-\lambda)\alpha)\pi_P(x)
\ge(1-\lambda)\alpha\pi_P(C(B)).
\end{align*}
If $\pi_P(C(B))\ge\pi_P^*-\eta$, the pair $(x,\alpha)$ satisfies the relaxed constraint above. As in part (b), the responder's actual payoff is at most the objective at this pair if $\pi_R(x)>0$, and is nonpositive otherwise. Thus, her payoff is at most $\overline\Pi_R^\infty(\lambda)+\varepsilon$ on these baskets, including when the proposer mixes over best replies. On the remaining baskets, her payoff is at most $\max_{x'\in\policy}|\pi_R(x')|$. Therefore,
\begin{align*}
\overline\Pi_R^d(\lambda)
\le\mu_d\left(\pi_P(C(B))\geq \pi_P^*-\eta\right)\left(\overline\Pi_R^\infty(\lambda)+\varepsilon\right)+\mu_d\left(\pi_P(C(B))<\pi_P^*-\eta\right)\max_{x'\in\policy}|\pi_R(x')|.\notag
\end{align*}

Because $\phi\left(\{x'\in\policy:\pi_P(x')>\pi_P^*-\eta\}\right)$ is a non-empty open set, it follows from \Cref{Lemma-LargeBasketAbsence} that as $d\rightarrow \infty$, $\mu_d\left(\pi_P(C(B))\geq \pi_P^*-\eta\right)\to 1$. Therefore, taking $d\to\infty$ and then $\varepsilon\to0$ delivers the desired upper bound. 
\end{proof}

\subsection{Proposer Commitment}\label{Appendix-ProposerCommitment}

\begin{proof}[Proof of \Cref{Theorem-ProposerCommitment}]
Fix $\lambda\in(0,1)$. Part (a) uses the baseline assumptions for a fixed basket-size distribution $\psi_d$. Suppose first that $\lambda<\overline\lambda_d$, and let $\sigma^*$ denote the proposer's equilibrium strategy. We know from  \Cref{Theorem-UniqueEqm} that the uninformed responder mixes and therefore her expected payoff $\E_d[\pi_R(\sigma^*)]$ equals zero. Now suppose that the proposer commits to the mixed strategy that proposes $\sigma^*(\cdot\mid B)$ with probability $(1-\epsilon)$ and proposes $M(B)$ with complementary probability, where $\varepsilon>0$ is small. Because $\E_d[\pi_R(M)]>0$---see \Cref{Lemma-ModerationPositive}---the uninformed responder strictly prefers to accept. Taking $\varepsilon\to0$, the proposer's gain relative to equilibrium converges to
\begin{align*}
(1-\lambda)(1-\alpha_\lambda^d)\E_d\left[\sum_{x\in B}\sigma^*(x\mid B)\pi_P(x)\right]>0.
\end{align*}
Therefore, the proposer strictly benefits from commitment.

Suppose instead that $\lambda\ge\overline\lambda_d$. The uninformed responder already accepts for sure in equilibrium, so the proposer selects a maximizer of $\hat\pi_P$ in each basket. Under any commitment, his payoff from a selected policy $x$ is at most $\hat\pi_P(x)$, since uninformed acceptance cannot exceed one and $\pi_P(x)\ge0$. Hence, no commitment gives him more than $\E_d\max_{x\in B}\hat\pi_P(x)=\Pi_P^d(\lambda)$, which is his equilibrium payoff.

For parts (b) and (c), we additionally assume compactness, \Cref{Assumption-Conflict,Assumption-BreakEven}, and the boundary condition $\policy\setminus\conflict\subseteq\cl(\congruence)$. For part (b), suppose that all policies are viable. The boundary condition and continuity imply that every policy with $\pi_R(x)=0$ gives the proposer at most $\pi_P^+$. Since $\policy\setminus\conflict$ is compact and contains $\congruence$, $\max_{x\in\policy\setminus\conflict}\pi_P(x)=\pi_P^+>0$.
Therefore, the proposer's best pure-strategy commitment gives him a payoff of $\pi_P^+$. Now take any lottery $\nu$ satisfying (P), and let the proposer commit to it. The uninformed responder accepts, so the proposer obtains $\int_\policy\hat\pi_P(x)\,\nu(dx)$, the program's objective. By contrast, a lottery that the uninformed responder rejects gives him at most $\lambda\pi_P^+<\pi_P^+$. Thus, no commitment gives him more than the program's value.

We now establish that an optimal solution exists and that it can take the form described in the theorem. Observe that the feasible set, $\{\nu\in \Delta(\policy):\int_\policy \pi_R(x)\nu(dx)\geq 0 \}$, is nonempty and weakly compact because $\policy$ is compact and $\pi_R$ is continuous. The function $\hat\pi_P$ is bounded and upper semicontinuous. Therefore, the Portmanteau theorem implies that the objective is upper semicontinuous, so the program attains a maximum.

Because the objective $\int_{\policy} \hat\pi_P(x) \nu(dx)$ is affine in $\nu$ and there is only one persuasion constraint, we can choose an optimal lottery supported on at most two policies.\footnote{By Bauer's maximum principle, we can choose an optimal lottery that is an extreme point of the feasible set. Because (P) is a single moment inequality, \citet[Theorem~2.1]{winkler1988extreme} implies that this lottery is supported on at most two policies.} If the value equals $\pi_P^+$, the pure maximizer above is optimal. If the value exceeds $\pi_P^+$, an optimal two-policy lottery includes a harmful policy $x_{-}$ such that $\hat\pi(x_{-})>\pi_P^+$ and a mutually beneficial policy $x_{+}$ to satisfy (P). At an optimum, (P) binds, proving part (b).

For part (c), suppose that the basket grows large. Observe that $\widehat\Pi_P^d(\lambda)\le\widehat\Pi_P^\infty(\lambda)$: every lottery induced by a commitment strategy under distribution $\psi_d$ can also be chosen when all policies are viable. Therefore, it suffices to show that large baskets allow the proposer to approximate the value in (b). Fix a small $\epsilon >0$. If the commitment payoff in (b), $\widehat\Pi_P^\infty(\lambda)$, equals $\pi_P^+$, choose a policy $x_{+}\in\congruence$ that gives the proposer more than $\pi_P^+-\varepsilon$ and assign it probability $p_{+}=1$; set $p_{-}=0$ and choose $x_{-}\in\conflict$ arbitrarily. Otherwise, use the two policies from the optimal lottery in (b), and slightly increase the weight on $x_{+}$; denote the resulting probabilities by $p_{+}$ and $p_{-}=1-p_{+}$. In either case, the resulting commitment strategy when all policies are viable gives the responder a strictly positive payoff from accepting and the proposer more than $\widehat\Pi_P^\infty(\lambda)-\varepsilon$. By continuity of both payoff functions, there exists $\eta >0$ and open neighborhoods $U_{+}\subseteq\congruence$ and $U_{-}\subseteq\conflict$ of $x_{+}$ and $x_{-}$, respectively, such that every choice of $x'_{+}\in U_{+}$ and $x'_{-}\in U_{-}$ satisfies
\begin{align*}
p_{+}\pi_R(x'_{+})+p_{-}\pi_R(x'_{-})>\eta\quad \text{ and }\quad
p_{+}\hat\pi_P(x'_{+})+p_{-}\hat\pi_P(x'_{-})>\widehat\Pi_P^\infty(\lambda)-2\varepsilon.
\end{align*}

Let $\mathcal E\equiv\{B\in\mathcal B:B\cap U_{+}\ne\varnothing\text{ and }B\cap U_{-}\ne\varnothing\}$ be the event that the basket intersects both neighborhoods. Consider the following commitment strategy $\sigma$: on event $\mathcal E$, propose $x'_{+}\in B\cap U_{+}$ with probability $p_{+}$ and $x'_{-}\in B\cap U_{-}$ with probability $p_{-}$; on the remaining baskets, propose $C(B)$. Both $U_{+}$ and $U_{-}$ are nonempty and open, so full support and \Cref{Lemma-LargeBasketAbsence} imply that $\mu_d(\mathcal E)\to1$. The responder's expected payoff from accepting satisfies
\begin{align*}
\E_d[\pi_R(\sigma)]
&\ge\eta\,\mu_d(\mathcal E)
-\max_{x\in\policy}|\pi_R(x)|\,\mu_d(\mathcal E^c),
\end{align*}
whose right-hand side converges to $\eta>0$ as $d\to\infty$. The uninformed responder therefore accepts for all sufficiently large $d$. Given that $\hat\pi_P(x)\ge0$ for all $x$, $\widehat\Pi_P^d(\lambda)
\ge\mu_d(\mathcal E)\left(\widehat\Pi_P^\infty(\lambda)-2\varepsilon\right)$.
Taking $d\to\infty$ and then letting $\varepsilon\to0$ yields $\liminf_{d\to\infty}\widehat\Pi_P^d(\lambda)\ge\widehat\Pi_P^\infty(\lambda)$.

\end{proof}
\subsection{Menus}\label{Appendix-Menus}
Given a basket that has $K$ policies, where the proposer must offer $n_K$ of them, we use $\E_K$ to denote the expectation over the draws and any randomization over menus. For a nonempty menu $S\subseteq B$, let $\bar\pi_i(S)\equiv |S|^{-1}\sum_{x\in S}\pi_i(x)$ denote player $i$'s average payoff from items in that menu. The uninformed responder's payoff from accepting is $\E_K[\bar\pi_R(S)]$. Let $I(S)$ be the informed responder's choice from $S$ and the status quo. If the uninformed responder accepts with probability $\alpha$, the proposer chooses a menu to maximize $\lambda\pi_P(I(S))+(1-\lambda)\alpha\bar\pi_P(S)$. 

\subsubsection{Proof of \Cref{Theorem-Menu}}\label{Appendix-MenuTheorem}

We begin with a claim about the proposer's best reply when the uninformed responder accepts with positive probability. For this claim, fix $K$ and $n\in\{2,\ldots,K\}$, and let $T_n(B)$ contain his favorite $n$ policies. Choose $\tilde\pi_P$ and $\delta>0$ as in \Cref{Assumption-Conflict}, and put $D\equiv\{x:\pi_P(x)\ge\tilde\pi_P\}$ and $p\equiv\phi(D)$. Every policy in $D$ gives the proposer more than any policy outside $D$ and gives the responder at most $-\delta$. Observe that $p\in (0,1)$.

\begin{claim}\label{Claim-MenuChoice}
For every $\alpha>0$ and every menu $S$ of size $n$ that maximizes the proposer's expected payoff on a regular basket, the following hold:
\begin{enumerate}[label=\emph{(\alph*)}]
\item The responder's average payoff satisfies
\begin{align}\label{Equation-MenuFavoriteBound}
\bar\pi_R(S)\le\bar\pi_R(T_n(B))+\frac2n\max_{x\in B}|\pi_R(x)|.
\end{align}
If the proposer's favorite $n$ policies are all harmful, then $S=T_n(B)$ or $S=T_{n-1}(B)\cup\{M(B)\}$, where the latter requires a mutually beneficial policy in $B$.
\item For any proposer strategy selecting such menus, the responder's expected gain from choosing when informed satisfies
\begin{align}\label{Equation-MenuSurplusGap}
\E_K[\pi_R(I(S))-\bar\pi_R(S)]\ge\frac{\delta p}{2}.
\end{align}
Consequently, in every PBE with positive uninformed acceptance, the responder's ex ante payoff is at least $\lambda\delta p/2$.
\end{enumerate}
\end{claim}
\begin{proof}
For part (a), suppose first that $T_n(B)$ contains a mutually beneficial policy. We show that $\bar\pi_R(S)\le\bar\pi_R(T_n(B))$. The conclusion is immediate if $S=T_n(B)$. Otherwise, because $T_n(B)$ uniquely maximizes the proposer's payoff when an uninformed responder accepts, an optimal $S$ must give him a strictly higher informed payoff. This requires $I(S)\in T_n(B)$, since every policy outside $T_n(B)$ gives him less than $I(T_n(B))$, and the status quo gives him zero. If an omitted policy $y\in T_n(B)\setminus S$ gave the responder less than $I(S)$, replacing any $x\in S\setminus T_n(B)$ with $y$ would preserve her informed choice and strictly increase the proposer's payoff from uninformed acceptance. Therefore, every removed policy gives her more than $I(S)$, whereas every added policy gives her less. Since the swapped sets have the same size, $\bar\pi_R(S)\le\bar\pi_R(T_n(B))$.

Suppose instead that $T_n(B)\subseteq\conflict$. We show that an optimal menu can differ from $T_n(B)$ only by replacing its least profitable policy with $M(B)$. If the informed responder rejects $S$, optimality requires $S=T_n(B)$. If she accepts, the other $n-1$ entries must be the proposer's favorite $n-1$ policies. Otherwise, swapping a less profitable entry other than $I(S)$ with an omitted favorite would preserve her informed choice, since the favorite is harmful, and strictly increase his payoff from uninformed acceptance. The sole mutually beneficial policy must then be $M(B)$, which maximizes both his informed payoff and its contribution to the menu average. This replaces one entry of $T_n(B)$, changing the responder's average by at most $2\max_{x\in B}|\pi_R(x)|/n$ and establishing \eqref{Equation-MenuFavoriteBound}.

For part (b), write $N_K^D\equiv|B\cap D|$. The menu must include at least $\min\{n-1,N_K^D\}$ policies in $D$. Otherwise, an omitted policy in $D$ could replace an entry outside $D$ other than the informed choice. This replacement preserves that choice and strictly raises the proposer's payoff from uninformed acceptance. Since $\pi_R(I(S))\ge\max\{0,\pi_R(x)\}$ for every $x\in S$ and $\pi_R(x)\le-\delta$ for $x\in S\cap D$, her gain from choosing when informed satisfies
\begin{align*}
\pi_R(I(S))-\bar\pi_R(S)
&\ge\delta\frac{|S\cap D|}{n}
\ge\frac{\delta N_K^D}{2K},
\end{align*}
where the last inequality uses $\min\{n-1,N_K^D\}/n\ge((n-1)/n)(N_K^D/K)\ge N_K^D/(2K)$. Taking expectations and using $\E_K[N_K^D]=Kp$ gives \eqref{Equation-MenuSurplusGap}. In a PBE with $\alpha>0$, the uninformed responder's willingness to accept implies $\E_K[\bar\pi_R(S)]\ge0$, so her ex ante payoff satisfies
\begin{align}\label{Equation-MenuPositiveSurplus}
\lambda\E_K[\pi_R(I(S))]+(1-\lambda)\alpha\E_K[\bar\pi_R(S)]
\ge\frac{\lambda\delta p}{2}>0.
\end{align}
\end{proof}

\begin{proof}[Proof of \Cref{Theorem-Menu}]
For $b\in(0,1]$, let $H(b)$ be the responder's average payoff from the proposer's favorite fraction $b$ of policies, and let $L(b)$ be her average payoff from her own worst fraction $b$. Formally,
\begin{align*}
H(b)\equiv\frac1b\int_{A_b}\pi_R(x)\,\phi(dx),\qquad
L(b)\equiv\frac1b\int_{D_b}\pi_R(x)\,\phi(dx),
\end{align*}
where, for $b<1$, $A_b\equiv\{\pi_P\ge q_P(b)\}$ and $D_b\equiv\{\pi_R\le q_R(b)\}$, with the cutoffs chosen so that $\phi(A_b)=\phi(D_b)=b$. Our assumptions guarantee that these cutoffs are unique. Set $A_1=D_1=\policy$, so $H(1)=L(1)=\E_\phi[\pi_R(X)]>0$.

The sets $A_b$ and $D_b$ expand with $b$, and the probability of the added policies equals the increase in $b$. Integrability therefore implies that both averages are continuous on $(0,1]$. Expanding $D_b$ adds policies that give the responder strictly more than those already included, so $L$ is strictly increasing. Moreover, $L(b)\le H(b)$ because $D_b$ contains the policies that give her the least. Recall that every policy in $D$ gives her at most $-\delta$. Because $A_b\subseteq D$ for $b\le p$, we have $L(b)\le H(b)\le-\delta$ on $(0,p]$. Since $L(1)>0$, continuity and strict monotonicity imply that $L$ has a unique zero $\overline b\in(p,1)$. Define $\underline b\equiv\inf\{b\in(0,1):H(b)\ge0\}$. Because $H$ is continuous, is negative on $(0,p]$, and satisfies $H(\overline b)\ge L(\overline b)=0$, this cutoff satisfies $p<\underline b\le\overline b<1$ and $H(\underline b)=0$. By definition, $H(b)<0$ for every $b<\underline b$. 

To relate these averages to finite baskets, let $W_{n_K}(B)$ contain the $n_K$ policies with the lowest responder payoffs. If $n_K/K\to b\in(0,1]$, then
\begin{align}\label{Equation-MenuSampling}
\E_K[\bar\pi_R(T_{n_K}(B))] \rightarrow H(b), \quad \E_K[\bar\pi_R(W_{n_K}(B))] \rightarrow L(b), \quad \frac1K\E_K\max_{x\in B}|\pi_R(x)| \rightarrow 0.
\end{align}
The selected policies differ from those above or below the population cutoff by a fraction that vanishes in expectation. Truncating payoffs and using integrability gives the first two limits and the bound on the maximum.\footnote{For $(E_b,Q_{n_K})=(A_b,T_{n_K})$ or $(D_b,W_{n_K})$, let $N_K$ count draws in $E_b$. The two selections are nested and differ by $|N_K-n_K|$ policies. For every $c>0$,
\begin{align*}
\E_K\left|\frac1K\sum_{x\in Q_{n_K}(B)}\pi_R(x)-\frac1K\sum_{j=1}^K\pi_R(X_j)\1_{\{X_j\in E_b\}}\right|
\le c\left(\sqrt{\frac{b(1-b)}K}+\left|b-\frac{n_K}K\right|\right)+\E_\phi[|\pi_R(X)|\1_{\{|\pi_R(X)|>c\}}].
\end{align*}
Taking $K\to\infty$ and then $c\to\infty$, and dividing by $n_K/K\to b>0$, gives the first two limits. The last follows from $K^{-1}\E_K\max_{x\in B}|\pi_R(x)|\le c/K+\E_\phi[|\pi_R(X)|\1_{\{|\pi_R(X)|>c\}}]$.}

For part (a), choose $\overline b<b<\liminf_K n_K/K$. Every menu gives the responder at least the average payoff of her worst menu of the same size. This worst average increases with menu size, so for all sufficiently large $K$,
\begin{align*}
\E_K[\bar\pi_R(S)]
&\ge\E_K[\bar\pi_R(W_{n_K}(B))]
\ge\E_K[\bar\pi_R(W_{\lceil bK\rceil}(B))]\rightarrow L(b)>0,
\end{align*}
where the limit follows from \eqref{Equation-MenuSampling}. The uninformed responder therefore accepts with probability one. A proposer best reply to $\alpha=1$, together with optimal informed choice and the induced beliefs, constitutes a PBE.

For part (b), suppose first that $n_K/K\to b\in(0,\underline b)$. By \eqref{Equation-MenuFavoriteBound} and \eqref{Equation-MenuSampling}, every best reply to $\alpha>0$ satisfies
\begin{align}\label{Equation-MenuNarrowLimit}
\E_K[\bar\pi_R(S)]
\le\E_K[\bar\pi_R(T_{n_K}(B))]+\frac2{n_K}\E_K\max_{x\in B}|\pi_R(x)|
\rightarrow H(b)<0.
\end{align}
The upper bound is independent of the acceptance probability and the menu rule. Therefore, for all sufficiently large $K$, the uninformed responder strictly prefers to reject any best reply to positive acceptance, ruling out $\alpha>0$ in equilibrium.

Suppose instead that $n_K/K\to0$. For any best reply to positive uninformed acceptance, \Cref{Claim-MenuChoice}(b) gives a strictly positive lower bound on the responder's expected gain from choosing when informed. We show that her expected payoff when informed converges to zero, so accepting when uninformed must give her a strictly negative payoff. For this bound, fix $K$ and any menu size $2\le n\le pK/2$. Recall that $m_K\equiv\E_K[(\pi_R(M(B)))^+]\to0$ by \Cref{Lemma-ModerationSurplus}. When $N_K^D\ge n$, the proposer's favorite $n$ policies lie in $D$. By \Cref{Claim-MenuChoice}(a), an optimal menu is either entirely harmful or consists of $M(B)$ and his favorite $n-1$ policies. Hence $\pi_R(I(S))\le(\pi_R(M(B)))^+$ on these baskets. On the remaining baskets, her informed payoff is at most the sum of the positive responder payoffs from all $K$ policies. Therefore,
\begin{align*}
\E_K[\pi_R(I(S))]
&\le m_K+\E_K\left[\sum_{j=1}^K(\pi_R(X_j))^+\1_{\{N_K^D<n\}}\right]\\
&=m_K+K\E_\phi[(\pi_R(X))^+]\Pr\{\operatorname{Bin}(K-1,p)<n\}\\
&\le m_K+K\E_\phi[(\pi_R(X))^+]e^{-(K-1)p^2/2}\rightarrow0,
\end{align*}
uniformly over $2\le n\le pK/2$. The equality uses independence and the fact that a draw giving the responder a positive payoff lies outside $D$. The final bound follows from Hoeffding's inequality.\footnote{Since $n-1\le p(K-1)/2$, \citet[Theorem~1]{hoeffding1963probability} gives $\Pr\{\operatorname{Bin}(K-1,p)<n\}\le e^{-(K-1)p^2/2}$.} Since $n_K\le pK/2$ for all sufficiently large $K$, \eqref{Equation-MenuSurplusGap} implies
\begin{align*}
\E_K[\bar\pi_R(S)]
\le\E_K[\pi_R(I(S))]-\frac{\delta p}{2}<0
\end{align*}
for all sufficiently large $K$, ruling out positive uninformed acceptance.

Now suppose that $\limsup_K n_K/K<\underline b$. If positive acceptance occurred in equilibrium for infinitely many $K$, those menu fractions would have a convergent subsequence with limit $b\in[0,\underline b)$. This contradicts \eqref{Equation-MenuNarrowLimit} if $b>0$, and the preceding bound if $b=0$. Thus, every PBE has $\alpha=0$ for all sufficiently large $K$.

To establish existence, choose $S^0(B)$ to maximize $\pi_P(I(S))$ and, among those maximizers, to maximize $\bar\pi_P(S)$. This is a best reply to $\alpha=0$ and, on each regular basket, to all sufficiently small positive $\alpha$.\footnote{There are finitely many menus, so a strict loss in informed proposer payoff outweighs any gain from uninformed acceptance when $\alpha$ is sufficiently small. This bound may depend on the basket.} The menu characterization and pointwise bounds in the proof of \Cref{Claim-MenuChoice} therefore apply to $S^0$ on each regular basket. Taking expectations yields \eqref{Equation-MenuSurplusGap} for $S^0$, and the preceding bounds imply $\E_K[\bar\pi_R(S^0)]<0$ for all sufficiently large $K$. The rule $S^0$, uninformed rejection, optimal informed choice, and the induced beliefs constitute a PBE.

Finally, if rankings oppose, $A_b$ and $D_b$ coincide up to null sets. Otherwise, their equal probabilities give policies $x\in A_b\setminus D_b$ and $y\in D_b\setminus A_b$ with both $\pi_P(x)>\pi_P(y)$ and $\pi_R(x)>\pi_R(y)$, contradicting opposing rankings. Hence $H=L$ and $\underline b=\overline b$.
\end{proof}

\subsubsection{Proof of \Cref{Proposition-MenuWelfare}}\label{Appendix-MenuWelfare}

\begin{proof}
Write $\Pi_i^K(\lambda)$ and $\alpha_\lambda^K$ for the single-proposal equilibrium payoffs and uninformed acceptance with a basket of size $K$. By \Cref{Corollary-LargePayoff}(a), $\Pi_R^K(\lambda)\to0$.

For part (a), \Cref{Theorem-Menu}(b) implies that every menu PBE has $\alpha=0$ for all sufficiently large $K$. The informed responder either rejects or chooses a mutually beneficial policy. Therefore, the proposer's equilibrium payoff cannot exceed that from proposing $M(B)$. In the single-proposal game, proposing $M(B)$ also gives him the strictly positive payoff $(1-\lambda)\alpha_\lambda^K\E_K[\pi_P(M(B))]$ from uninformed acceptance, since $\alpha_\lambda^K>0$ by \Cref{Theorem-UniqueEqm}. Because this is a feasible proposal strategy, his equilibrium payoff strictly exceeds his payoff from the menu.

For part (b), \Cref{Claim-MenuChoice}(b) implies that, in every menu PBE with $\alpha>0$, the responder's payoff is at least $\lambda\delta p/2$. Since $\Pi_R^K(\lambda)\to0$, she strictly prefers such an equilibrium to the single-proposal benchmark for all sufficiently large $K$. By \Cref{Theorem-Menu}(a), every menu PBE has $\alpha=1$ for all sufficiently large $K$ under condition (i), establishing that case.

Under condition (ii), suppose that $\odds_\lambda<t_\infty$. It remains to consider menu PBE with $\alpha=0$. By \Cref{Theorem-UniqueEqm,Theorem-LargeBasket}, $\alpha_\lambda^K\in(0,1)$ for all sufficiently large $K$. The proof of \Cref{Theorem-ResponderCommitment}(a) then gives $\Pi_R^K(\lambda)<\lambda m_K$.

We claim that $\pi_R(I(S))\ge(\pi_R(M(B)))^+$ on almost every basket. If this inequality failed, $M(B)$ would be mutually beneficial and would give the responder more than every policy in $S$. Replacing her informed choice, or any entry if she rejects, with $M(B)$ would make it her informed choice. This strictly benefits the proposer because $M(B)$ is his unique favorite mutually beneficial policy and gives him more than the status quo, contradicting optimality. Therefore, the responder's menu payoff satisfies
\begin{align*}
\lambda\E_K[\pi_R(I(S))]\ge\lambda m_K>\Pi_R^K(\lambda).
\end{align*}
Under condition (ii), the required basket size is common to all $n\in\{2,\ldots,K\}$ and all menu PBE for the fixed $\lambda$.
\end{proof}

\subsection{Effort}\label{Appendix-Effort}

\begin{proof}[Proof of \Cref{Proposition-Effort}]
We first claim that in every PBE, $q>0$. Sequential rationality requires the informed responder to accept every mutually beneficial proposal, including after a deviation. Therefore, the proposer can obtain a strictly positive payoff from  proposing $M(B)$ whenever a mutually beneficial policy is viable. His expected payoff from doing so is at least $\lambda\E\bigl[\pi_P(M(B))\1_{\{\pi_R(M(B))>0\}}\bigr]>0$. Because $c(q)/q\to0$ as $q\rightarrow 0$, a sufficiently small positive effort gives him a strictly positive payoff, which implies the claim.

Conditional on identifying a basket, its distribution is $\mu$. In the continuation play, the proposer chooses a best response in every basket, and the responder chooses a best response to the resulting proposals. By \Cref{Theorem-UniqueEqm}, the continuation payoffs are $\Pi_i(\lambda)$. The proposer's effort choices consequently solve $\max_{q\in[0,1]}\{q\Pi_P(\lambda)-c(q)\}$, and every solution, followed by the equilibrium continuation play, constitutes an equilibrium of the effort game.\footnote{If effort is observed, the belief restriction in PBE implies that the conditional basket distribution remains $\mu$ after every positive effort, including deviations. \Cref{Theorem-UniqueEqm} therefore delivers the same continuation payoffs.} Continuity of $c(\cdot)$ ensures that a solution exists.

For part (a), \Cref{Proposition-Education} and \eqref{Equation-Proportionality} imply that $q\Pi_P(\lambda)-c(q)$ has strict increasing differences in $(q,\lambda)$ on $(0,\overline\lambda]$ and in $(q,-\lambda)$ on $[\overline\lambda,1)$. Thus, every selection of optimal effort increases below the cutoff and decreases above it. 

For part (b), fix $\lambda$ and $\lambda'$ such that $\Pi_P(\lambda')>\Pi_P(\lambda)$, and let $q$ be any optimal effort at $\lambda$. Because $q>0$, the proposer gains $q[\Pi_P(\lambda')-\Pi_P(\lambda)]>0$ by choosing the same effort at $\lambda'$ as at $\lambda$. Because he can retain this effort at $\lambda'$, his optimal effort choice must deliver a strictly higher payoff than at $\lambda$. By \Cref{Proposition-Education} and \eqref{Equation-Proportionality}, it follows that his equilibrium payoff strictly increases on $(0,\overline\lambda]$ and strictly decreases on $[\overline\lambda,1)$.

Finally, the responder's payoff is $q^*(\lambda)\Pi_R(\lambda)$, where $q^*(\lambda)>0$ is expected equilibrium effort. Part (a) implies that $q^*(\lambda)$ increases on $(0,\overline\lambda]$, while \Cref{Proposition-Education} and \eqref{Equation-Proportionality} establish that $\Pi_R(\lambda)$ is strictly positive and strictly increasing on this interval. Their product therefore strictly increases, proving part (c).
\end{proof}

\subsection{Search}\label{Appendix-Search}

\begin{proof}[Proof of \Cref{Proposition-OptimalStopping}]
We first determine the proposer's stopping rule for a fixed acceptance probability, then find the uninformed responder's best response to the resulting proposals. Fix the uninformed responder's acceptance probability $\alpha\in[0,1]$, and let $v>0$ solve \eqref{Equation-DiscountReservation}. This solution exists uniquely because the left side strictly increases from zero, while the right side is continuous, nonincreasing, positive at zero, and converges to zero as $v\rightarrow\infty$. Let $y$ be the highest expected payoff from proposing a policy he has retained, with $y=0$ before the first arrival. We show that his continuation value is $\max\{y,v\}$.

The reservation equation implies
\begin{align}\label{Equation-SearchWaiting}
\begin{aligned}
\frac{\eta}{\eta+r}\E\max\{y,v,\hat\pi_P(X,\alpha)\}
&=v &&\text{if }y\le v,\\
\frac{\eta}{\eta+r}\E\max\{y,v,\hat\pi_P(X,\alpha)\}
&\le\frac{\eta y+rv}{\eta+r}<y &&\text{if }y>v.
\end{aligned}
\end{align}
The left side is the expected discounted payoff from waiting for the next arrival and then receiving the larger of $v$ and the payoff from his best retained policy after that arrival. This payoff is at most $\max\{y,v\}$ in both cases. Induction and a limiting argument show that no stopping strategy gives him more than $\max\{y,v\}$, including strategies that stop between arrivals or randomize over delays.\footnote{Suppose he plans to propose his best retained policy after a delay $h\in[0,\infty]$ if no policy arrives, and his continuation payoff after an arrival is bounded by the larger of $v$ and his best retained payoff. His payoff is then at most a weighted average of $y$ and the left side of \eqref{Equation-SearchWaiting}, with weights $e^{-(\eta+r)h}$ and $1-e^{-(\eta+r)h}$, respectively. Both terms are at most $\max\{y,v\}$. Averaging covers random delays, and abandoning gives zero. Induction on the number of arrivals he may wait for gives the bound for every strategy that stops by a fixed arrival. For an arbitrary strategy, truncate it by abandoning at the $k$th arrival if search continues. Let $T_j$ and $X_j$ denote future arrival times, measured from the current history, and their characteristics. On every path of arrivals, the discounted expected payoff from any proposal is at most $y+\sum_{j\ge1}e^{-rT_j}\hat\pi_P(X_j,\alpha)$, whose expectation is at most $y+\eta/r$ because $\hat\pi_P(X,\alpha)\le X$ and $\E[X]=1$. As $k\to\infty$, the truncated payoffs converge almost surely to the original payoff, so dominated convergence gives the same bound.}

The proposer attains this bound by proposing his best retained policy immediately if $y\ge v$ and otherwise waiting for the first arrival in $A\equiv\{x:\hat\pi_P(x,\alpha)>v\}$, which has positive probability by \eqref{Equation-DiscountReservation}. Writing $T$ for that arrival time and $X_T$ for its characteristic, independence and Poisson arrivals give
\begin{align}\label{Equation-SearchProposalLaw}
\Pr(T\in dt,X_T\in dx)
&=\eta e^{-\eta\phi(A)t}e^{-x}\1_A(x)\,dt\,dx,\\
\E[e^{-rT}\hat\pi_P(X_T,\alpha)]
&=\frac{\eta\int_A\hat\pi_P(x,\alpha)e^{-x}\,dx}{r+\eta\phi(A)}=v.\notag
\end{align}
The last equality follows from the reservation equation. Thus, when $y<v$, waiting gives him $v$ and is strictly better than proposing immediately. When $y>v$, \eqref{Equation-SearchWaiting} makes the delay bound strictly less than $y$ for every positive delay, so he strictly prefers to propose immediately. Since a draw gives expected proposal payoff exactly $v$ with probability zero, every optimal strategy starting with no policy proposes the first policy in $A$, almost surely. In particular, the proposer never abandons search.

We now turn to the responder's inference. By \eqref{Equation-SearchProposalLaw}, the conditional density of the proposed characteristic at every positive proposal time is $e^{-x}\1_A(x)/\phi(A)$. Receiving a proposal therefore tells her that its characteristic lies in $A$, but elapsed time conveys no further information. Her payoff from accepting when uninformed has the sign of $\int_A(2-x)e^{-x}\,dx$.

The uninformed responder cannot always reject in equilibrium. If $\alpha=0$, the reservation equation implies $v<2\lambda$, and the proposer selects only policies in $(v/\lambda,2]$. These proposals give her a strictly positive expected payoff, so she would prefer to accept. For $\alpha>0$, define
\begin{align*}
x_+=\frac{v}{\lambda+(1-\lambda)\alpha},\qquad
x_-=\frac{v}{(1-\lambda)\alpha}>x_+.
\end{align*}
We must have $x_+<2$: otherwise every proposed policy would harm the responder, so she would reject, contradicting $\alpha>0$. Up to endpoints of probability zero, the proposer stops on $[x_+,\infty)$ if $x_-<2$ and on $[x_+,2]\cup[x_-,\infty)$ if $x_-\ge2$. Her payoff from accepting therefore has the sign of
\begin{align}\label{Equation-SearchAcceptance}
\int_A(2-x)e^{-x}\,dx
=\begin{cases}
(1-x_+)e^{-x_+},&x_-<2,\\
e^{-2}+(1-x_+)e^{-x_+}-(x_--1)e^{-x_-},&x_-\ge2.
\end{cases}
\end{align}

For part (a), suppose the uninformed responder accepts for sure. We first show that $x_-<2$. Suppose instead that $\alpha=1$ and $x_-\ge2$. Since $\lambda<1/2$, we have $1<2(1-\lambda)\le x_+<2$ and $x_-=x_+/(1-\lambda)<2x_+$. For $1<x_+<2$,\footnote{For $z\in[1,2]$, let $h(z)=e^{z-2}-(z-1)-(2z-1)e^{-z}$. Since $h''(z)=e^{z-2}+(5-2z)e^{-z}>0$, $h(1)=0$, and $h(2)=-3e^{-2}$, convexity gives $h(z)<0$ for $1<z\le2$. Multiplying by $e^{-z}$ gives the inequality.}
\begin{align}\label{Equation-SearchCutoffComparison}
e^{-2}+(1-x_+)e^{-x_+}<(2x_+-1)e^{-2x_+}.
\end{align}
Because $x_-<2x_+$ and $(x-1)e^{-x}$ strictly decreases on $[2,\infty)$, \eqref{Equation-SearchAcceptance} is strictly negative. The uninformed responder would then prefer to reject, a contradiction.

Thus, $x_-<2$, and her payoff from accepting has the sign of $1-x_+$. Since $\alpha=1$ gives $x_+=v$, she is willing to accept if and only if $v\le1$. The proposer then offers the first policy with $x\ge v$, and the reservation equation becomes $rv=\eta[e^{-v}-3\lambda e^{-2}]$. For $v\in(0,1]$, the implied productivity $rv/(e^{-v}-3\lambda e^{-2})$ strictly increases from zero to $\underline\eta\equiv re^2/(e-3\lambda)$. Hence, full acceptance occurs exactly when $\eta\le\underline\eta$, with a unique cutoff $x_+(\eta)=v\le1$.

For part (b), suppose instead that the uninformed responder mixes and $x_-<2$. The proposer offers the first policy with $x\ge x_+$. Since $X$ is unit exponential, its conditional mean above $x_+$ is $x_++1$, so the responder earns $1-x_+$ on average from accepting. Her indifference requires $x_+=1$, giving $v=\lambda+(1-\lambda)\alpha$. The conditions $\alpha<1$ and $x_-<2$ are then equivalent to $2\lambda<v<1$. The reservation equation gives
\begin{align*}
rv=\eta[ve^{-1}-3\lambda e^{-2}],\qquad
v=\frac{3\eta\lambda}{e(\eta-re)},\qquad
\alpha=\frac{v-\lambda}{1-\lambda}.
\end{align*}
These values satisfy $2\lambda<v<1$ exactly when $\underline\eta<\eta<\overline\eta$, where $\overline\eta\equiv2re^2/(2e-3)>\underline\eta$ because $\lambda<1/2$. Moreover,
\begin{align*}
\frac{dv}{d\eta}=-\frac{3\lambda r}{(\eta-re)^2}<0.
\end{align*}
Since $v$ is the proposer's equilibrium payoff and $\alpha=(v-\lambda)/(1-\lambda)$, both his payoff and the uninformed responder's acceptance probability strictly fall in $\eta$.

For part (c), suppose $x_-\ge2$. Part (a) rules out full acceptance, so the uninformed responder must mix. By \eqref{Equation-SearchAcceptance}, her indifference requires
\begin{align}\label{Equation-SearchIndifference}
e^{-2}+(1-x_+)e^{-x_+}=(x_--1)e^{-x_-}.
\end{align}
The right side is at most $e^{-2}$, so $x_+\ge1$. Both sides strictly decrease continuously from $e^{-2}$ to zero on their respective domains, $x_+\in[1,2)$ and $x_-\in[2,\infty)$. Thus, each $x_-\ge2$ determines exactly one $x_+\in[1,2)$, increasing continuously from one to two as $x_-$ increases from two to infinity. The definitions of the cutoffs then give $\alpha=\frac{\lambda x_+}{(1-\lambda)(x_--x_+)}$ and $v=\frac{\lambda x_+x_-}{x_--x_+}$. By \eqref{Equation-SearchCutoffComparison} and \eqref{Equation-SearchIndifference}, $x_->2x_+$ whenever $x_+>1$, so $0<\alpha<\lambda/(1-\lambda)<1$. At $x_-=2$, indifference gives $x_+=1$, with $v=2\lambda$ and $\alpha=\lambda/(1-\lambda)<1$.

It remains to show that each $\eta\ge\overline\eta$ determines exactly one such pair of cutoffs. Dividing the reservation equation by $\eta v$ yields
\begin{align*}
\frac r\eta
&=\int_{x_+}^2\left(\frac{x}{x_+}-1\right)e^{-x}\,dx
+\int_{x_-}^\infty\left(\frac{x}{x_-}-1\right)e^{-x}\,dx.
\end{align*}
Each integral strictly decreases as its cutoff increases. Because indifference requires $x_+$ to rise with $x_-$, the right side strictly decreases continuously from $e^{-1}-3e^{-2}/2=r/\overline\eta$ to zero as $x_-$ increases from two to infinity. Consequently, each $\eta\ge\overline\eta$ determines a unique pair $x_+(\eta)<2\le x_-(\eta)$ and a unique acceptance probability.

We have shown that every stationary equilibrium must have the stated form. To establish existence, consider the strategies derived in each case, with the informed responder accepting if and only if $x\le2$ and the uninformed responder holding the beliefs given by \eqref{Equation-SearchProposalLaw} at every positive proposal time. The stopping argument shows that the proposer has no profitable deviation after any private history, and the payoff comparisons above show that the responder's behavior is optimal at every proposal time. Thus, these strategies and beliefs constitute a stationary PBE, which is essentially unique.
\end{proof}
\end{document}